\documentclass[structabstract]{aa}   
\usepackage{amssymb,amsmath,mathabx}
\usepackage{graphicx}  
\usepackage{txfonts}
\usepackage{natbib}
\usepackage{color}   
\usepackage{threeparttable}  
\usepackage{booktabs}
\usepackage{wasysym}
\bibpunct{(}{)}{;}{a}{}{,} 

 \def\ProDiMo{{\sc ProDiMo\ }}

\def\Circum{C$_{54}$H$_{18}$}
\def\ra{\rightarrow}
\def\rra{$\rightarrow$}
\def\rla{\rightleftarrows}
\def\Hg{\rm H}
\def\H2{H$_{\rm 2}$}
\def\Dg{\rm D}
\def\elec{\rm e^-}
\def\UV{\rm UV}
\def\nd{n_{\rm d}}
\def\Td{T_{\rm d}}

\def\Tg{T_{\rm g}}
\def\Tq{T_{\rm q}}
\def\Ns{N_{\rm surf}}
\def\surf{_{\rm surf}}
\def\nsite{nb_{\rm site}}
\def\diff{^{\rm diff}_i}
\def\diffj{^{\rm diff}_j}
\def\diffH{^{\rm diff}_{\rm H\varhash}}
\def\diffHc{^{\rm diff}_{\rm *H\varhash}}
\def\diffD{^{\rm diff}_{\rm D\varhash}}
\def\diffDc{^{\rm diff}_{\rm *D\varhash}}

\def\adsH{^{\rm ads}_{\rm H}}
\def\des{^{\rm des}_i}

\def\despg{^{\rm des,pg}_i}
\def\descg{^{\rm des,cg}_i}
\def\desHp{^{\rm des}_{\rm p}}
\def\desDp{^{\rm des}_{\rm D_p}}
\def\desHc{^{\rm des}_{\rm c}}

\def\silHp{^{\rm sil}_{\rm H\varhash}}
\def\actdes{^{\rm des, act}_i}

\def\therdiff{^{\rm diff, th}_i}
\def\Bell{_{\rm Bell}}
\def\Eb{E^{\rm b}_i}
\def\Ec{E^{\rm c}_i}
\def\ED{$\Delta E^{\rm H-D}$}
\def\ECD{$\Delta E^{\rm CH-CD}_{\rm PAH}$}
\def\EHDHH{\Delta E^{\rm H_2-HD}}
\def\EbHp{$E^{\rm b}_{\rm H\varhash}$}
\def\EbDp{$E^{\rm b}_{\rm D\varhash}$}
\def\EbHc{$E^{\rm b}_{*{\rm H\varhash}}$}
\def\EbHcf{E^{\rm b}_{*{\rm H\varhash}}}
\def\EbDc{$E^{\rm b}_{*{\rm D\varhash}}$}
\def\EbPAHH{$E^{\rm b}_{\rm PAH-H}$}
\def\EbPAHD{$E^{\rm b}_{\rm PAH-D}$}
\def\EdH2PAH{$E^{\rm diss}_{\rm H_2,PAH}$}
\def\EdHDPAH{$E^{\rm diss}_{\rm HD,PAH}$}
\def\EdHHchem{$E^{\rm diss}_{\rm H_2,*}$}
\def\EdHDchem{$E^{\rm diss}_{\rm HD,*}$}
\def\Ef{E^{\rm b}_f}
\def\layer{_{\rm layer}}
\def\PAH{{\rm PAH}}

\def\act{^{\rm act}_i}
\def\actH{^{\rm act}_{\rm H}}

\def\actPAHH{^{\rm act}_{\rm PAH-H_x,H}}
\def\actPAHD{^{\rm act}_{\rm PAH-H_x,D}}
\def\actPAHpH{^{\rm act}_{\rm PAH-H_x^{n+},H}}
\def\actPAHpD{^{\rm act}_{\rm PAH-H_x^{n+},D}}
\def\actHpHp{^{\rm act}_{\rm H\varhash,H\varhash}}
\def\actHcHc{^{\rm act}_{\rm *H\varhash,*H\varhash}}
\def\actHcDc{^{\rm act}_{\rm *H\varhash,*D\varhash}}
\def\actHpDp{^{\rm act}_{\rm H\varhash,D\varhash}}
\def\actHpHc{^{\rm act}_{*{\rm H\varhash,H\varhash}}}
\def\actDpHc{^{\rm act}_{*{\rm H\varhash,D\varhash}}}

\def\Eactc{E^{\rm act}_{\rm *}}

\def\Nact{N_{\rm act}}
\def\gc{^{\rm gc}_i}
\def\gcH{^{\rm gc}_{\rm H}}
\def\gcD{^{\rm gc}_{\rm D}}

\def\gp{^{\rm gp}_i}
\def\cg{^{\rm cg}_i}

\def\pgtherm{^{\rm pg, th}_i}
\def\pgph{^{\rm pg, ph}_i}
\def\pgCR{^{\rm pg, CR}_i}
\def\cgtherm{^{\rm cg, th}_i}
\def\cgph{^{\rm cg, ph}_i}
\def\cgCR{^{\rm cg, CR}_i}

\def\ppH{^{\rm pp}_{\rm H\varhash}}
\def\pgH{^{\rm pg}_{\rm H\varhash}}
\def\pcH{^{\rm pc}_{\rm H\varhash}}
\def\pcHmol{^{\rm pc}}
\def\ppHmol{^{\rm pp}}

\def\pcD{^{\rm pc}_{\rm D\varhash}}
\def\cpH{^{\rm cp}_{\rm *H\varhash}}
\def\freq{\nu_{0,i}}

\def\CRdiff{^{\rm diff, CR}_i}

\def\CRdiff{^{\rm diff, CR}_i}
\def\mi{m_i}
\def\rate{\ {\rm s}^{-1}}
\def\ratecoeff{\ {\rm cm}^{3}\ {\rm s}^{-1}}
\def\ratedens{\ {\rm cm}^{-3}\ {\rm s}^{-1}}

\def\nice{n_{\varhash,i}}
\def\nchem{n_{*,i}}
\def\nHchem{n_{*{\rm H\varhash}}}
\def\Hp{_{\rm H\varhash}}

\def\Hpr{{\rm H\varhash}}
\def\Hcr{{\rm *H\varhash}}

\def\HpHc{_{\rm H\varhash,*H\varhash}}

\def\HgHc{_{\rm H,*H\varhash}}
\def\HcHc{_{\rm *H\varhash,*H\varhash}}
\def\Hmolcc{^{\rm cc}_{\rm H_2}}
\def\Hmolpp{^{\rm pp}_{\rm H_2}}

\def\nicetot{n_{\varhash,{\rm tot}}}
\def\nH{n_{\rm H}}
\def\mH{m_{\rm H}}
\def\nHp{n_{\rm p}}
\def\nHc{n_{\rm c}}
\def\nHtot{n_{\langle {\rm H} \rangle}}
\def\Hmol{_{{\rm H}_2}}  
\def\nHmol{n_{{\rm H}_2}}

\def\cm3{\ \rm cm^{-3}}  
\def\sil{^{\rm sil}}  
\def\surfchem{_{\mathrm{surf,chem}}}
\def\Rpc{\alpha_{\rm pc}}
\def\Cedge{${\rm C}_{\rm edge}$}
\def\Cgraph{${\rm C}_{\rm graph}$}
\def\Yes{$\surd$}
\def\1M{{\sc MC-Analytic}}
\def\2M{{\sc MC-Phys}}
\def\3M{{\sc MC-Phys-Chem-PAH}}
\def\4M{{\sc Disk-analytic}}
\def\5M{{\sc Disk-Phys}}
\def\6M{{\sc Disk-Phys-Chem-PAH}}
\def\Ereac{$\Delta E^{\rm reac}$}
\def\a2{r^2}
\def\UVr{h\nu}
\def\UV{$h\nu$} 
\begin{document}

\title{Warm dust surface chemistry}
\subtitle{\H2 and HD formation}
\author{W. F. Thi\inst{1}, S. Hocuk\inst{1,6}, I. Kamp\inst{2}, P. Woitke\inst{3,7}, Ch. Rab\inst{4,1}, S. Cazaux\inst{5}, P. Caselli\inst{1}}
\institute{
Max Planck Institute for Extraterrestrial Physics, Giessenbachstrasse, 85741 Garching, Germany 
\and 
Kapteyn Astronomical Institute, University of Groningen, Postbus 800,
             NL-9700 AV Groningen, The Netherlands
\and             
SUPA, School of Physics \& Astronomy, University of St. Andrews, North Haugh, St. Andrews, KY16 9SS, UK
\and
Institute for Astrophysics, T\"{u}rkenschanzstr.17, A-1180 Vienna, Austria
\and
Faculty of Aerospace Engineering, Delft University of Technology, Delft, The Netherlands
\and
CentERdata, Tilburg University, P.O. Box 90153, 5000 LE, Tilburg, The Netherlands
\and
Centre for Exoplanet Science, University of St Andrews, St Andrews, UK
}
  \abstract
  {Molecular hydrogen (\H2) is the main constituent of the gas in the planet-forming disks that surround many pre-main-sequence stars. \H2 can be incorporated in the atmosphere of the nascent giant planets in disks. Deuterium hydride (HD) has been detected in a few disks and can be considered the most reliable tracer of \H2, provided that its abundance throughout the disks with respect to \H2 is well understood.}
  {We wish to form \H2 and HD efficiently for the varied conditions encountered in protoplanetary disks: the densities  vary from 10$^4$ to 10$^{16}$ $\cm3$; the dust temperatures range from 5 to 1500~K, the gas temperatures go from 5 to a few 1000 Kelvin, and the ultraviolet field can be10$^7$ stronger than the standard interstellar field.}
  {We implemented a comprehensive model of \H2 and HD formation on cold and warm grain surfaces and via hydrogenated polycyclic aromatic hydrocarbons in the physico-chemical code \ProDiMo ({\sc PRO}toplanetary {\sc DI}sk {\sc MO}del). The \H2 and HD formation on dust grains can proceed via the Langmuir-Hinshelwood and Eley-Ridel mechanisms for physisorbed or chemisorbed H (D) atoms.
  \H2 and HD also form by H (D) abstraction from hydrogenated neutral and ionised PAHs and via gas phase reactions.} 
  {\H2 and HD are formed efficiently on dust grain surfaces from 10 to $\sim$~700~K. All the deuterium is converted into HD in UV shielded regions as soon as \H2 is formed by gas-phase D abstraction reactions. The detailed model compares well with standard analytical prescriptions for \H2 (HD) formation. At low temperature, \H2 is formed from the encounter of two physisorbed atoms. HD molecules form on the grain surfaces and in the gas-phase. At temperatures greater than 20~K, the meeting between a weakly bound H- (or D-) atom or a gas-phase H (D) atom and a chemisorbed atom is the most efficient \H2 formation route. H$_2$ formation through hydrogenated PAHs alone is efficient above 80~K. However, the contribution of  hydrogenated PAHs to the overall \H2 and HD formation is relatively low if chemisorption on silicate is taken into account and if a small hydrogen abstraction cross-section is used. The \H2 and HD warm grain surface network is a first step in the construction of a network of high-temperature surface reactions.}   
  {}

   \maketitle
\section{Introduction}\label{introduction}
Molecular hydrogen is the most abundant molecule in virtually every interstellar environment from the Milky Way to high-redshift objects.  Molecular mass estimates are uncertain because direct observations of \H2 are hampered by its homonuclear nature. In addition, the lowest pure-rotation transition of \H2 is not observable from the ground. 
\H2 is also important as one of the major coolants of the warm gas \citep{Shull1978ApJ...220..525S,Flower1986MNRAS.218..729F,Shaw2005_0004-637X-624-2-794,Hollenbach1999_RevModPhys.71.173}. In cold regions, the interactions of \H2 with cosmic rays initiate the efficient ion-neutral chemistry.  

Protoplanetary disk masses derived from CO and isotopologue observations tend to give values that are lower than estimates from dust observations \citep{Miotello2016A&A...594A..85M, Thi2001ApJ...561.1074T}. HD has been used to infer high gas masses in protoplanetary disks (Bergin et al. 2013; McClure et al. 2016). Detailed modeling of the \H2 and CO chemistry is required to determine protoplanetary disk masses.

\H2 is mostly formed on grain surfaces and thorough H-abstraction of hydrogenated polycyclic aromatic hydrocarbons  \citep{WAKELAM20171,Gould1963ApJ...138..393G,Vidali2013_doi:10.1021} because the low density of most interstellar environments excludes \H2 formation by three-body reactions. When a hydrogen atom collides with the dust, it can weakly bound to the surface by the van der Waals forces (a few meV), the interaction is  then of the physisorption type. It can also strongly bound to the surface by the covalent force (eV), and the interaction is called chemisorption. In cold molecular cloud conditions, the dust grains are mostly below 15~K, a temperature low enough for two weakly bound physisorbed hydrogen atoms to stay on the grain surfaces. The surface hydrogens scan the surfaces,  meet each other, and recombine into \H2. In regions with dust temperatures higher than 20~K such as at the surface of photodissociation regions, only chemisorbed hydrogen atoms can remain on the grain surfaces. Hydrogen atoms can also chemically attach to polycyclic aromatic hydrocarbons (PAHs) to form hydrogenated PAHs or to amorphous carbon grains to form hydrogenated amorphous carbon.
\H2 are subsequently formed by abstracting the hydrogen from those species \citep{Duley1996,Pirronello1999A&A344,Mennella1999ApJ...524L..71M}.

To account for the wide range of physical conditions (densities, gas and dust temperatures, radiation field) that occurs in planet-forming protoplanetary disks \citep{Dutrey2014prpl.conf..317D,Woitke2016A&A...586A.103W}, many physico-chemical disk models have implemented the \H2   formation model of \cite{Cazaux2002ApJ...575L..29C, Cazaux2004ApJ...604..222C} through with an efficient \H2 formation on warm dust grains. The \H2  formation rate in the Interstellar medium (ISM) is constrained by precise observations obtained by the {\it Far Ultraviolet Spectroscopic Explorer} (FUSE) satellite at low density \citep{Gry2002A&A...391..675G} and by the near-infrared \H2   emissions for dense Photodissociation Regions (PDRs) \citep{Habart2004A&A...414..531H}. 

\H2 formation models mostly focus on specific environments. Detailed Monte-Carlo kinetic simulations have been used to model \H2   formation in the diffuse cloud environment \citep{Chang2006A&A...458..497C,Iqbal2014ApJ...784..139I,Iqbal2012_0004-637X-751-1-58}.\cite{Hincelin2015A&A...574A..24H} proposed a method to model \H2   formation on low temperature grain surfaces. \cite{LeBourlot2012A&A...541A..76L} implemented a detailed model of \H2   formation for photodissociation regions including Langmuir-Hinshelwood and Eley-Rideal mechanisms and  \citet{Bron2014A&A...569A.100B} studied \H2 formation on stochastically-heated small grains. \cite{Boschman2015AA...579A..72B} considered PAHs as an efficient medium for H$_2$ formation.

We present in this paper an \H2 formation model on warm dust grains and on hydrogenated PAHs. We have not considered \H2 formation on carbonaceous grains, which will be included in a future study. The model was designed to model \H2 formation for a large variety of physical conditions as found in protoplanetary disks. The results are compared to the formation rates computed using semi-analytical \H2 formation formulations. Our model relies on a set of measured and theoretical data.

Experimental studies on \H2   formation have been performed on cold ($T_{\rm d}<$20~K) polycrystalline \citep{Pirronello1997_1538-4357-475-1-L69} and amorphous silicates \citep{Katz1999ApJ...522..305K,Perets2007_1538-4357-661-2-L163,Vidali2007_doi:10.1021,He2011_C1CP21601E,Gavilan2012_doi:10.1111}. Physisorption of atomic hydrogen proceeds without a barrier \citep{Downing20110592,Navarro-Ruiz2014_C4CP00819G}.  \cite{He2011_C1CP21601E} concluded that the desorption energy distribution of the newly-formed HD is much broader if HD forms on an amorphous rather than on a crystalline silicate surface. 

Other studies concentrate on the \H2   formation \citep{Perets2005ApJ...627..850P,Roser2002_0004-637X-581-1-276,Manico2011_1538-4357-548-2-L253} on amorphous water ice since water ice mantles cover most grains at A$_{\rm V}\geq$3 mag \citep{Boogert2015ARA&A..53..541B}.

Once on a grain surface, an atomic hydrogen can diffuse and meet another atom.  H atom diffusion on amorphous ice has been studied experimentally \citep{Matar2008A&A...492L..17M,Masuda1998A&A...330..773M,Watanabe2010_2041-8205-714-2-L233,Hama2012ApJ...757..185H,Dupuy2016ApJ...831...54D} and theoretically \citep{Al-Halabi2007MNRAS.382.1648A,Veeraghattam2014ApJ...790....4V}. The encounter between two atoms results in the \H2   formation. This reaction occurs with no or a small activation barrier of $\sim$250~K \citep{Navarro-Ruiz2014_C4CP00819G}. This mechanism, which involves two adsorbed species, is called the Langmuir-Hinshelwood mechanism \citep{Hama2013_ChemRev}. An adsorbed H-atom can also react with an impinging gas-phase H atom to form \H2 in the so-called Eley-Rideal process.

At dust temperatures above $\sim$20~K, the surface residence time of physisorbed H atoms is so low that \H2   is not produced efficiently anymore. Only chemisorbed atoms remain on the surfaces long enough for \H2   to form. The formation of chemisorption bonds on graphitic and silicate surfaces implies overcoming activation barriers \citep{Jeloaica1999_157,Sha2005doi:10.1063/1.1827601,Bonfanti2015,Goumans2009_doi:10.1111,Martinazzo2006doi:10.1063}. Theoretical works on the chemisorption of hydrogen atoms on silicate surfaces seem to show at first large discrepancies in the energy barrier and binding energy. \cite{Garcia-Gil2013_doi:10.1021/jp4025365} computed via a first-principle computation the interaction of H with the (010) surface of forsterite (Mg$_2$SiO$_4$). By overcoming an activation barrier, H can attach  either to a shallow chemisorption (1880~K, 162 meV) site after overcoming a small barrier of 290~K (25 meV), or to a deep chemisorption site with an absorption energy of 7775~K (670 meV) and a barrier of  1880~K (162 meV). \cite{Oueslati2015_C4CP05128A} performed density functional calculations with different Mg-rich olivine nano-clusters and concluded that there is a distribution of chemisorption sites with binding energy ranging from 8000 up to 30000~K, see also \cite{Kerkeni2013_doi:10.1093}. They demonstrated a linear dependency (Bell-Evans-Polanyi principle) between the activation energy for chemisorption $E_{\rm act}$ and the binding energy ($E_{\rm b}$) and found a relationship between the \H2   reaction barrier for the Langmuir-Hinshelwood mechanism and the binding energy, independent of silicate dust grain shape, size, crystallinity and composition.  Different types of nano-crystals and crystal surfaces have different chemisorption sites. \cite{Oueslati2015_C4CP05128A}  findings can explain the large range of binding and activation energies found in various researches \citep{Goumans2009_doi:10.1111,Kerkeni2013_doi:10.1093,Navarro-Ruiz2014_C4CP00819G,Navarro-Ruiz2015doi:10.1093}. \H2   formation via the encounter of a gas-phase H atom and a chemisorbed atom (Eley-Rideal process) proceeds without activation barrier \citep{Navarro-Ruiz2015doi:10.1093}.

\H2 can also form by the abstraction of an H-atom from hydrogenated PAHs with a small or no activation barrier \citep{Farebrother2000CPL...319..303F, Rutigliano2001CPL340_13R,Bachellerie2007_223,Ivanovskaya2010,Hirama2004_307,Skov2014FaDi..168..223S,Pasquini2016C5CP07272G}. The sticking of H-atoms on graphitic surfaces and PAHs is the first step in the \H2 formation process and has been studied since the 1960s \citep{Xianwei2005doi:10.1063,Bonfanti2015,Ferullo2016}. \cite{Dumont2008PhysRevB.77.233401} studied the \H2 formation with a kinetic Monte-Carlo model.

The paper is organised as follows: in Sect.~\ref{H2_formation_model} we describe our \H2   and HD formation model. The analytical \H2 formation models are presented in \ref{analytical_models}. In Sect.~\ref{analytical_models} we describe the grid of cloud models and the standard {\sc DIANA} model for the comparison between our model and the analytical formulations. In Sect.~\ref{Results} we present the results and a discussion in our grid of models and conclude in Sect.~\ref{Conclusions}.

\section{\H2  and HD formation model}~\label{H2_formation_model}
The \H2 and HD formation model on  cold and warm dust grains follow that of \citet{Cazaux2002ApJ...575L..29C,Cazaux2004ApJ...604..222C}. \H2 formation on grain surfaces should occur for a wide range of grain temperatures. The binding energies of physisorbed species are between few 100~K  up to few 1000~K. Chemisorbed species have binding energies of few 10,000~K.  We take \H2 formation on PAHs (Sect.~\ref{PAH_H2}) and in the gas-phase (Sect.~\ref{H2_gas_phase}) into account.

Surface chemistry concepts are reviewed in \cite{Bonfanti2016QUA:QUA25192}.  We introduce our notations and the notion of pseudo-species. Chemisorption sites are modeled by a pseudo-element named *, whose total elemental abundance is set by the number density of surface sites and the number of available grain surfaces, which itself depends on the grain size distribution. The pseudo-element * has also a pseudo-species counterpart *. The pseudo-species do not migrate nor desorb. 
Atomic hydrogen atoms (and deuterium atoms) can adsorb weakly on a physisorption site, hop to a neighboring site or attempt to overcome a barrier to reach a strongly bound chemisorption site. Atoms can also diffuse between chemisorption sites or go to a physiorption site if they can overcome the activation barriers.

\H2 and HD formation occur via the Langmuir-Hinshelwood (LH) and Eley-Rideal (ER) mechanisms on dust grains and PAHs for physisorbed and chemisorbed species.

We assume that each physisorption site is associated with a chemisorption site. The physisorbed atoms are designated by H\# and D\#. The model includes the most abundant gas- and surface-species as listed in Table~\ref{tab_gas_reactions}. A chemisorbed atom is assigned a star * in front of it. Chemisorbed H and D are thus called *H\# and *D\# respectively.  The pseudo-elemental conservation reads [*]+[*H\#]+[*D\#]=$n_*$+$n_{* \mathrm{H}\varhash}$+$n_{* \mathrm{D}\varhash}$ = $n\surfchem$, where $n
\surfchem=4\pi \Ns \a2 \nd$ is the number density of chemisorption sites ($N_{\rm surf}$ is the surface density of physisorption and chemisorption sites and is equal to 1.5$\times$10$^{15}$ cm$^{-2}$, $\nsite = 4\pi \Ns \a2$ is the number of adsorption sites per monolayer, $r$ is the grain mean radius, and $\nd$ is the number density of dust grains in cm$^{-3}$). We adopted a treatment by rate reactions since the densities in disks are high ($n_{\mathrm{H}}>10^4$ cm$^{-3}$).  The main surface reactions are listed in Table~\ref{tab_dust_reactions}. Table~\ref{tab_variables} and ~\ref{tab_variables2} list the main variables used in this study.

\subsection{Adsorption and sticking}~\label{sect_adsorption}

Most of the processes in our model follow the work of \cite{Hasegawa1992ApJS...82..167H}. The first process is the adsorption and desorption of atomic hydrogen or deuterium from the grain surfaces. A gas-phase atom or molecule $i$ (in this paper $i$=H, D, PAH, ...) adsorbs on a surface physisorption (unlimited number) or a chemisorption (limited number) site at the general rate
\begin{equation}
R^{\rm ads}_i = 4 \pi \a2 \bar{v}_i f_{\rm avail} \nd S_i Q\Bell(a\act,E\act,\Tg)\rate,~\label{eq_Rads}
\end{equation}
where $4\pi \a2$ in cm$^{2}$ is the grain surface area ($r$ is the grain radius), $\bar{v}_i\!=\!(k\Tg/(2\pi \mi)^{1/2}$ is the thermal speed in cm s$^{-1}$,  $\nd$ is the number density of dust grains in cm$^{-3}$, and $S_i$ is the sticking coefficient. The fraction of available sites $f_{\rm avail}$ is unity for physisorption and equal to the fraction of unoccupied sites for chemisorption ($f_{\rm avail}=n_*/n\surfchem$). $Q\Bell$ is the transfer function (see Sect.~\ref{sect_ER_chemisorption}). The treatment of the physisorption and chemisorption differs. The sticking coefficient depends on the surface and type of adsorption \citep{Jones1985MNRAS.217..413J}.

\subsubsection{Physisorption}

There is no activation energy for  physisorption ($Q\Bell$=1). Hollenbach \& McKee (1979) proposed for the sticking coefficient of H the following formula
\begin{multline}
S_{\rm phys}^{-1}=1+0.4\times\left(\frac{\Tg+\Td}{100}\right)^{0.5}+
\\0.2\times\frac{\Tg}{100}+0.08\times\left(\frac{\Tg}{100} \right)^2.
\end{multline}
The adsorption rate on physisorption sites (eq.~\ref{eq_Rads}) simplifies to
\begin{equation}
R\gp=4 \pi \a2 v^{\rm th}_i \nd S_{\rm phys} \rate.
\end{equation}
We assume that the number of physisorption sites per grain remains constant as the ice mantle grows and that the mean grain radius $r$ is not changed. The sticking of atomic hydrogen on water ice has been recently studied \citep{Veeraghattam2014ApJ...790....4V} while \cite{Chaabouni2012A&A538A} computed the sticking coefficient of H and D atoms onto silicate surfaces. 

\subsubsection{Eley-Rideal chemisorption}~\label{sect_ER_chemisorption}

In the precursor-mediated adsorption mechanism, an atom adsorbs first without barrier to a weak physisorption site. This atom can subsequently diffuse to a deeper chemisorption site with small activation barrier (\#H + * \rra \#H*). The direct mechanism involves the gas-phase hydrogen (deuterium) atoms impinging on the surface and overcoming directly the activation barrier ($E\gc >$1000~K). Theoretical studies on these two mechanisms do not show actual decrease of the activation barrier when the atom is already physisorbed for silicate grains \citep{Navarro-Ruiz2014_C4CP00819G}. The chemisorption is of C1- type whereby the H$^+_{\mathrm{chem}}$  is bonded to an O anion and a negative charge is transfered to a nearby Mg cation \citep{KERKENI20171}.
 We treat the formation of a chemisorption bond from an impinging gas-phase H-atom (D-atom) as an activated Eley-Rideal process with rate $R^{\mathrm{gc}}_i$
\begin{equation}
R\gc= \sqrt{ \frac{k\Tg}{2\pi m_i}} \frac{Q\gc}{\Ns}  n_{*} S_{\rm chem}\rate,\label{eq_ER_chemisorption}
\end{equation}
with the Bell's rate being $Q\gc=Q\Bell(a\gc,E\gc,\Tg)$, $\Ns$ is the number density of surface sites (in cm$^{-2}$), and $n_*$ the number density of chemisorption sites (cm$^{-3}$). The probability to overcome an adsorption activation barrier for chemisorption is described by a tunnelling-corrected Arrhenius formula called Bell's formula \citep{Bell1980}:
\begin{equation}
Q\Bell (a\gc,E\gc,\Tg)=\frac{\beta\exp(-\alpha)-\alpha\exp(-\beta)}{\beta-\alpha}, \label{Bell_kappa}
\end{equation}
where
\begin{equation}
\alpha = E\gc /k\Tg,
\end{equation}
and
\begin{equation}
\beta = \frac{4 \pi a\gc}{h} \sqrt{2 \mi E\gc}.
\end{equation}
$\mi$ is the mass of the impinging species ($m_{\rm{H}}$ for H or $m_{\rm{D}}$ for D). $a\gc$ is the width of the activation barrier assuming a rectangular-shaped barrier of height $E\gc$, and $h$ is the Planck constant. The Bell's formula is non-dimensional and reduces to the thermal term when tunnelling does not occur ($a\gc \ra \infty$). It has been used successfully to model gas-phase rates when direct tunnelling effect occurs.

One can determine a temperature $\Tq$ below which quantum tunnelling effect dominates by equating $\alpha$ with $\beta$ for a general process with an activation barrier:
\begin{equation}
\Tq=\left(\frac{E_{\rm act}}{k}\right)\left(\frac{h} {4\pi a \sqrt{2 m E_{\rm act}}}\right)=2.46243\left(\frac{E_{\rm act}}{a_{\rm \AA}^2m_{\rm amu}}\right)^{1/2},~\label{eqn_quantum_temperature}
\end{equation}
where $E_{\rm act}$ is the activation energy in Kelvin, $a_{\rm \AA}$ is the barrier width in \AA, and $m_{\rm amu}$ is the mass of the tunnelling species in atomic mass units. The assumed value of the barrier width $a_{\mathrm{\AA}}$ has a stronger impact on the species diffusion rate than the mass or diffusion energy, for which only the square root of the value counts, at low surface temperature. 

The barrier width for H-atom (and D-atom) Eley-Rideal chemisorption is assumed to be 0.5 \AA. The sticking coefficient $S$ plays a major role at high dust temperature in controlling the chemisorption \citep{Cazaux2011A&A...535A..27C}.
Both sticking coefficients account for the effect of gas and dust temperatures. 

The inverse of the surface density of sites $\Ns$ is a typical surface site cross-section ($\sigma_{\mathrm{surf,site}}=1/\Ns \simeq$1/1.5$\times$10$^{15}$ cm$^{2}$), $n_{*}=4 \pi \a2 f_{\rm avail} \nd \Ns \le n\surfchem$ is the number denisty of unoccupied chemisorption sites. When the surface  is covered by more than $N\layer$ layers, the ER chemisorption rate is set to zero ($R\gc=0$). The standard value used in our models for $\Nact$ is two. 

For chemisorption, we use the sticking coefficient of \cite{Cuppen2010doi:10.1111/j.1745-3933.2010.00871.x,Sha2005doi:10.1063/1.1827601}
\begin{equation}
S_{\rm chem} = \left({1+5\times10^{-2}\sqrt{\Tg+ \Td}+ 1\times10^{-14}\Tg^4}\right)^{-1}.
\end{equation} 
with temperatures in K. 

\subsubsection{Adsorption rate coefficients}

The rate coefficients are
\begin{equation}
d\nice/dt = R\gp n_i \ratedens,
\end{equation}
for physisorption and
\begin{equation}
d\nchem/dt = R\gc n_i \ratedens
\end{equation}
for Eley-Rideal chemisorption.

\subsection{Desorption}~\label{sect_desorption}

An adsorbed species can thermally desorb, or desorb after the absorption of a UV photon, or after a cosmic ray has crossed the grain and deposited some energy. Desorption can be an activated process and is endothermic. The desorption energy is defined as
\begin{equation}
E\des=\Eb + E\actdes\ \ {\rm ergs}.
\end{equation}
For physisorption, which is non dissociative, there is no barrier ($E\actdes$=0) and the desorption and binding energy $\Eb$ are equal.
The binding energy $\Eb$ is in principle not unique but follows a distribution of values. Breaking of a chemisorption bond involves an activation energy equal to the chemisorption activation energy.
The thermal desorption rate is
\begin{equation}
R\cgtherm=\freq Q\Bell (a\actdes,E\actdes,\Td)e^{-\Eb/k\Td}\ {\rm \ in \ s}^{-1},
\end{equation}
The frequency is given by the formula for a rectangular barrier of width $a_{\rm c}$ and height $E_{\rm act}$
\begin{equation}
\freq=\sqrt{\frac{2 \Ns \Eb}{\pi^2 m_i}}.
\end{equation}
We derived a frequency $\freq$ of (1--10) $\times$ 10$^{12}$ Hz assuming a surface site density $\Ns$ = 1.5 $\times$ 10$^{15}$ sites cm$^{-2}$. Only desorption from chemisorption sites are activated $Q\Bell (a\actdes,E\actdes,\Td)=Q\cg=Q\gc$. For desorption from physisorption sites $Q\Bell (a\actdes,E\actdes,\Td)$=1, thus
\begin{equation}
R\pgtherm=\freq e^{-\Eb/k\Td}\ {\rm \ in \ s}^{-1},
\end{equation}
The total desorption rate for species $i$ includes cosmic-ray hit induced desorption and is assumed to follow a first-order desorption process. It reads for the physisorbed species
\begin{equation}
R\despg = R\pgtherm+R\pgph +R\pgCR,
\end{equation}
and for the chemisorbed species
\begin{equation}
R\descg =  R\cgtherm+R\cgph +R\cgCR \rate.
\end{equation}
The desorption for species $i$ reads
\begin{equation}
dn_i/dt = R\despg n\act + R\descg\nchem \ratedens.
\end{equation}
For physisorbed species the concentration of active surface species $i$ is
\begin{equation}
\begin{array}{lcll}
n\act & = &  n_i& \mathrm{if}\ \nicetot \leq \nsite   \nd \\
                   & = & n_i (\Nact /N\layer )  & \mathrm{if} \ \nicetot> \nsite  \nd,\\
\end{array}
\end{equation}
where the number of physisorbed layers on a dust grain is $N\layer  = \nicetot/(n_{\mathrm{d}} \nsite )$. $\nicetot\!=\!\sum_j n_{\varhash,j}$ is the total number density of physisorbed species and
$\Ns $ is the number of sites per monolayer. $\Nact$ is the number of chemically physisorption active layers and is a free parameter of the model. All chemisorbed species can desorb since we restrict the maximum number of chemisorption layers to one. \\

Photodesorption is accounted for either through a factor that scales with the interstellar UV field (0D model) or using the actual computed UV field obtained by detailed radiative transfer and photodissociation cross-sections (2D disk models). The photodesorption rate of physisorbed species $i$ is given by
\begin{equation}
  R\pgph = \pi  \a2 \frac{\nd}{n_{\rm act}}
                      \,Y_i\,\chi F_{\rm Draine}\rate,
\end{equation}
where $Y_i$ is the photo-desorption yield, $\chi F_{\rm Draine}=$1.9921$\times$ 10$^8$ photons/cm$^2$/s is a flux-like measure of the local UV energy density computed from continuum radiative transfer (e.g., \citealt{Woitke2009A&A...501..383W}). We assume that photodesorption affects chemisorbed species the same way with rate $R \cgph$.

Cosmic-ray induced desorption follows the treatment of \citet{Hasegawa1993MNRAS.261...83H}.
\begin{equation}
  R\pgCR = f(70K)\,R\pgtherm(70K)\,
                      \frac{\zeta_{\rm CR}}{5\times 10^{-17}}\rate,
\end{equation}
where $\zeta_{\rm CR}$ is the cosmic ray ionisation rate of H$_2$,
$f(70K)\!=\!3.16 \times 10^{-19}$ the 'duty-cycle' of the grain
at 70\,K and $R^{\rm des, th}_i(70\,\rm K)$ the thermal desorption rate
for species $i$ at temperature $\Td\!=\!70$\,K. The adopted
value for $f(70K)$ is strictly valid only for 0.1 $\mu$m grains
in dense molecular clouds. Explosive desorption were not considered and will be included in future works \citep{Ivlev2015ApJ...805...59I}. The dust temperature
after a Cosmic-Ray hit is not high enough to desorb a species in a chemisorption site.

\subsection{Thermal and tunnelling surface diffusion}~\label{sect_diffusion}

On grain surfaces the diffusive movement of H atoms from one site to another site occurs either by thermal hopping when there is sufficient energy to overcome the energy barrier or by tunnelling. Diffusion can be viewed as a random walk process. Atoms on a physisorption site can hop to another such site or to a deeper chemisorption site after overcoming an activation barrier \citep{Barlow1976ApJ...207..131B,Leitch-Devlin1984MNRAS.210..577L,Tielens1987ASSL..134..397T}. A H-atom in a chemisorption site needs to overcome the energy difference between a physisorption and a chemisorption site in addition to the activation barrier in order to move to a physisorption site \citep{Aronowitz1980ApJ...242..149A}. We use again the Bell's formula to model the surface diffusion tunnelling effects
\begin{equation}
R\therdiff = \freq Q\diff (a\diff,E\diff )e^{-\Delta E_{ij}/kT_{\rm d}}/\nsite\ {\rm in}\rate.
 \end{equation}
The factor $Q\diff $ is the Bell formula (eq.~\ref{Bell_kappa}) with
\begin{equation}
\alpha = E\diff /kT
\end{equation}
and 
\begin{equation}
\beta = \frac{4 \pi a\diff}{h} \sqrt{2 m E\diff }.
\end{equation}
$\Delta E_{if}$ is the binding energy difference between the two adsorption sites:
\begin{equation}
\begin{array}{rcll}
\Delta  E_{if} & =  & 0                                 & {\rm\  if}\ \Eb \le \Ef,\\
\Delta E_{if}  & = & \Eb-\Ef & {\rm otherwise}.\\
\end{array}
\end{equation}
Thus $\Delta E=0$ for hopping between two physisorption sites or between two chemisorption sites.
$m$ is the mass of the diffusing species. $\alpha$ corresponds to the thermal diffusion (hopping) while $\beta$ refers to the quantum tunnelling. The diffusion time $t\diff $ is the inverse of $R\diff  $. 
The surface diffusion rates for species $i$ are defined as the combination of thermal, tunnelling and cosmic-ray induced diffusion
\begin{equation}
R\diff  = R\therdiff +R\CRdiff\rate ,
\end{equation}
where the cosmic-ray induced diffusion rate is \citep{Hasegawa1993MNRAS.261...83H,Reboussin2014MNRAS.440.3557R}
\begin{equation}
  R\CRdiff = f(70K)\,R\therdiff(70K)\,
                      \frac{\zeta_{\rm CR}}{5\times 10^{-17}}\rate.
\end{equation}
An atom bound at a physisorption site can diffuse to another physisorption site, desorb, or land on the chemisorption site associated with the current physisorption site \citep{Cazaux2002ApJ...575L..29C,Cazaux2004ApJ...604..222C}. This view is valid when the current physisorption site is related to a silicate or carbonaceous surface, i.e. when the surface has fewer than a monolayer of ice. In denser regions, multi-layer ice mantle can be built rapidly. When the grain has one or more monolayers, the H-atom can only physisorb on the ice mantle, e.g. water ice. In this case we assume that a physisorbed H/D-atom can still diffuse through the bulk of the ice mantle but at a lower diffusion rate than on the ice mantle surface and overcome the barrier to land on an available chemisorption site. The diffusion rates state that bulk diffusion is permitted so that as the ice mantle grows, the number of sites available for scanning also increases, independently of the assumed number of active layers. The reactions compete explicitly with the desorption processes (thermal, cosmic-ray induced, and photodesorption). Their respective rates concern the active species. The process is formally represented as
\begin{equation}
\rm{H}\varhash + * \ra *\rm{H}\varhash,
\end{equation}
and
\begin{equation}
\rm{D}\varhash + * \ra *\rm{D}\varhash.
\end{equation}
The diffusion of species $i$ is restricted to the physisorption active layers similar to the desorption
\begin{equation}
dn(i)/dt = R\diff n\act(i)\ratedens.
\end{equation}
An alternative interpretation of our assumption is that the average diffusion rate for all the species on the surface and in the bulkis lower by a factor $\Nact /N\layer$ when there is more than one monolayer.

\subsection{The rate equation treatment}~\label{sect_rate_equations}

Since we focus on \H2 and HD formation in dense cold and warm regions, we adopted the
rate equation treatment for the surface chemistry.

\subsubsection{Langmuir-Hinshelwood reactions}~\label{sect_LH}

The Langmuir-Hinshelwood surface reaction prescription follows the implementation \citep{Hasegawa1992ApJS...82..167H}. The surface reaction rate coefficient $k_{ij}$ (in cm$^{3}$ s$^{-1}$) between surface adspecies $i$ and $j$ with respective number density $n_i$ and $n_j$ is the probability of reaction per encounter ($\kappa_{ij}$) times the rate of encounter between the two species scanning the surface:
\begin{equation}
k_{ij} = \kappa_{ij} (R\diff+R\diffj)/n_d\ratecoeff, \label{surface_rate}
\end{equation}
where $\kappa_{ij}$ is the probability for the reaction to occur upon encounter between species $i$ and $j$ after both have diffused on the grain mantle,
$R\diff$ and $R\diffj$ (in s$^{-1}$) are the diffusion rates for species $i$ and $j$, and $n_d$ (in cm$^{-3}$) is the number density of dust grains. We assume that the newly-formed \H2   desorb immediately because of the high exothermicity of the reaction. The probability for the reaction to occur 
$\kappa_{ij}$ follows a competition between association of the two species, modeled by the Bell formulation to account for thermal crossing and tunnelling of potential activation barrier, and diffusion \citep{Bonfanti2016QUA:QUA25192,Garrod2011ApJ...735...15G,Ruaud2016MNRAS.459.3756R}:
\begin{equation}
\kappa_{ij} = \frac{Q\Bell (a^r_{ij},E\act )}{Q\Bell (a_r,E\act)+P\diff+P\diffj},
\end{equation}
where $a^r_{ij}$ is the width of the barrier and $E\act$ the activation barrier height (energy), $P\diff=R\diff/\freq$. The mass used in the Bell formula is that of the lighter species. When no competition is accounted for the probability is
\begin{equation}
\kappa_{ij}'= \freq Q\Bell (a^r_{ij},E\act ),
\end{equation}
For barrierless reactions, $\kappa_{ij}$ = 1 and $\kappa_{ij}'$ = 1, and the rate becomes
\begin{equation}
k_{ij} \simeq (R\diff+R\diffj)/n_d\ratecoeff, ~\label{eqn_diffusion_limited_rate}
\end{equation}
When $E\act << E\diff, E\diffj$ and without tunnelling effects $\kappa_{ij} \rightarrow$ 1 and the rate tends towards the diffusion-limited rate
\begin{equation}
k_{ij} \simeq (R\diff+R\diffj)/n_d\ratecoeff,
\end{equation}
while without competition the rate is 
\begin{equation}
k_{ij}' =\freq Q\Bell (a^r_{ij},E\act) (R\diff+R\diffj)/n_d\ratecoeff.  
\end{equation}
In that case, $k_{ij}>>k_{ij}'$. On the other hand, when $E\act >> (E\diff, E\diffj$), the diffusion terms ($R\diff$ and $R\diffj$) dominate and cancel out in the rate coefficient (since the diffusion terms are present as numerator and denominator), which becomes
\begin{equation}
k_{ij} \simeq  \freq Q\Bell (a^r_{ij},E\act )/n_d \ratecoeff,
\end{equation}
and again $k_{ij}>>k_{ij}'$. The diffusion is so fast that the reactants are always in the situation where the recombination (association) can occur. 

The barrier for H diffusion from a physisorption site to another one in Kelvin is between 256~K \citep{Kuwahata2015PhRvL.115m3201K} to 341~K \citep{Congiu2014FaDi..168..151C} whereas the barrier for \H2 formation is 250~K or less \citep{ Navarro-Ruiz2014_C4CP00819G}. We adopted a barrierless \H2 recombination between two physisorbed H atoms.

At 10 ~$<\Td<$~20~K and assuming comparable diffusion and association (reaction) activation energy and weak photodissociation, the probability of reaction  becomes $\kappa_{ij}\simeq 1/3$. At low temperature ($\Td<$ 10~K), thermal processes become insignificantly slow and both the diffusion and reaction rates are dominated by tunnelling. Diffusion tunnelling of H-atoms has a large barrier width of $\sim$~3.9 \AA\ compared to $\sim$0.5 \AA\ for reaction barrier \citep{Limbach2006} such that the reactive tunnelling rate dominates at $\Td<$10~K, the probability of reaction  becomes $\kappa_{ij}\simeq 1$. Therefore the \H2 formation rate below 20~K by recombination of physisorbed H-atom is diffusion limited even if a small barrier exists.
We do not restrict \H2 formation by recombination of physisorbed H-atoms (LH mechanism) even when  the surface is covered by a layer of physisorbed H-atom \citep{Gavilan2012_doi:10.1111}.

\subsubsection{Eley-Rideal reactions}~\label{sect_ER}

According to the semi-equilibrium theory, the probability of a gas-phase radical recombining with an atom located in an adsorption site is equal to the probability of the gas atom directly impinging on the occupied  site multiplied by the probability of the gas atom having enough kinetic energy to overcome the reaction barrier, if any, with the possibility to tunnel through the barrier. Due to the long-range attractive potential, the impinging species has an energy of $1/2k\Tg + \Eb$ relative to the surface species ($\Eb$ is the binding energy). Part of this energy can be used to overcome a reaction barrier.

Laboratory and theoretical works suggest that the formation of \H2 via ER process is barrierless (or has a very small barrier) both on silicate and carbonaceous surfaces. The Eley-Rideal formation of \H2 from an impinging H on a physisorbed or chemisorbed H-atom has a rate of
\begin{equation}
R^{\rm gp}_{\rm H_2} = \sqrt{ \frac{k\Tg}{2\pi m_{\rm{H}}}} \frac{1}{N_{\rm surf}} n_{\mathrm{H}\varhash}= 2.425\times 10^{-12} \sqrt{\Tg}n_{\mathrm{H}\varhash}\rate.
\end{equation}
and
\begin{equation}
R^{\rm gc}_{\rm H_2} = \sqrt{ \frac{k\Tg}{2\pi m_{\rm{H}}}} \frac{1}{N_{\rm surf}} n_{* \mathrm{H}\varhash}\rate.
\end{equation}
For HD, the rate is composed of two terms, one is the rate when a gas phase atom hits a chemisorbed D-atom and the second when a D-atom hits a chemisorbed H atom. The rate $R^{\rm gc}_{\rm H_2}$ is null when there is more than $N\layer$ of ice, i.e. the water-ice mantle shield the chemisorbed H-atoms from being directly hit by a gas-phase atom.f

\subsection{Diffusion-mediated chemisorption}\label{sect_diffusion_chemisorption}

As the direct Eley-Rideal chemisorption is hampered by the presence of water ice mantle below the water sublimation (between 90 and 150~K depending on the gas density), chemisorption mostly occurs after a H-atom has diffused through the mantle and reached the interface between the mantle and the refractory surface. The diffusion rate in the bulk is decreased by the total number of ice layers. This is an extremely simple  method to model a slower diffusion in the ice mantle compared to the diffusion at the ice surface. At the interface, the H-atom can overcome the barrier to chemisorption. The rate for this reaction is
\begin{equation}
R\pcH =  \kappa\pcH R\diffH \left(n_*/ \nd\right)\rate.~\label{eqn_rpc}
\end{equation}
This rate should be compared to the thermal-dominated desorption rate of the physisorbed H-atom
\begin{equation}
R\pgH = \nu_{\rm pg}e^{-E^{\rm b}\Hp/kTd}\rate.
\end{equation}
In the thermal regime ($\Td>\Tq$=78~K), the lowest activation barrier for chemisorption is $E^{\rm act}\Hp \sim$ 900--1000~K on silicate \citep{Oueslati2015_C4CP05128A} compared to a diffusion energy of $\simeq$406~K \citep{Perets2007_1538-4357-661-2-L163} and an adsorption energy of $\simeq$~510~K. The diffusion-mediated chemisorption rate is reaction-limited 
\begin{equation}
R\pcH =  \nu\pcH Q\pcH (n_*/ \nd) \simeq \nu\pcH e^{-E^{\rm act}\Hp /k\Td} (n_*/ \nd). \rate
\end{equation}
The activation barrier to form \H2 from the encounter of a physisorbed and a chemisorbed atom is the same as the barrier to form a chemisorption bond because both processes involve the breaking of the H chemical bond to the surface. The rate per gas volume is 
\begin{equation}
d\nHchem/dt =  R\pcH n^{\rm act}_{\rm H}\ratedens.
\end{equation}
The reverse mechanism is a chemisorbed H-atom escaping the deep well to reach a physisorption site $R\cpH$.
The formation of \H2   can happen after a H-atom (H\#)  hops from a physisorption site to a site occupied by a chemisorbed H-atom with number density $n_{*\rm{H}\varhash}$, the rate is
\begin{equation}
R\HpHc =R\diffH(\nHchem/\nd)\rate\\ \label{eqn_Hrpc}.
\end{equation}

\subsection{Reactions involving PAHs}~\label{PAH_reactions} 
PAHs are not formed or destroyed in our chemical network and only
exchange charges with other positively-charged species (for examples
H$^+$, He$^+$, Mg$^+$, Fe$^+$, C$^+$, Si$^+$, S$^+$, HCO$^+$, ...) or can be hydrogenated (PAH-H$_x$, PAH-H$_x^+$, PAH-H$_x$D, PAH-H$_x$D$^+$). PAH-H$_x^+$ ($x$=0, 1, ..., 18). The ionised PAH-H$_x$s can recombine with a free electron.
Chemical reaction rates involving PAHs are highly uncertain. Most of the rates
are extrapolations from a few existing laboratory or theoretical rates
and are discussed in Appendix~\ref{PAH_chemistry}.

\subsection{\H2 and HD formation on neutral and cationic PAHs}~\label{PAH_H2}

Experimental and theoretical studies on neutral hydrogenated PAHs (called here PAH-H$_x$, with $x$=1, 2, ..., 18) suggest  that \H2   can form through barrierless Eley-Rideal abstractions  \citep{Bauschlicher1998ApJ...509L.125B,Mennella2012ApJ...745L...2M,Rauls2008ApJ...679..531R,Thrower2012ApJ...752....3T}. \cite{Morisset2008doi:10.1063/1.2952685} computed quantum dynamically  the sticking of a H atom on a graphite surface.

The \H2 formation proceeds in two steps. The first step is the hydrogenation of the PAHs or ionised PAHs, followed by H-abstraction. The adopted PAH is the large compact circumcoronene (C$_{54}$H$_{18}$), which has a peri-condensed stable structure \citep{Tielens1987ASIC..191..273T}. 
{Although the carbon backbone fragmentation efficiency upon absorption of a UV photon increases with the degree of hydrogenation \citep{Wolf2016ApJ...832...24W}, we modeled hydrogenated PAHs up to PAH-H$_x$ with $x$=18  equal to the number of edge carbon for the circumcoronene.
\begin{table}
\begin{center}
\caption{PAH successive hydrogenation energy barriers ($E\actPAHH/k$) in Kelvin. The adopted values are shown in bold face.\label{tab_PAH_data}}
\vspace*{-0.5mm}          
\begin{tabular}{lllll}
\toprule
  Hydrogenation  level      &  outer edge &  edge  & center & Ref.\\
\noalign{\smallskip}
\hline
\noalign{\smallskip}         
 PAH-H                                      &   116   &  1740    & 2553   &  $^a$\\
                                            &   692   &              &            &  $^b$\\
                                            &   {\bf 324}   &        &            &  $^c$\\
 PAH-H$_2$                                      &   348   & 2669    & 3365   & $^a$\\
                                            &      0    &              &           &   $^b$\\
                                            &   {\bf   0}    &              &           &   $^c$\\
 PAH-H$_3$                                      &   348   &              &          &   $^b$\\
                                            &  {\bf 533}    &              &          &   $^c$\\
 PAH-H$_4$                                       &      0   &              &          &   $^b$\\  
                                             &  {\bf 0}      &              &          &   $^c$\\
 PAH-H$_5$                                     &      0   &              &          &   $^b$\\  
                                            &  {\bf 742}     &              &          &   $^c$\\
 PAH-H$_6$                                     &      0   &              &          &   $^b$\\  
                                            &  {\bf 0}      &              &          &   $^c$\\
 PAH-H$_7$                                       &      0   &              &          &   $^b$\\  
                                            &  {\bf 406}      &              &          &   $^c$\\
 PAH-H$_8$                                       &      0   &              &          &   $^b$\\                                           
                                            &  {\bf 0}      &              &          &   $^c$\\
 PAH-H$_9$                                    &  {\bf 348}  &              &          &   $^c$\\         
 PAH-H$_{10}$                                         &  {\bf 0}      &              &          &   $^c$\\
PAH-H$_{11}$                                      &  {\bf 603}  &              &          &   $^c$\\   
PAH-H$_{12}$                                 &  {\bf  0}  &              &          &   $^c$\\       
PAH-H$_{13}$                                       &  {\bf 382}      &              &          &   $^c$\\
 PAH-H$_{14}$                                      &  {\bf  0}  &              &          &   $^c$\\    
PAH-H$_{15}$                     &  {\bf 382}      &              &          &   $^c$\\
 PAH-H$_{16}$                                 &  {\bf 0}      &              &          &   $^c$\\
 PAH-H$_{17}$                                          &  {\bf 452}      &              &          &   $^c$\\
 PAH-H$_{18}$                                          &  {\bf 0}      &              &          &   $^c$\\
\noalign{\smallskip}     
\bottomrule
\end{tabular}
\\
\resizebox{90mm}{!}{
\begin{minipage}{100mm}{{\bf Ref.} $^a$\cite{Cazaux2016NatSR...619835C} for PAH cations; $^b$ \cite{Rauls2008ApJ...679..531R}; $^c$ \cite{Boschman2015AA...579A..72B}, the values are for coronone C$_{24}$H$_{12}$.}
\end{minipage}}
\vspace*{-1mm}
\end{center}
\end{table}

We adopted a cross section of 1.1 \AA$^{2}$ per reactive carbon atom for radiative hydrogen association \citep{Boschman2015AA...579A..72B} together with a barrier $E\act $
 \begin{equation}
 k_{\mathrm{PAH-H_x,H}} = 2.78 \times10^{-11}(T_g/300)^{1/2} N_C Q_{\mathrm{Bell}}(E\actPAHH )\ratecoeff,
\end{equation} 
where $Q_{\mathrm{Bell}}(E\actPAHH)$ is a Bell's formula (eq.~\ref{Bell_kappa}), i.e. we consider that H-tunnelling is possible and $x \ge 0$. It is clear that hydrogenation of neutral PAHs is an activated process because the formation of a C-H bond requires a rehybridisation (sp2 to sp3) of the carbon orbitals. \cite{ Rauls2008ApJ...679..531R,Karlicky2014doi:10.1063/1.4867995,Allouche20061402-4896-2006-T124-018} (see also \citealt[]{Ferullo2016,Klose1992A&A...260..321K}) obtained a value of $E\actPAHH/k$=692~K (0.06 eV). Other studies \citep[][]{Sha2005doi:10.1063/1.1827601} found barrier of 2321~K (0.2 eV). \cite{Areou2011doi:10.1063/1.3518981} found experimental evidence of a barrier.

\cite{Cazaux2016NatSR...619835C} and \cite{ Rauls2008ApJ...679..531R} have studied the successive PAH hydrogenation barriers. The barrier energies depend on the type of attachment sites (outer edge site, an edge ste, or a center site). Table~\ref{tab_PAH_data} provides a summary of the values present in the literature. Computations \citep{ Rauls2008ApJ...679..531R} suggest that the barrier vanishes for high levels of hydrogenation.  \cite{Boschman2015AA...579A..72B} modeled the PAH hydrogenation with alternate high and low barriers. We adopted the series of barrier energies from \cite{ Boschman2015AA...579A..72B}. Low energy barriers are central to permit \H2 formation at intermediate dust temperatures ($\Td$=20--100~K) when the chemisorption on silicate grains may be inefficient.

De-hydrogenation of PAH-H$_x$ ($x$$\ge$1), occurs mostly by photodissociation. The photodissociation threshold for hydrogenated circumcoronene ($E_{th}$) is  equal to the binding energy \citep{Andrews2016AA...595A..23A}. Computations of the binding energies depend on whether H is chemisorbed to an edge carbon (\Cedge) or not (\Cgraph), see e.g., \cite{Ferullo2016,Rasmussen2013JPCA..117.4279R}.
When it is attached to a \Cgraph \ atom, the binding energy is $\sim$~0.6~eV. An atom attached to an edge carbon is more strongly bound (1--2 eV). The binding energy was found for H and D on graphite to be 0.6 (6963~K) and 0.95 eV (11024~K)\citep{Zecho2002JChPh.117.8486Z} and \cite{Ferro2003_609}. The chemisorption site for H atom, located on the top of a \Cgraph\ carbon atom, has an energy of 0.57 eV (6847~K) for coronene C$_{24}$H$_{12}$ \citep{Jeloaica1999_157} and weak binding site with 0.040 eV (464~K) may exist  \citep{Ma2011_doi:10.1063}. \cite{Haruyama2011} found a binding energy of pyrene C$_{16}$H$_{10}$ of  0.6 eV. 
\cite{Ferullo2016} used an improved density function theory model and computed for a chemisorption binding on an edge carbon of 2 eV for anthracene. In this study the adsorption (binding) energies of an H-atom on a \Cedge\ atom of a PAH are taken from \cite{Klaerke2013AA...549A..84K,Bauschlicher2014}. For circumcoronene the binding energy $E_0$ of an extra H-atom in a C-H bound is  $\sim$~1.4~eV (16250~K).

The unimolecular thermal dissociation rate of a PAH-H$_x$ at an effective temperature $T_e$ follows an Arrhenius approximation to the  Rice-Ramsperger-Kassel-Marcus (RKKM) model \citep{Jochims1994ApJ...420..307J}:
\begin{equation}
R_{\mathrm{PAH-H_x,T_e}} = R_0(T_e) \exp\left(-E_0/kT_e\right) \rate,
\end{equation}
where $T_e$ is an effective temperature for a PAH with $N_C$ carbon atoms upon absorption of a photon of energy $\UVr$ in eV \citep{Tielens2005pcim.book.....T}:
\begin{equation}
T_e \simeq 2000 \left( \frac{\UVr}{N_C}\right)^{0.4} \left( 1-0.2 {\left(\frac{E_0}{\UVr}\right)}\right).
\end{equation}
The pre-exponential factor $R_0(T_e) = (kT_e/h)\exp{(1+(\Delta S/R))}$ s$^{-1}$ \citep{Reitsma2014_PhysRevLett.113.053002},
where $\Delta S$ is the entropy change assumed to be 55.6 J K$^{-1}$ mol$^{-1}$
\citep{Ling1998} and $R$ is the gas constant.
  
Unimolecular dissociation competes with relaxation by the emission of infrared photons with a typical rate $R_{\mathrm{IR}}$ of 1 s$^{-1}$. The yield for photodissociation for $\UVr>E_0$ reads:
\begin{equation}
Y_{\mathrm{PAH-H_x,UV}} = \frac{R_{\mathrm{PAH-H_x,T_e}}}{R_{\mathrm{PAH-H_x,T_e}}+R_{\mathrm{IR}}}.
\end{equation}
The yield is null for $\UVr<E_0$. The yield is used together with the PAH cross-section \citep{Draine2001ApJ...551..807D,Li2001ApJ...554..778L} and the local UV field spectrum to compute the actual photodissociation rate.

PAHs and hydrogenated PAHs can exchange IR photons with the dust grains and
reach an average temperature $T_{\mathrm{PAH}}$. In radiative thermal equilibrium $T_{\mathrm{PAH}}$ is equal to the dust grain temperature $\Td$  in the optically thick midplane of protoplanetary disks \citep{Woitke2016A&A...586A.103W}. Hydrogenated PAHs can undergo thermal unimolecular dissociation with the rate
\begin{equation}
R_{\mathrm{PAH-H_x,therm}} = R_0(T_{\mathrm{PAH}}) \exp\left(-E_0/kT_{\mathrm{PAH}}\right) \rate.
\end{equation}
An impinging H-atom can abstract the dangling H from a hydrogenated PAH-H$_x$ to form \H2 (or PAH-H$_x$D to form HD)  via a barrierless Eley-Rideal mechanism \citep{Rauls2008ApJ...679..531R,Bauschlicher1998ApJ...509L.125B}. \cite{Cuppen2008doi:10.1063/1.2913238} model the \H2 formation by abstraction from hydrogenated graphite using the kinetic Monte-Carlo technique.

The cross-section for this reaction is 0.06 \AA$^2$ per reactive carbon atom \citep{Mennella2012ApJ...745L...2M}  for neutral PAHs:
 \begin{equation}
k_{\mathrm{PAH-H_x,H}} = 1.5 \times10^{-12}(T_g/300)^{1/2} N^{\rm reac}_C \ratecoeff. \label{PAH_Habstracton_rate}
\end{equation}
$N^{\rm reac}_C$=x for PAH-H$_x$, i.e. the rate scales with the number of extra hydrogens attached to the PAH. A small barrier ($\sim$10 meV, or $\sim$ 115K) may be present, but we chose to neglect it \citep{Casolo2009doi:10.1021/jp9040265}.
\cite{Petucci2018_doi:10.1063/1.5026691} computed a high energy of 1150 K for the barrier. 
\citet{Zecho2002_188} found that the D abstraction on low D-covered graphite bombarded with H-atoms proceeds with a cross-section of up to 17 \AA$^2$ (and 4 \AA$^2$ at high coverage). Eley-Rideal cross sections around 4 \AA$^2$ have been also computed by \cite{Pasquini2016C5CP07272G} using the quasi-classical trajectory method. The cross-sections do not show isotopic dependencies. Therefore, we adopted the same cross-section for H and D formation on hydrogenated PAHs using the cross-section measured in the experiments on amorphous carbon (a:C-H) by \cite{Mennella2012ApJ...745L...2M}, which is however much lower than the values measured by \citet{Zecho2002_188} or comptued by \cite{Pasquini2016C5CP07272G}. \cite{Duley1996} adopted a cross section of 10 \AA$^2$. On the other extreme, \cite{Skov2014FaDi..168..223S} estimated an extremely low cross-section of 0.01 $\AA^2$. We tested the effect of choosing a higher abstraction cross-section in appendix \ref{appendix_PAH}.

PAH and hydrogenated PAHs can be ionised at low $A_V$ and ionisation competes with photodissociation. A lower value of 0.02 \AA$^2$ has been reported by \cite{Oehrlein2010_doi:10.1063}.
The hydrogenation of PAH cations 
\begin{equation}
\PAH^+ + {\rm H} \ra (\PAH-{\rm H})^+
\end{equation}
proceeds without activation barrier or with a small barrier. \cite{Cazaux2016NatSR...619835C} computed a small barrier of 116~K (0.01 eV) for the first hydrogenation of coronene cation, consistent with the value of \cite{Hirama2004_307}. The rate is quasi-independent on the size of the PAH \citep{Demarais2014ApJ...784...25D,Snow1998Natur.391..259S}. We adopt therefore a size-independent rate:
\begin{equation}
 k_{\mathrm{PAH-H_x^{n+},H}}=2\times10^{-10} (T_g/300)^{-3/2}Q_{\mathrm{Bell}}(E\actPAHpH) \ratecoeff,
\end{equation}
with $E\actPAHpH$=116~K. We further assume that the photodissociation of ionised hydrogenated PAHs follow the same rate as for the neutral PAHs. 

H-abstraction reaction with cationic hydrogenated PAHs is a barrierless ion-neutral reaction. Therefore, the rate should be in the order of magnitude of a Langevin rate. We choose to use the scaling law of \citet{Montillaud2013AA...552A..15M}:
\begin{equation}
k_{\mathrm{(PAH-H_x)^{n+}}} = 1.4\times10^{-10} \left(\frac{N_H}{12}\right)\left(\frac{N_C}{24}\right)^{-1} \ratecoeff,
\end{equation}
where $N_H$ and $N_C$ are the number of hydrogen and carbon atoms that constitute the PAH respectively and $n+$ is the charge of the PAH. The standard interstellar abundance of PAHs is $3\times10^{-7}$ \citep{Tielens2005pcim.book.....T}. In protoplanetary disks, a fraction $f_{\PAH}$ is still present. The \H2 formation efficiency depends on the charge of the PAH and hydrogenated PAHs.The recombination of PAH-H$^+$ follows the same rate as for PAH$^+$ apart that the recombination is dissociative $\PAH-\mathrm{H}^+ + \elec \ra PAH + \Hg$.}

Hydrogen atoms can also physisorb on PAHs. The \H2 formation can be theoretically more efficient than formation from chemsisorbed H atoms for graphite \citep{Casolo2009doi:10.1021/jp9040265}. We have not considered \H2 formation from photodissociation of PAHs \citep{Castellanos2018arXiv180602708C,Castellanos2018arXiv180602703C}.
H diffusion can compete with desorption \cite{Borodin2011PhysRevB.84.075486}. In  our model we assumed that H-atoms are too strongly bound for an efficient diffusion to another site for chemisorption.

\subsection{\H2 gas-phase formation and destruction}~\label{H2_gas_phase}

We incorporated in our \H2 and HD formation model the major formation and destruction routes for \H2.  Gas-phase \H2 reactions have been discussed by \cite{Glover2003ApJ...584..331G,Galli1998A&A...335..403G}. The formation occurs  via H$^-$, whose electron can be ejected, carrying with it the excess heat of formation of 4.8~eV:
\begin{equation}
\begin{array}{rcl}
\Hg + \elec   & \ra & \Hg^- + \UVr,\\
\Hg^-+\Hg     & \ra &\Hg_2+\elec,\\
\end{array}
\end{equation}
with rate for the later reaction from \cite{Launay1991A&A...252..842L,Martinez2009ApJ...705L.172M} and H$^+$:
\begin{equation}
\begin{array}{rcl}
\Hg + \Hg^+ &\rightarrow&  \Hg_2^+ + \UVr,\\
\Hg_2^+ + \Hg & \ra &\Hg_2 + \Hg^+.
\end{array}
\end{equation}
In both cases, the first step is a slow radiative recombination reactions. The hydrogen anions and cations are destroyed by mutual neutralisation \citep{Moseley1970PhRvL..24..435M}
\begin{equation}
\Hg^+ + \Hg^- \ra \Hg + \Hg,
\end{equation}
and H$^+$ can recombine with electron
\begin{equation}
\Hg^+ + \elec \ra \Hg,
\end{equation}
or the electron can be photodetached
\begin{equation}
\Hg^-  + \UVr \ra \Hg + \elec.
\end{equation}
Protons are formed by charge exchange with He$^+$ or by ionisation by X-ray or cosmic rays. They can recombine or exchange the charge with a species X with ionisation potential lower than 13.6 eV:
\begin{equation}
\Hg^+ + \elec \ra \Hg +\ \UVr ,
\end{equation}
or the electron can be photodetached
\begin{equation}
\Hg^+  + {\rm X} \ra \Hg + {\rm X}^+.
\end{equation}
At density $>$10$^{8}$ cm$^{-3}$, three body reactions become important, when the third body carries the excess heat of formation in form of kinetic energy \citep{Palla1983ApJ...271..632P}
\begin{equation}
\begin{array}{rcl}
\Hg + \Hg + \Hg  &\rightarrow&  \Hg_2 + \Hg,\\
\Hg + \Hg + \Hg_2  &\rightarrow&  \Hg_2 + \Hg_2,\\
\end{array}
\end{equation}
The rate coefficient for the first reaction is controversial \citep{Forrey2013PhRvA..88e2709F,Forrey2013ApJ...773L..25F,Jacobs1967JChPh..47...54J,Abel2002Sci...295...93A,Flower2007MNRAS.377..705F,Palla1983ApJ...271..632P} as it is a prime path to form H$_2$ in the Early Universe. Rates for the second reactions have been measured by \cite{Trainor1973JChPh..58.4599T} or modeled by \citet{Whitlock1974JChPh..60.3658W,Schwenke1988JChPh..89.2076S,Schwenke1990JChPh..92.7267S}.

At very high gas densities and temperatures collision-induced dissociations (collider reactions) can occur \citep{Ohlinger2007PhRvA..76d2712O}.

\subsection{\H2 destruction by dissociative chemisorption}~\label{H2_dissociatve_chemisorption}

At high gas temperatures, \H2 impinging on bare silicate grain surfaces can dissociatively chemisorb. The model the process as two reactions
\begin{equation}
\Hg_{2} + * \ra \Hcr + \Hpr,
\end{equation}
where we adopted a barrier height of 5802~K \citep{Song2016_166} for the first reaction with the rate using formula~\ref{eq_ER_chemisorption}. The dissociative chemisorption of \H2 on PAH edges proceeds as
\begin{equation}
\Hg_{2} + {\rm PAH-H_x} \ra {\rm PAH-H_{x+1}} + \Hg,
\end{equation}
with a barrier height of 3481~K \citep{Dino2004_713}.

\subsection{HD formation and destruction}\label{HD_gas_phase}

HD formation occurs both on grain surfaces, by abstraction of hydrogenated PAHs and by deuterium substitutions in the gas-phase \citep{Cazaux2009A&A...496..365C}. \H2 has a zero-point energy of 3135.5~K (2179.3~$\pm$~0.1 cm$^{-1}$) and HD of 2719.7~K (1890.3~$\pm$~0.2 cm$^{-1}$) resulting in an energy difference of $\EHDHH$~=~415.8~K \citep{Irikura2007JPCRD..36..389I}.
The formation of HD occurs also in the gas-phase where exchange reactions can be efficient \citep{Watson1973ApJ...182L..73W,Brown1986MNRAS.223..429B}. 
Gas phase HD formation reactions have been discussed in the context of dust free or low-metallicity early Universe \citep{Stancil1998ApJ...509....1S,Galli1998A&A...335..403G,Cazaux2009A&A...496..365C}, cold molecular cloud by \cite{Roueff2007ApJ...661L.159R}  and for PDR regions by \cite{LePetit2002A&A...390..369L}. The lower zero-point energy of the deuterated species translates into larger reaction activation energies.
The radiative association 
\begin{equation}
\Hg + \Dg \ra \Hg \Dg + \UVr,
\end{equation}
has an extremely low rate of $8\times 10^{-27}$ cm$^{3}$s$^{-1}$ at 100~K \citep{Stancil1997ApJ...490...76S}. In the gas-phase, HD can be formed efficiently at high temperature by the substitution reaction once \H2 has been formed \citep{Mitchell1973JChPh..58.3449M,Sun1980JChPh..73.6095S,Garrett1980JChPh..72.3460G,Simbotin2011PCCP...1319148S} by the reaction
\begin{equation}
\Hg_2 + \Dg \rla \Hg \Dg + \Hg,
\end{equation}
which has a barrier of 3820~K. The backward reaction is endothermic by 420~K in the literature. We adopted an endothermicity of 415.8~K. The ion-neutral reaction
\begin{equation}
\Hg_2 + \Dg^+\rla \Hg \Dg + \Hg^+,
\end{equation}
behaves unexpectly with temperature \citep{Smith1982ApJ...263..123S,Honvault2013JPCA..117.9778H,Gonzalez2013JChPh.139e4301G,Lara2015JChPh.143t4305L} and follows a Langevin rate of $\sim$2.1$\times$10$^{-9}\ratecoeff$  ($k=2.05 \times 10 ^{-9} (T/300) ^{0.2417} \exp(-3.733/\Td)\ratecoeff$). 
The backward reaction is endothermic by 462~K. D$^+$ is formed from H$^+$ via
\begin{equation}
\Hg^+ + \Dg \rla\Hg + \Dg^+,
\end{equation}
with an endothermicity of 41~K \citep{Watson1976RevModPhys.48.513}. In region of low ionisation (low Cosmic Ray flux), this route may become inefficient.
At low temperature, the rates of neutral-neutral reactions are negligible without tunnelling effects. Forward deuteration fractionation reactions are reversible with the forward reaction favoured because of the difference in zero-point energies. 

 \H2   and HD can be photodissociated and a self-shielding factor applies to both molecules with \H2 being much more shielded than HD \citep{LePetit2002A&A...390..369L,Thi2010MNRAS.407..232T}.
 
\subsection{Adopted data}~\label{molecular_data}

The conclusions from our modeling depend on the choice of molecular data. It is believed that amorphous silicates can react with H (D) atoms  more effectively and/or rapidly than crystalline silicates because they are thermodynamically unstable. 
Silicate surfaces show a distribution of chemisorption sites with binding energy ranging from $\sim$1000~K to $\sim$ 20000~K \citep{Oueslati2015_C4CP05128A}. The activation energy to overcome the barrier and chemisorb follows the Bell-Evans-Polanyi principle. As the dust grain temperature increases chemisorption sites with deeper potential are open so that we assume that a binding energy is 35$\times \Td$, knowing that the typical desorption occurs at \EbHc/30~K \citep{Luna2017ApJ...842...51L}:
\begin{equation}
\EbHcf/k = 35 \times \Td \ {\rm K},
\end{equation}
with $min$(\EbHc$/k$)~=~10,000~K and $max$(\EbHc$/k$)~=~25,000K. We adopted a simple relation between the activation energy $\Eactc$ and the binding energy  \EbHc
\begin{equation}
\Eactc/k = \EbHcf/k -9100 \ {\rm K},
\end{equation}
which gives an activation energy of $E\actH/k$=~900~K for a binding of \EbHc$/k$~=~10,000~K. The lowest activation energy is consistent with the value of \cite{Cazaux2002ApJ...575L..29C}. Both paths to form a chemisorbed H have the same barrier energy ($E\gcH=E\pcH=E\actH$). The barrier energy for the reaction between a physisorbed and a chemisorbed H-atom is $E\HpHc$=$E\actH$. There is no barrier for the Eley-Rideal \H2 formation ($E\HgHc$=0) and the barrier is $E\HcHc=2E\actH$ for \H2 formation from two chemisorbed atoms because it implies the breaking of two chemisorption bonds.
\begin{table}
\begin{center}
  \caption{Surface molecular data for physisorption processes.\label{tab_molecular_data}}
\vspace*{-0.5mm}          
\begin{tabular}{llll}
    \toprule
  Surface  & $\Eb/k$ &  $E\diff /k$  & $a_{\mathrm{db}}$ \\
                 &    (K)   & (K)    &   (\AA)  \\
\noalign{\smallskip}   
\hline
\noalign{\smallskip}    
\multicolumn{4}{c}{H atom}\\ 
\noalign{\smallskip}    
 Amorph. silicate   & 510$^a$   & 406$^a$ &  3.8$^a$  \\
 Amorph. ice          &  650$^b$  & 341$^c$ &  3.9$^c$  \\
                              &  607$^d$  & 516$^d$        &            \\
 Poly-crystalline ice        &                 & 256$^e$ & \\     
 \noalign{\smallskip}   
 \hline
 \noalign{\smallskip}   
 \multicolumn{4}{c}{D atom}\\ 
\noalign{\smallskip}   
 Amorph. sillicate   &    569$^a$        &  406$^a$   &   3.8$^a$     \\
 Amorph. ice          &   708$^{b,f}$   &   341$^{g}$ &        \\ 
 Amorph. ice          &   665$^d$        &   415.8$^d$&        \\
                              &                         &  255$^{h}$  & \\
Poly-crystalline ice  &                       &   267$^e$ & \\                   
\noalign{\smallskip}     
\bottomrule
\end{tabular}
\\
\resizebox{90mm}{!}{
\begin{minipage}{100mm}{$^a$exp.:\cite{Perets2007_1538-4357-661-2-L163};
$^{b}$theor.:\cite{Al-Halabi2007MNRAS.382.1648A}; $^{c}$exp.: from a fit to the data from \cite{Congiu2014FaDi..168..151C};$^d$exp.:\cite{Perets2005ApJ...627..850P}; exp.:$^e$\cite{Kuwahata2015PhRvL.115m3201K} ; $^h$exp.:\cite{Matar2008A&A...492L..17M}}; 
\end{minipage}}
\vspace*{-1mm}
\end{center}
\end{table}
For physisorption, there is also a variety of experimental and theoretical energies, which are summarized in Table~\ref{tab_molecular_data}. The diffusion barrier and the desorption energies for H atoms on amorphous silicate were taken from \cite{Perets2007_1538-4357-661-2-L163}. \cite{Hama2012ApJ...757..185H} showed that there are many physisorption sites  with a central energy of 22meV. \cite{Asgeirsson2017_doi:10.1021} have theoretically shown that a distribution of desorption (32--77 meV or 371--894~K) and diffusion (1--56
meV or 11--650~K) energies exist for H in an amorphous water ice matrix with a linear dependency of $E\diffH=0.68$ \EbHp. The theoretical results confirm the experiments where more than one energies have been found \citep{Hama2012ApJ...757..185H}. \cite{Asgeirsson2017_doi:10.1021} provide a possible explanation on the discrepancy between different experimental results \citep{Manico2011_1538-4357-548-2-L253,Perets2005ApJ...627..850P,Horneker2003_r1943,Matar2008A&A...492L..17M}. At low H-coverage H is mostly adsorbed on the deep sites, while at high coverage the deep sites are occupied and H atoms seat on shallower sites. The distribution is narrower for crystalline ice. 

The greater mass of D compared to H results in a smaller zero-point-energy and therefore in a larger binding energy. The binding energy for D atom is $\simeq$~58~K (5 meV) higher than the value for H atom \citep{LePetit2009A&A...505.1153L}. The diffusion energy barrier is not affected by the difference in zero-point-energy and $E\diffD=E\diffH$ since the zero-point-energy is accounted for in both the initial and final site.

Our choice of energies for physisorption is guided by: (1) a unique relatively high binding energy of $E\desHp/k$=\EbHp$/k$=~600~K whether H is attached on amorphous silicate or on amorphous water ice corresponding to a low coverage situation; (2) we adopted the scaling law of \cite{Asgeirsson2017_doi:10.1021} and obtained $E\diffH/k$=~408~K; (3) $E\desDp/k$=$E\desHp/k$+58~=~658~K; (3)  
$E\diffD=E\diffH$=~408~K. The chemisorption binding energies for HD are also increased by 58~K from the values for \H2. 

We illustrate the application of the Bell's formula to the tunnelling diffusion of physisorbed H-atoms. The physisorbed H-atom diffusion behaviour on cold grain surfaces is discussed in details in \citet{Congiu2014FaDi..168..151C}. In the Bell's formula, $E\diff $ is the surface thermal diffusion in ergs. For physisorbed H-atoms we adopted a diffusion energy of 408~K. The barrier width to tunnelling $a\ppH$ on amorphous water ice surface is found by matching the experimental data \citep{Congiu2014FaDi..168..151C} and a reasonable value is $a\ppH=$ 3.9 $\times$ 10$^{-8}$ cm (3.9~\AA). $h$ is the Planck constant in erg s and $m$ is the mass  of the diffusing species in grams. 
\begin{figure}[!ht]  
  \centering  
   \resizebox{\hsize}{!}{\includegraphics[angle=0,width=10cm,height=9cm,trim=20 20 70 200, clip]{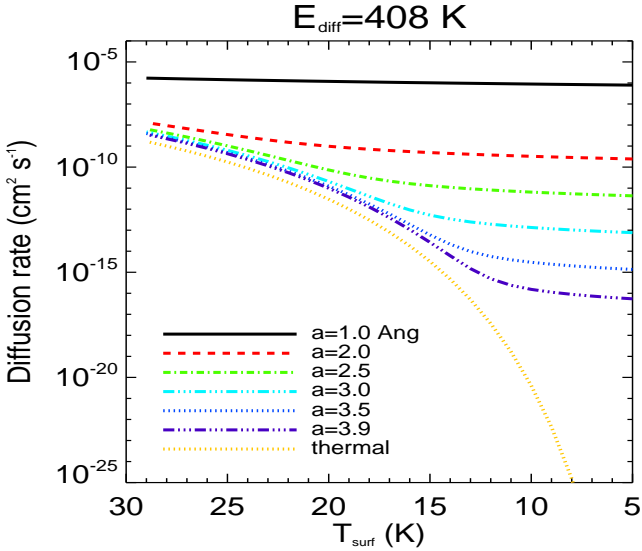}}
\caption{Atomic hydrogen grain surface diffusion rate as function of the surface temperature for different values of the barrier width $a$ in \AA. In our model, we chose a value of 3.9 \AA. The thermal diffusion rate is also shown.}
  \label{fig_Hdiffusion}            
\end{figure}  
 The surface diffusion rate for H-atom $k\diff =\nu_0 Q\diff  / N_{\rm surf}$ (cm$^{-2}$ s$^{-1}$) is shown for different choices of the barrier width in Fig.~\ref{fig_Hdiffusion}. The figure also shows a purely thermal surface diffusion rate, which drops dramatically at low dust temperatures below $\Tq$=11~K from eq.~\ref{eqn_quantum_temperature}. 

\section{Previous analytical \H2 formation models on silicate dust grains}\label{analytical_models}

\subsection{Cazaux model}

The standard analytical model for \H2 formation follows the model of \cite{Cazaux2002ApJ...575L..29C,Cazaux2004ApJ...604..222C}, which has been used since \citet{Hollenbach1971ApJ...163..165H}. The rate is 
\begin{equation}
  R_{\rm H_2}^{\rm Cazaux} \equiv \frac{1}{2}\,\varepsilon
                \,4\pi  \a2 
                \,v^{\rm th}_{\rm H}(\Tg)\,S_{\rm H}\,n_d \equiv \frac{1}{2}R\adsH \rate, 
\label{H2form}
\end{equation}
where $v^{\rm th}_{\rm H}\!=\!(k\Tg/(2\pi\,m_{\rm H}))^{1/2}$ is the thermal
relative velocity of the hydrogen atom, $4\pi
\a2,\nd$ is the total surface of the dust component per volume, and
S$_{\rm H}$ is the sticking coefficient, and $\epsilon$ is the recombination efficiency. The formula can be derived from the steady-state balance between formation and destruction of physisorbed H-atom H\#
\begin{equation}
dn\Hp/dt=R\adsH\nH-R\desHp\nHp-2k\Hmolpp\nHp\nHp=0
\end{equation}
and the \H2 formation rate is
\begin{equation}
(d\nHmol/dt)_{\rm form}=k\Hmol\nHp\nHp \ratedens.
\end{equation}
At low temperatures and low UV field, we can neglect the desorption, and we obtain the formation rate density
\begin{equation}
(d\nHmol/dt)_{\rm form}=\frac{1}{2}R\adsH\nH \ratedens,
\end{equation}
where $\nH$ is the number density of atomic hydrogen in the gas (in cm$^{-3}$). The rate density is strictly speaking an upper limit since H atoms have to adsorb first.  For HD formation, we follow the model described in \cite{Cazaux2009A&A...496..365C} and the rate assumes the same efficiency as for \H2.

\subsection{Jura's empirical \H2 formation rate coefficient}\label{model_Jura}
The \H2 formation (eq.~\ref{H2form}) can be rewritten to recover the  standard rate for \H2   formation, which has been measured by \cite{Jura1974ApJ...191..375J,Jura1975ApJ...197..581J,Jura1975ApJ...197..575J}. The observed rate has been derived from observations obtained by the Copernicus satellite in diffuse clouds and confirmed by \cite{Gry2002A&A...391..675G} using FUSE observations. 
First we define the average number of dust grains as
\begin{equation}\label{eq_ndust}
n_{\mathrm{d}} = \frac{1.386\ \mathrm{amu}\ \nHtot \delta}{(4/3)\pi \rho_{\mathrm{d}} r^3} \ \cm3, 
\end{equation}
which can be approximated by
\begin{equation}
\nd \approx 1.83\times10^{-15}\, \nHtot \left(\frac{\mathrm{\mu m^3}}{r_{\mu m} ^3}\right)\left(\frac{\delta}{0.01}\right)\cm3,\label{nd_formula}
\end{equation}
where $\delta$ is the dust-to-gas mass ratio assumed to be 0.01 and the total gas number density is $\nHtot=n_{\rm H}+2 n_{{\rm H}_2}$. We have assumed a silicate mass density $\rho_{\rm d}$ of 3.0 g cm$^{-3}$. The number abundance of Helium is 0.096383, giving an extra mass to the gas of 0.386. For an average grain radius $r$ of 0.1 micron and an efficiency of  unity, we can find
\begin{equation}
R_{\mathrm{H}_2} = 4.18\times 10^{-18} S_{\rm H} \sqrt{\Tg}(0.1\  \mu{\rm m}/  r_{\mu m} )\nHtot \rate.
\end{equation}
The formation rate does not explicit dependent on the dust temperature. The implicit assumption ($\varepsilon$=1) is that all adsorbed H-atom will eventually leave the grains as \H2. 
The \H2 formation rate coefficient for a gas at 80~K as found in the diffuse interstellar medium,  $r_{\mu m} $ =0.1 $\mu$m,   and a sticking coefficient of unity, is 
\begin{equation}
k_{\mathrm{H_2}} = 3.74\times10^{-17}  \ratecoeff,
\end{equation}
which is consistent with the values found by \cite{Gry2002A&A...391..675G} between 3.1 $\times10^{-17}$ and 4.5 $\times10^{-17}\ratecoeff$. They confirmed the earlier results of \cite{Jura1975ApJ...197..581J}, who found a \H2 formation rate coefficient of 3.0 $\times10^{-17}\ratecoeff$. Therefore, Jura's \H2 formation rate coefficient is compatible with the highest possible rate for \H2 formation on silicate dust grains. The empirical \H2 formation rate coefficient relies on a detailed knowledge of  the \H2 photodissociation rate in the clouds.
\section{ProDiMo chemical models}\label{prodimo_models}
\begin{table}
\begin{center}
\caption{Cloud model parameters. \label{tab_cloud_models}}
\vspace*{-0.5mm}          
\resizebox{88mm}{!}{\begin{tabular}{lcc}     
\toprule
Parameter & symbol & values\\ 
\noalign{\smallskip}     
\hline
\noalign{\smallskip}  
Gas density & $\nHtot$        &   2$\times$10$^{4}$ $\cm3$\\
Temperature & $\Td=\Tg$     &  10 to 700~K\\
Extinction & A$_{\rm V}$             & 10\\ 
Strength of interstellar UV       & $\chi^{\rm ISM}$   & 1\\
Cosmic ray H$_2$ ionisation rate  & $\zeta_{\rm CR}$    & $\!\!1.7\times 10^{-17}$~s$^{-1}\!\!\!$\\
Mean grain radius & $r$  & 10$^{-5}$ cm\\
Dust-to-gas mass ratio & $\delta$ &  0.01\\
PAH abundance rel. to ISM         & $f_{\rm PAH}$         & 1\\
\noalign{\smallskip}
\bottomrule
\end{tabular}}
\vspace*{-2mm} 
\tablefoot{$\chi^{\rm ISM}$~=~1 is the ISM Draine UV field. $f_{\rm PAH}$~=~1 corresponds to a PAH abundance of 3 $\times$ 10$^{-7}$.}
\end{center}
\end{table}
{\sc ProDiMo} is a code built to model the gas and dust grain physics
and chemistry
\citep{Woitke2009A&A...501..383W,Kamp2010A&A...510A..18K,Woitke2016A&A...586A.103W}. It has been
used to model disk Spectral Energy Distributions
(SEDs, \citealt{Thi2011MNRAS.412..711T}), water deuteration chemistry
\citep{Thi2010MNRAS.407..232T} CO rovibrational emissions including
UV-fluorescence \citep{Thi2013A&A}, and many {\it
  Herschel} observations from the {\it GASPS} large programme \citep{Dent2013PASP..125..477D}.  X-ray
physics are implemented
\citep{Aresu2012A&A...547A..69A,Meijerink2012A&A...547A..68M,Aresu2011A&A...526A.163A}. A detailed discussion of the different physics and their implementations are given in the articles listed above. Here we summarize the main features. In our chemical modeling we included 116 gas and ice species,
PAHs.  Self-shielding against photodissociation for \H2 and HD is taken into account. 
Reaction rate coefficients that are not explicitly discussed in this paper are taken from UMIST2012 \citep{McElroy2013A&A...550A..36M}
The adsorption energies are mixed from various sources \citep{Aikawa1996ApJ...467..684A},\cite{Garrod2006A&A...457..927G}, and UMIST2012 \citep{McElroy2013A&A...550A..36M}. The network was complemented by reactions relevant to high temperature conditions \citep{Kamp2017A&A...607A..41K}.
We used the chemistry solver in the {\sc ProDiMo} code in a zero-dimensional model. Further modeling in the context of protoplanetary disks will be reported in subsequent articles. The assumptions are $\Td$=$\Tg$, a fixed UV field strength and extinction $A_{\rm V}$ for the zero-dimensional model. We expanded the so-called small disk chemical network by including the species require modeling \H2 and HD formation (see Table~A\ref{tab:standard-species}). Only relevant surface and gas-phase chemical reactions are included, and they are listed in Tables~\ref{tab_dust_reactions} and ~\ref{tab_gas_reactions}. The later table lists reactions with singly-hydrogenated PAHs. Similar reactions with multi-hydrogenated PAHs are used in the modeling but are not shown. The results of runs with multiply-hydrogenated PAHs are shown in appendix \ref{appendix_PAH}.
The elemental abundances are taken from \citet{Kamp2017A&A...607A..41K} and in addition the adopted deuterium elemental abundance is 1.5$\times$10$^{-5}$ \citep{Linsky1995ApJ...451..335L}.
\begin{figure}[!htbp]
  \centering
  \includegraphics[angle=0,width=9.0cm,height=7.5cm,trim=25 70  70 300, clip]{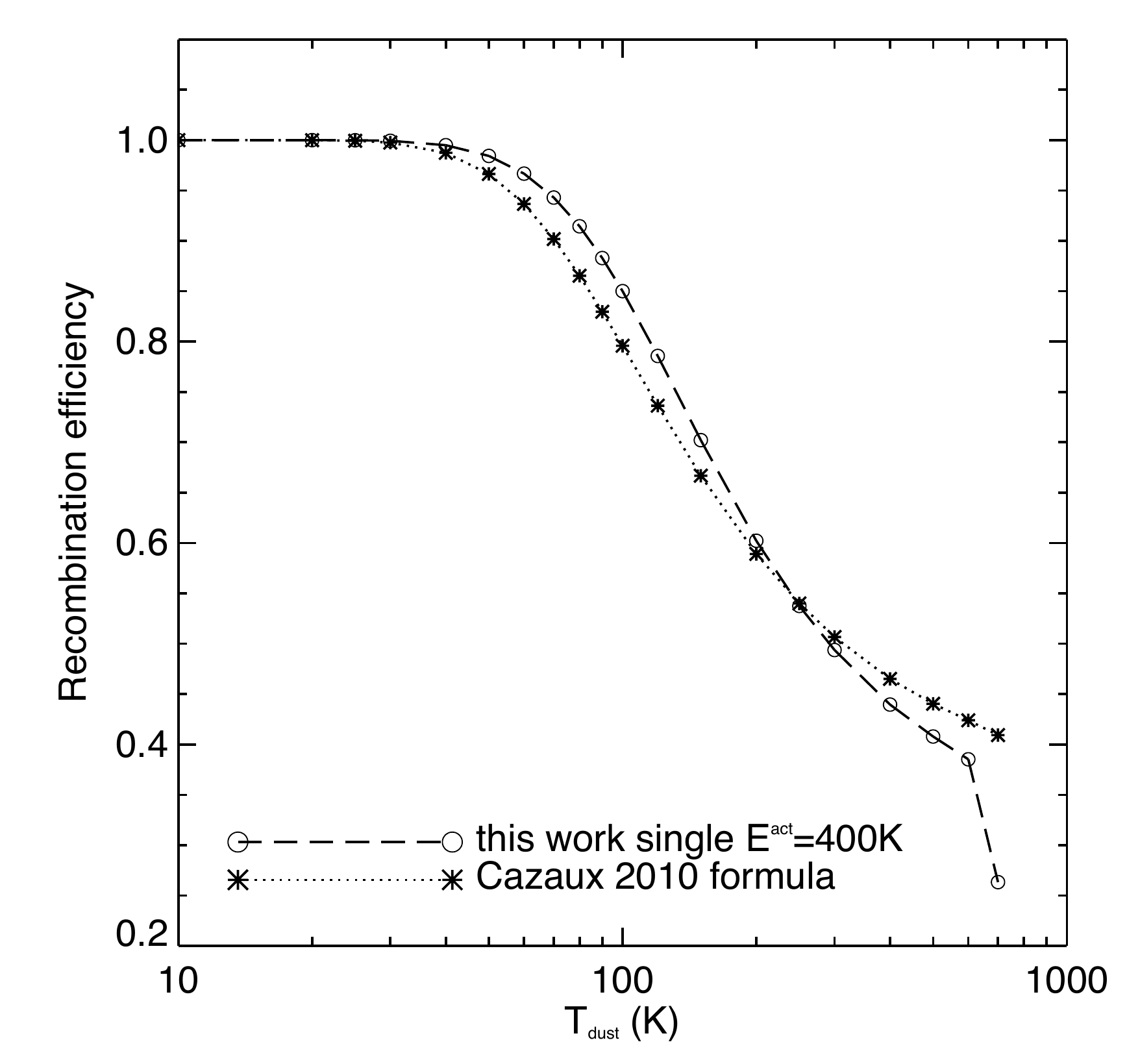}
  \caption{Comparison between the analytic \H2 formation efficiency using and the formula from \citet{Cazaux2004ApJ...604..222C} with the 2010 update (the binding energy for physisorbed H-atom $E_{\rm H_p}=600$~K, and the saddle point $E_{\rm S}=E_{\rm H_p}-E\rm{^{pc}}= 200$~K, with $E\mathrm{^{pc}}$ being the activation energy to overcome to go from a physisorption site to a chemisorption site) and our numerical code for a zero-dimensional model.}
  \label{fig_H2_formation_efficiency}          
\end{figure}  
\begin{table}[!htbp]
\begin{center}  
\caption{Cloud models. \label{tab_model_series}}
\vspace*{-0.5mm}          
\begin{tabular}{llccc}     
\toprule
\# & Model & Physi- & Chemi- & Formation\\ 
    &            &    sorption                    &       sorption                  & on PAHs\\
\noalign{\smallskip}     
\hline
\noalign{\smallskip}  
1 & \1M & &  &   \\
2 & \2M & \Yes &  &   \\
3 & \3M & \Yes & \Yes & \Yes\\
\noalign{\smallskip}
\bottomrule
\end{tabular}
\end{center}
\end{table}
\section{Results \& discussions}~\label{Results}
\begin{figure*}[!htbp]
  \centering
  \includegraphics[angle=0,width=9.0cm,height=8cm,trim=40 70  70 285, clip]
  {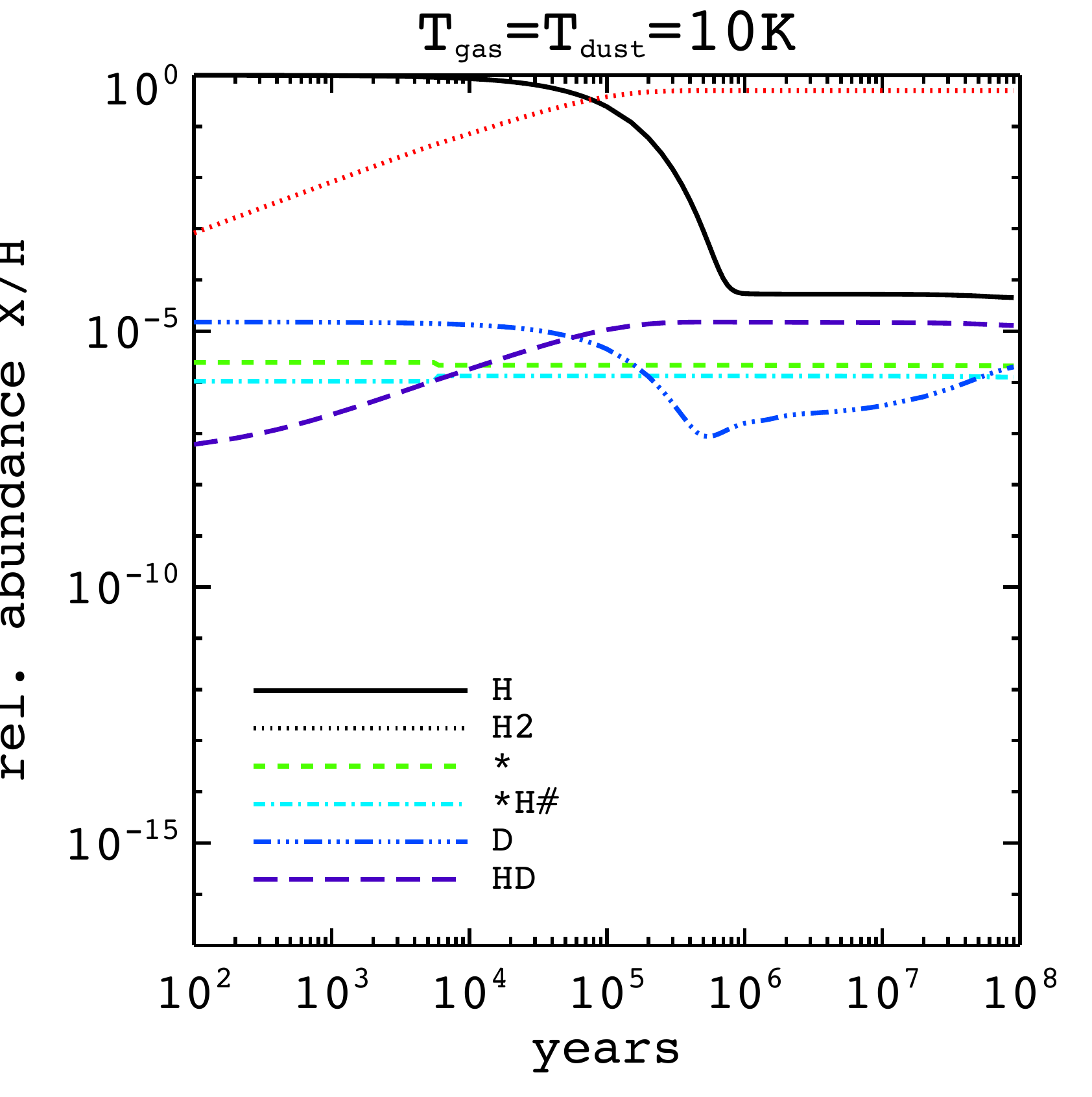}  
  \includegraphics[angle=0,width=9.0cm,height=8cm,trim=40 70  70 285, clip]{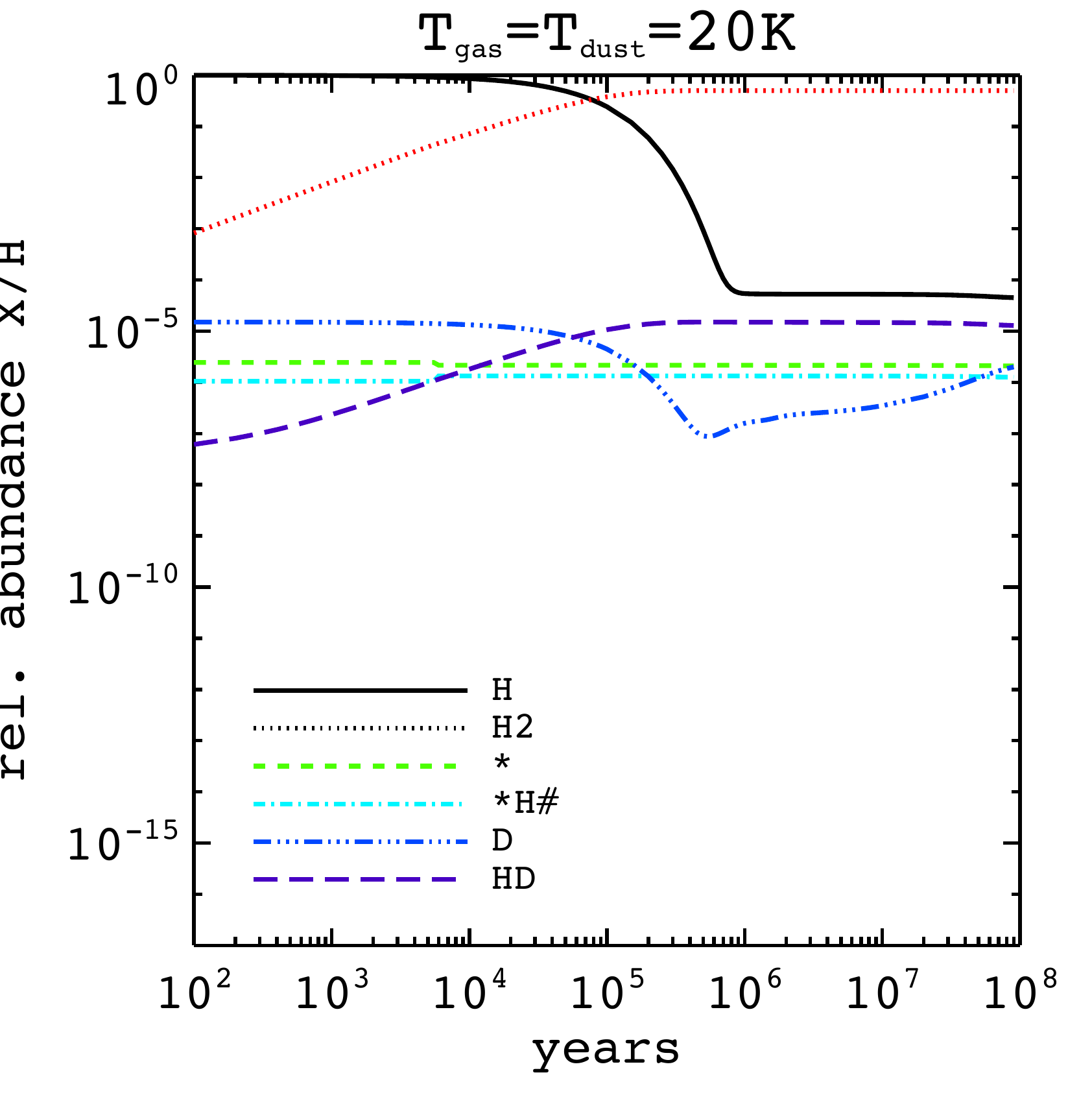}
 \includegraphics[angle=0,width=9.0cm,height=8cm,trim=40 70  70 285, clip]{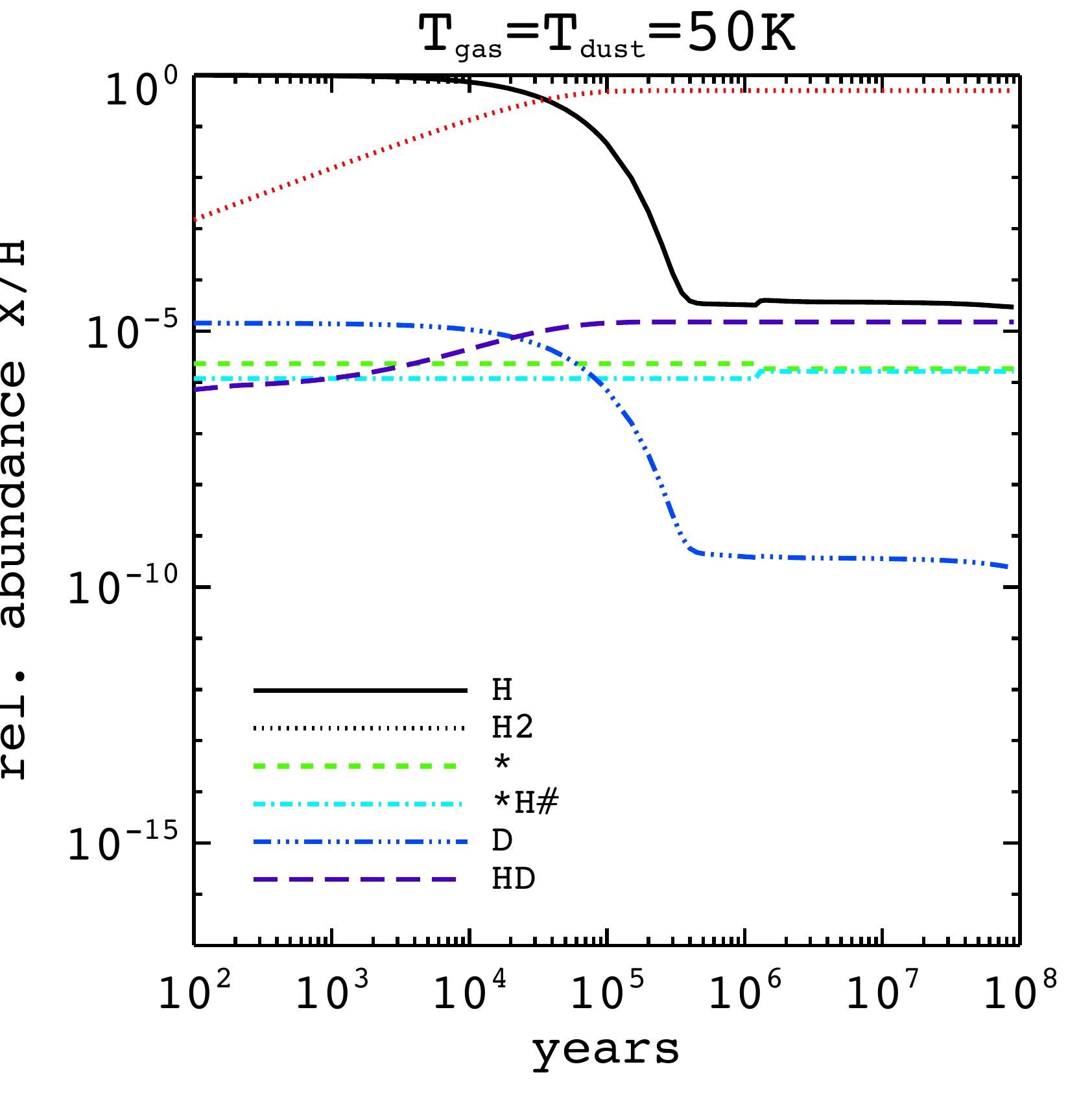}
\includegraphics[angle=0,width=9.0cm,height=8cm,trim=40 70  70 285, clip]{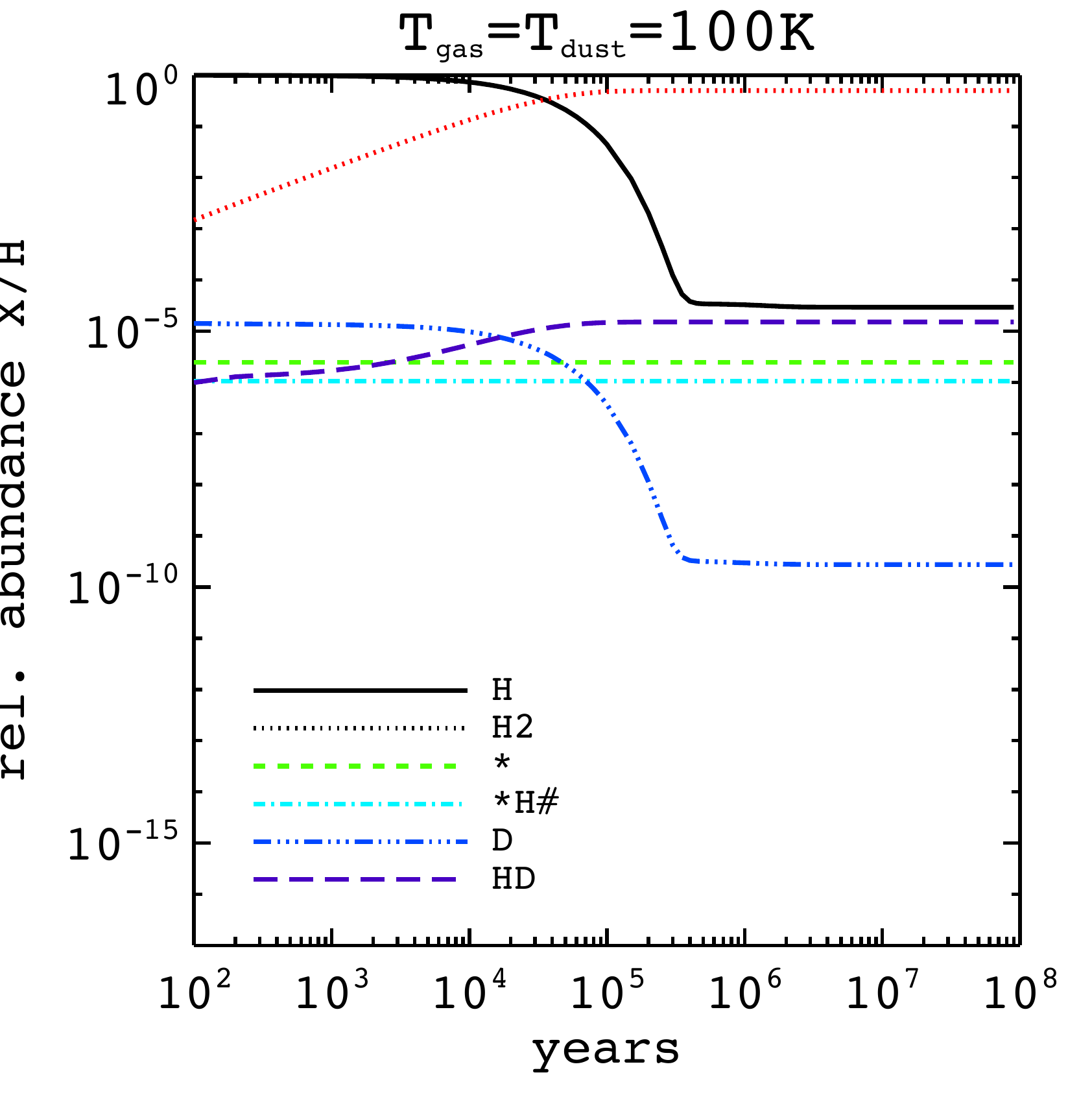}  
 \includegraphics[angle=0,width=9.0cm,height=8cm,trim=40 70  70 285, clip]{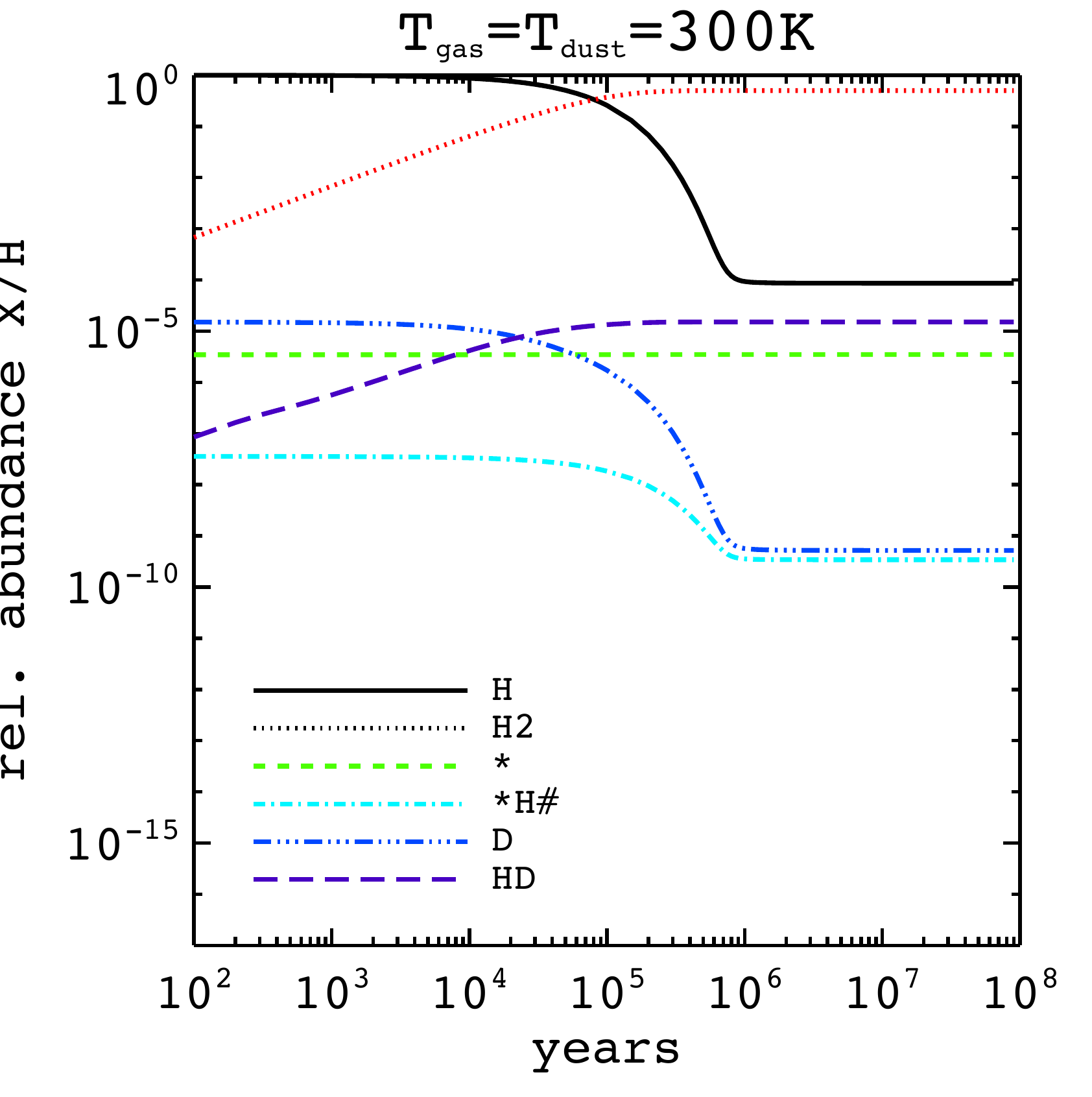}
 \includegraphics[angle=0,width=9.0cm,height=8cm,trim=40 70  70 285, clip]{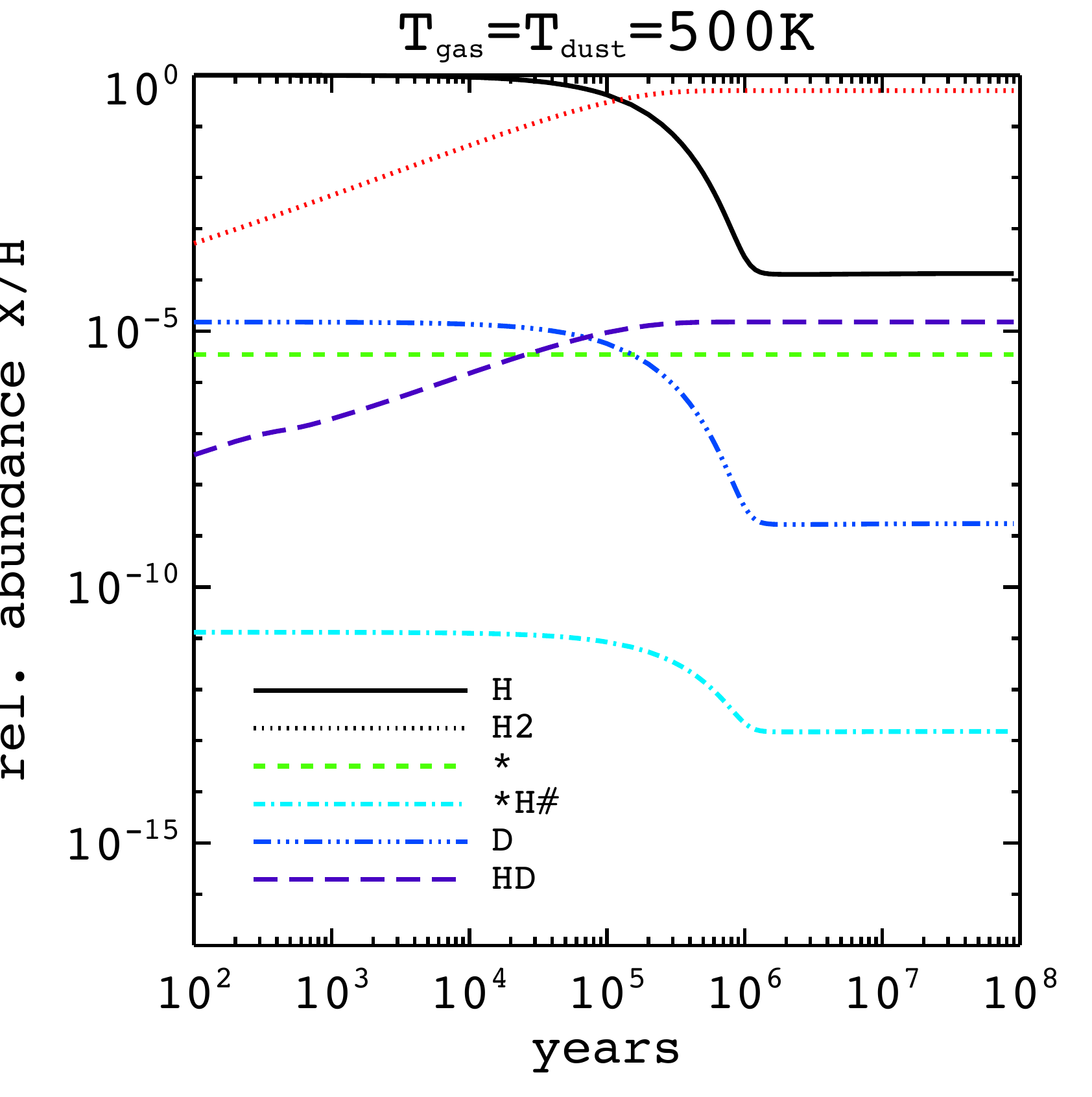}  
  \caption{H, \H2, D, HD, and physisorbed H as function of time using the surface chemistry model with physisorbed and chemisorbed species and a unique physisorption to chemisorption activation energy of 400~K. The first row are from left to right models at 10 and 20~K; the second row are models at 50 and 100~K. The bottom row shows models at 300 and 500~K. The * symbol corresponds to the abundance of unoccupied chemisorption sites.}
  \label{fig_mc_results1}          
\end{figure*}  
%
\subsection{Analytical \H2 formation efficiency}
The \H2 formation rate at all temperatures without considering the Eley-Rideal processes reads
\begin{equation}
\left(d\nHmol/dt \right)_{\rm LH}= k\ppHmol \nHp \nHp + k\pcHmol \nHp \nHc + k\Hmolcc\nHc\nHc.\label{eq_H2rate_LH}
\end{equation}
\begin{figure*}[!htbp]   
  \centering
  \includegraphics[angle=0,width=6.0cm,height=5cm,trim=40 70  70 285, clip]{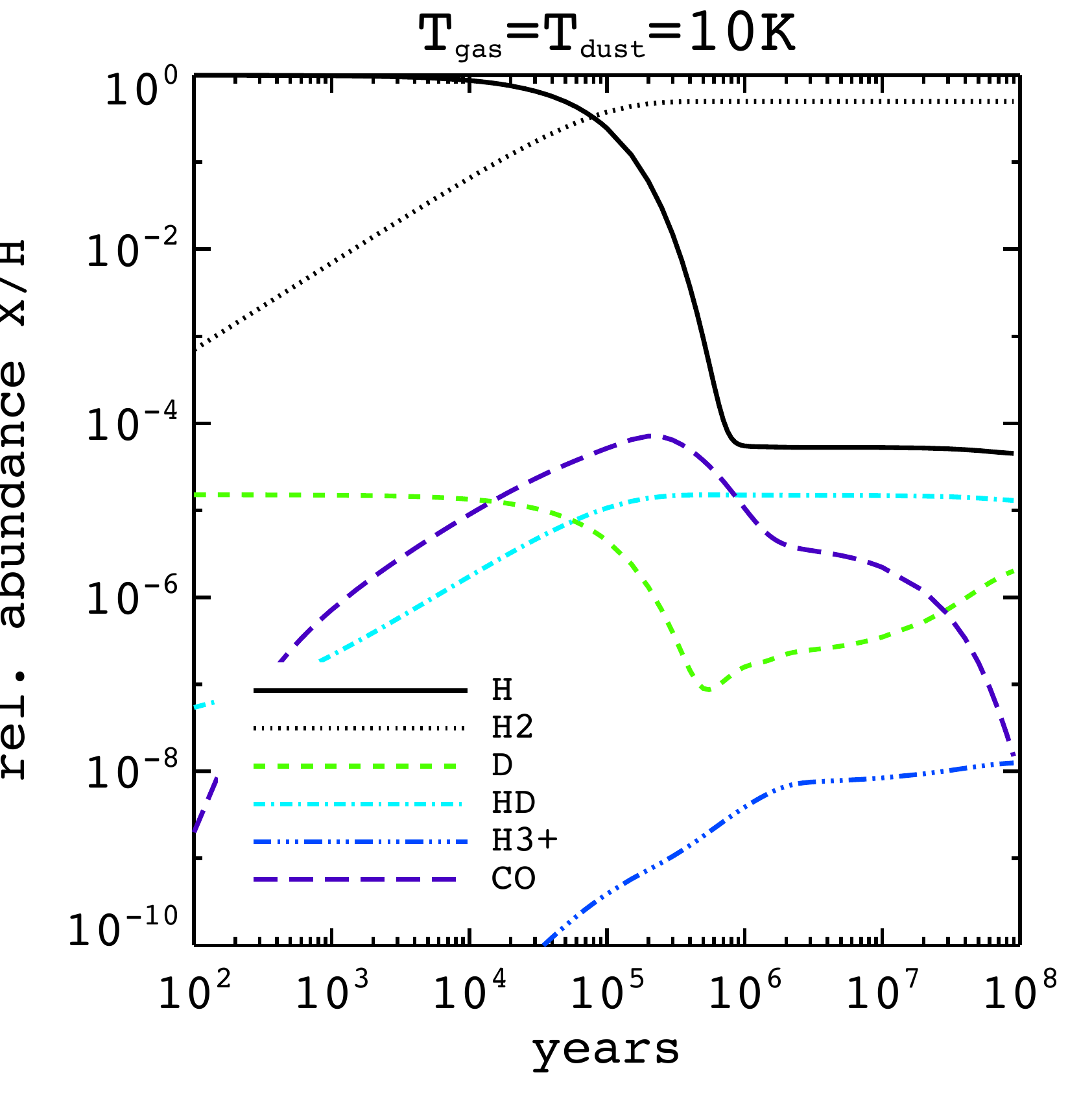}  
  \includegraphics[angle=0,width=6.0cm,height=5cm,trim=40 70  70 285, clip]{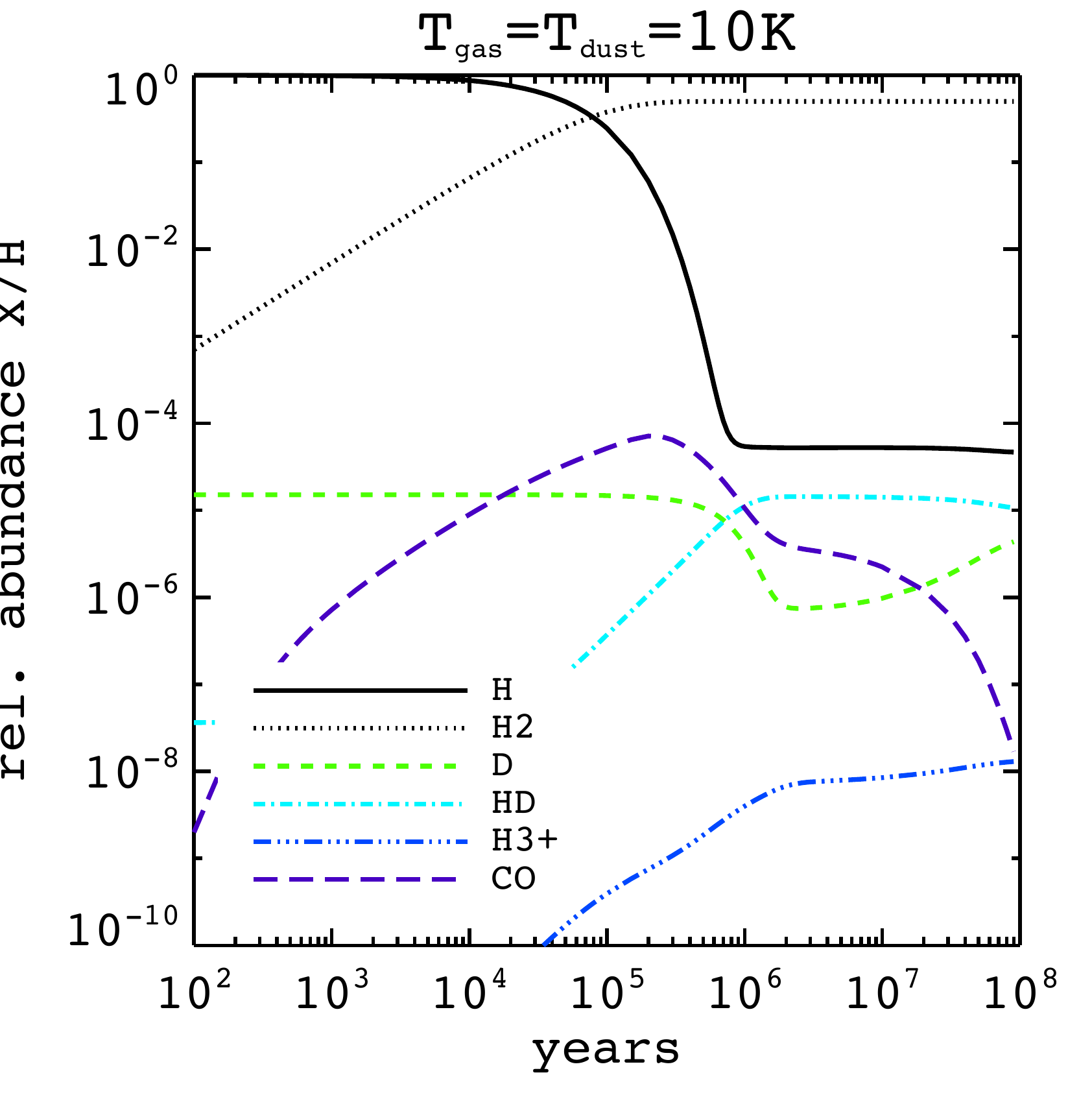}
  \includegraphics[angle=0,width=6.0cm,height=5cm,trim=40 70  70 285, clip]{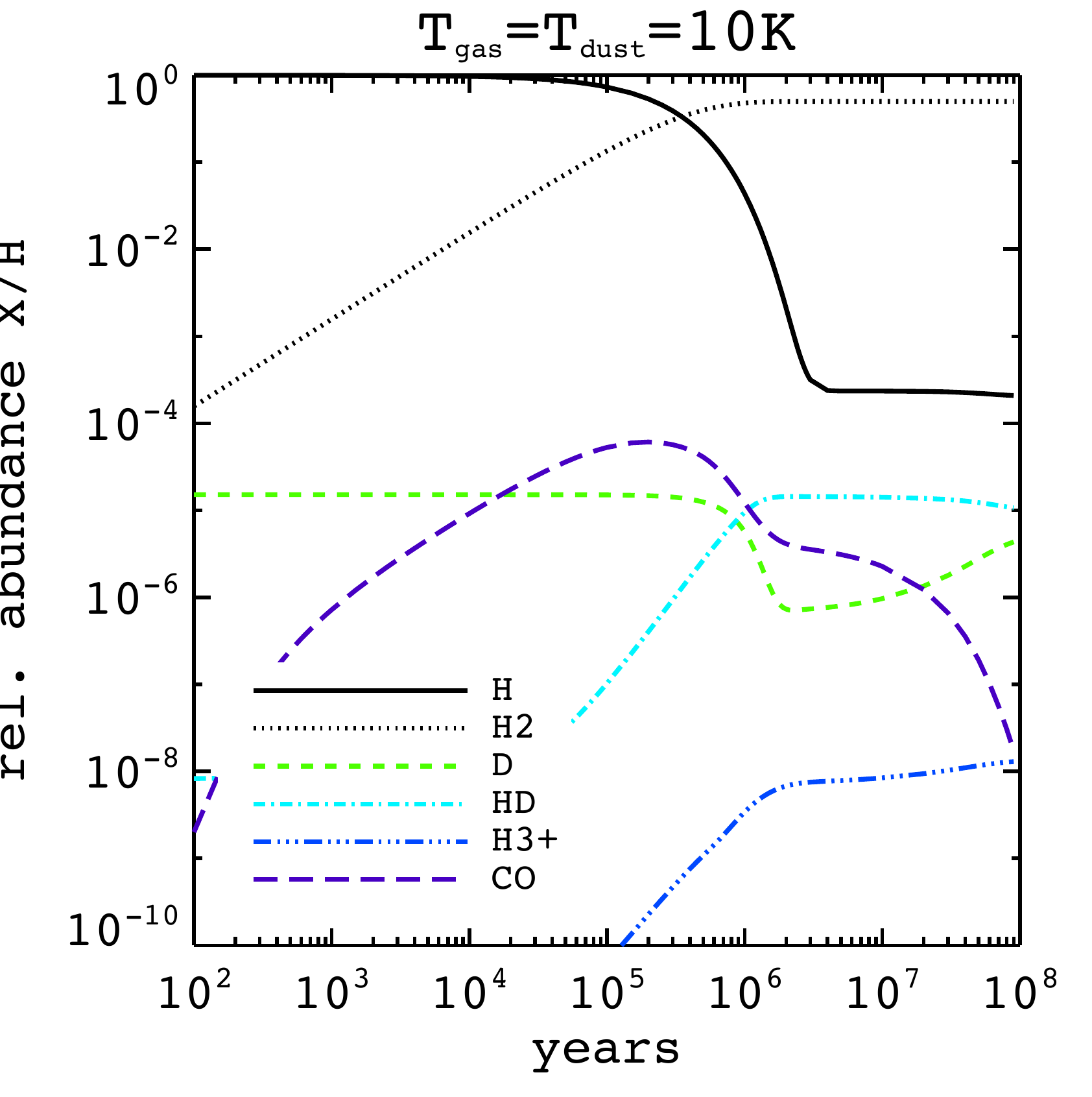}
  \caption{H, \H2, D, HD, and physisorbed H as function of time at 10~K for three models of \H2 and HD formation. The left model is a surface chemistry model without chemisorption species; the middle panel is the model with the analytical Cazaux formation model and the right panel utilises the \H2 formation model of \cite{Jura1974ApJ...191..375J,Jura1975ApJ...197..581J,Jura1975ApJ...197..575J}.}
  \label{fig_mc_results_10K}          
\end{figure*}    

 We assume that $E\HpHc =E\pcH={\rm min}(E\sil)=E\silHp=1000$~K and the barrier width $a\pcH$ is 1 \AA\ and obtain  $\Tq \simeq $ 78~K using eq.~\ref{eqn_quantum_temperature}. Here we did not study the effects of different values for the barrier width. In addition, the shape of the barrier has been assumed to be rectangular although the use of most realistic barrier profiles may affect the results  \citep{Taquet2012A&A...538A..42T}. Knowing that $n\surfchem=4\pi \Ns  \a2 \nd \simeq n_*+  \nHc$ (some chemisorbed sites can be occupied by D atoms), for $\Td$ below 78~K, we can use the quantum tunnelling transfer function in the recombination-limited approximation ($E\silHp \gg E\desHp$)
\begin{equation}
\Rpc \simeq \nu\pcH \exp{\left(-\frac{4 \pi a\pcH}{h} \sqrt{2 \mH E\silHp},\right)} \rate.
\end{equation}
For all $\Td<78$~K, $\Rpc \gg R\desHp$. The diffusion-chemisorption, i.e. when the rate coefficient is dominated by  the chemisorption barrier term, for $\Td>78$~K is  
\begin{equation}
k\pcH = \nu\pcH e^{-E\pcH/k\Td}/\nd \nsite\ratecoeff.
\end{equation}
The surface-mediated \H2 formation rate coefficient is in the thermal regime
\begin{equation}
k\HpHc \simeq \nu\HpHc e^{-E\HpHc/k\Td}/\nd\nsite \ratecoeff.
\end{equation}
The surface-mediated chemisorption processes rate becomes
\begin{multline}
\Rpc=k\pcH n_* + k\HpHc \nHc \simeq\\
 \nu\silHp e^{-E\silHp/\Td}\left(n\surfchem/\nd\right)/\nsite \rate,
\end{multline}
which arranges to
\begin{equation}
\Rpc \simeq \nu\silHp e^{-E\silHp/k\Td} \rate,
\end{equation}
In the thermal regime,
\begin{equation}
\frac{R\desHp}{\Rpc} \simeq  \frac{\nu\desHp e^{-E\desHp/k\Td}}{\nu\silHp e^{-E\silHp/k\Td}}
\end{equation}
\begin{equation}
\frac{R\desHp}{\Rpc} \simeq \sqrt{\frac{E\desHp}{E\silHp}}e^{-\left(E\desHp-E\silHp\right)/k\Td}.
\end{equation}
In the diffusion-limited regime ($E\silHp \ll E\desHp$), the ratio becomes
\begin{equation}
\frac{R\desHp}{\Rpc} \simeq \sqrt{\frac{E\desHp}{E\diffH}}e^{-\left(E\desHp-E\diffH\right)/k\Td}.
\end{equation}
The change of chemisorbed H-atom reads
\begin{equation}
d\nHc/dt=k\pcHmol\nHp(n_*-\nHc) -R\desHc\nHc- 2k\Hmolcc\nHc\nHc.
\end{equation}
In steady-state, the equation becomes
\begin{equation}
k\pcHmol\nHp(n\surfchem-2\nHc) -R\desHc\nHc- 2k\Hmolcc\nHc\nHc=0,
\end{equation}
which is a second degree equation in $\nHc$. The solution is
\begin{multline}
\nHc = 2\Rpc\nHp/\\
\left(-(R\desHc+2k\pcH\nHp) + \sqrt{(R\desHc+2k\pcH\nHp)^2+8 k\Hmolcc \Rpc \nHp} \right),\label{eq_nHc}
\end{multline}
where $ k\pcH\nHp n\surfchem = \Rpc\nHp$. If the desorption of chemisorbed H-atoms can be neglected
\begin{equation}
d\nHmol/dt \simeq \frac{1}{2}k\pcH\nHp n_*.
\end{equation}
Since $\nHp = R\adsH \nH / (\Rpc+R\desHp)\label{eq_nHp}$,
\begin{equation}
d\nHmol/dt \simeq \frac{1}{2} \left( { 1+ \frac{R\desHp}{\Rpc}} \right)^{-1} R\adsH \nH
\end{equation}
The combination of equations \ref{eq_H2rate_LH}, \ref{eq_nHc}, and \ref{eq_nHp} can be divided by the H-atom impinging rate to give the \H2 formation efficiency for the Langmuir-Hinshelwood processes. The \H2 formation efficiency has been derived by  \citep{Cazaux2002ApJ...575L..29C,Cazaux2004ApJ...604..222C} using a different model
\begin{equation}
\epsilon=\left( { 1+ \frac{R\desHp}{\Rpc}} \right)^{-1}\xi,
\end{equation}
with $R\desHp=\beta_{H_p}$, in \cite{Cazaux2002ApJ...575L..29C} and $\xi$ the correction factor at high temperatures, which reflects the H atoms desorbing from chemisorption sites. Recent developments have shown that a layer of \H2 does not prevent the further formation of more \H2, thus we can neglect the first term in the parentheses in the formula of \cite{Cazaux2004ApJ...604..222C}. Both analytical efficiencies are shown in Fig.~\ref{fig_H2_formation_efficiency}. Our curve has been obtained with a unique physisorption to chemisorption site activation energy of 400~K, which corresponds to a saddle point energy $E_{\rm s}$ of 200~K for the Cazaux's formula. Despite the differences in the treatment of many processes (presence of chemisorption sites, diffusion-mediated versus direct transfer from a physisorption site to an adjoining chemisorption site, ''blocking'' of chemisorption sites by ice layers, ...) between our model and the model from \cite{Cazaux2004ApJ...604..222C}, the agreement is remarkable for all temperatures apart in the upper end of the dust temperature range, i.e. when the efficiency drops faster than in the Cazaux's model due to the extra term $2k\pcH\nHp$ instead of only $R\desHc$ when the efficiency is limited by the desorption only. As the efficiency remains relatively high even at 600~K, the actual \H2 formation rate is limited by the sticking coefficient at high dust temperatures \citep{Cazaux2011A&A...535A..27C}. Part of the differences between the two efficiency curves can be ascribed to the use of different transmission function for $\Rpc$ (see Fig.~\ref{transmission_function}).
\begin{figure*}[!htbp]
  \centering
  \includegraphics[angle=0,width=9.0cm,height=7.5cm,trim=25 70  70 300, clip]{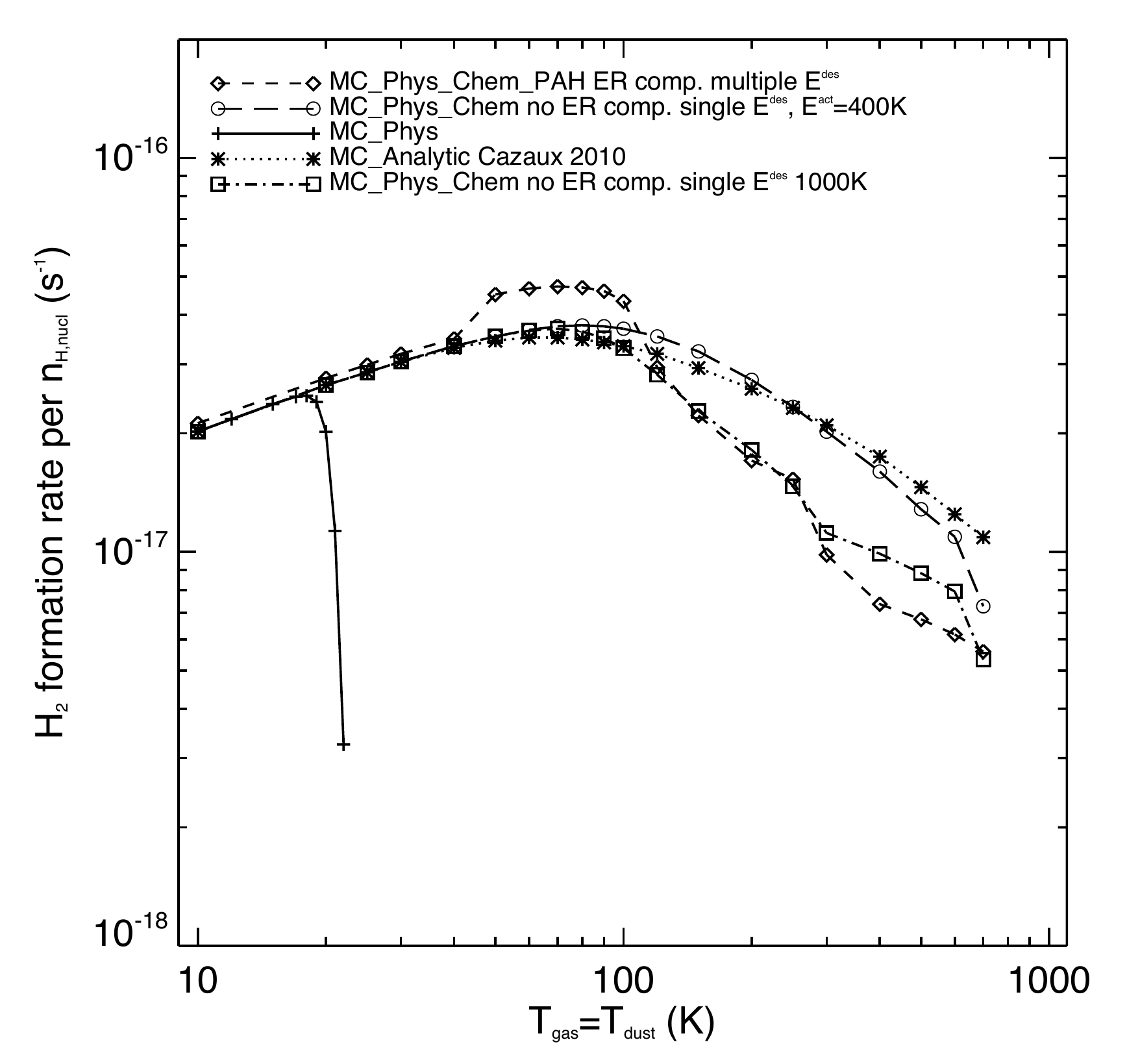}  
  \includegraphics[angle=0,width=9.0cm,height=7.5cm,trim=25 70  70 300, clip]{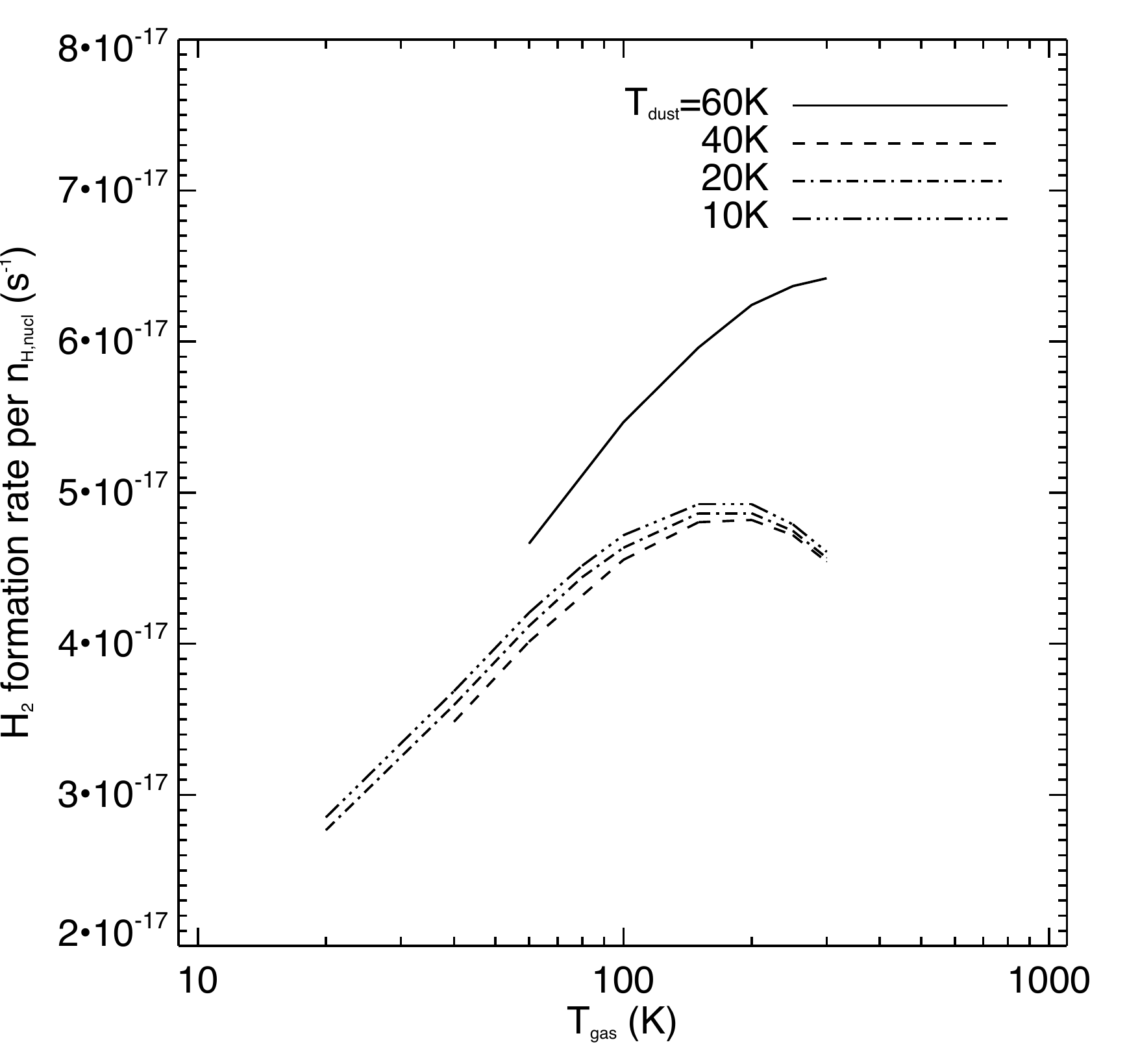}
  \caption{Effective \H2 formation rate. The left panel shows two \H2 formation models for a molecular cloud of $\nH=10^{4}$ cm$^{-3}$ and assuming $\Td=\Tg$ (\1M, \2M and \3M). We also show the effect of competition diffusion-reaction rates (no compet. \3M). The right panel show \H2 formation model \3M when $\Td$ and $\Tg$ can be different.}
  \label{fig_H2_formation_rate}          
\end{figure*}  
\subsection{Numerical models of molecular clouds}~\label{numerical_models}

The abundances of H, \H2, D, HD, H physisorbed and chemisorbed on grain surfaces  are shown for the model for a zero-dimensional model at several temperatures (10, 20, 50, 100, 300, and 500~K) with a single activation energy of 400~K from a physisorption to a chemisorption site at various grain temperatures in Fig.~\ref{fig_mc_results1}. The total conversion from H to \H2 is reached at all temperatures at (1--2)$\times$10$^{5}$ yrs. The number density of atomic H plateaus at $\sim$~1~cm$^{-3}$ after a few Myrs. At $\Td > 20$~K, HD abundance reaches its maximum faster than \H2. 

The abundance of surface hydrogen is predominately chemisorbed even at 10~K (see the top left panel of Fig.~\ref{fig_mc_results1}). Without chemisorption sites, the abundance of physisorbed H-atoms is 2.5 dex higher (see left panel of Fig.~\ref{fig_mc_results_10K}). This may affect the surface hydrogenation rates. The drop at time $>$ 10 Myrs of the HD abundance is due to the strong fraction of HDO gas and HDO ice. The reason for this high chemisorption site occupancy is the large number of available chemisorption sites and the high surface diffusion rate.

We also run series of models with only physisorption sites, a model with an activation barrier for chemisorption of 1000~K instead of 400~K, and a model with multiple barrier height and Eley-Rideal formation processes included. The \H2 formation rate for a 0.1 micron grain is shown for the various models in the left panel of Fig.~\ref{fig_H2_formation_rate}. The rates were computed from the outcomes of the models themselves by determining at each temperature the time $t_{\rm H_2}$ that the model needs to convert 2/3 of the hydrogen atoms into molecules, i.e. the time $t_{\rm H_2}$ such that $n_{\rm H_2}=n_{\rm H}$. The formation rate $R$ is computed as $R=1/(t_{\rm H_2} \nHtot)$.

Below 20~K, the \H2 formation is adsorption rate limited, independent on the actual \H2 formation rate on the grain surfaces as shown by the analytical treatment above. When the formation rate is low, the number density of H atom on the surface increases. This is the case when the rate is lower after the formation of H$_2$O ice layers.  As expected, the desorption of H\# above $\sim$~20~K prevents \H2 formation. The computed \H2 formation rate differences between the Cazaux model and the surface chemistry model with a single barrier for chemisorption corresponds to the differences seen in the analytical treatment (Fig.~\ref{fig_H2_formation_efficiency}). \H2 formation is efficient up to $\sim$~700~K (efficiency of 0.4), which corresponds to $\sim E\desHc/30$ \citep{Luna2017ApJ...842...51L}. Atomic D is only slightly more bound to the surface and will also form efficiently on grains up to  $\sim$~700~K. The building up of ice layers does not prevent surface H-atoms to reach the silicate surface and to adsorb on an empty chemisorption site. Both the diffusion and desorption rates decrease with increasing number of ice layers.

The model with a single barrier of 1000~K for chemisorption shows a drop in efficiency above 100~K. It reflects the decrease in the $R\desHp / \Rpc$ ratio with a higher energy barrier. In the model with varying barrier energy and chemisorption desorption energy, the efficiency is similar to the model with a single barrier at 1000~K. The higher \H2 formation rate is due to the \H2 formation by the Eley-Rideal mechanism, whose rate can be as high as the Langmuir-Hinshelwood rate \citep{LeBourlot2012A&A...541A..76L}. \H2 formation through single hydrogenated PAHs and cation PAHs is less efficient at all temperatures than on silicate grains because the rate is basically an Eley-Rideal mechanism rate. The models assume that only atomic H and D can chemisorb. In more realistic models where the chemisorption sites can be occupied by other species (e.g., -OH), the efficient of \H2 formation on silicate grains may decrease and the fraction of \H2 formed via hydrogenated PAHs increases.
The \H2 formation rate by H-atom abstraction of hydrogenated PAHs will increase if we adopt a larger abstraction cross-section (see appendix \ref{appendix_PAH}). Studies have suggested that cross-sections more than ten times higher than our value may be possible \citep{Zecho2002_188}.
The additional path to form \H2 through the hydrogenated PAHs is required to explain observed \H2 formation rates that are higher than the theoretical limit set by the H atom impinging rate on grain surfaces. In this study, we have not included \H2 formation via photodissociation of Hydrogenated Amorphous Carbon nano-grains (HACs). \citet{Jones2015A&A...581A..92J}  examined this route and concluded that this mechanism could be efficient for warm grains ($T$=50--100~K) in moderately UV-illuminated regions ($I_{\mathrm{UV}}=$1--100). \cite{Duley1986MNRAS223} has described a model of \H2 formation on carbonaceous grains at temperartures found in protoplanetary disks. The \H2 formation on HACs would be explored in future studies.

In Fig.~\ref{fig_mc_results_10K} we compare the abundances of the standard model with models without chemisorbed species, the analytical Cazaux model and the Jura model \citep{Jura1974ApJ...191..375J,Jura1975ApJ...197..581J,Jura1975ApJ...197..575J}. As expected the \H2 formation rates in the full surface chemistry and in the analytical model are quasi-identical. HD formation rate in the Cazaux model is slower than in the other models. 

On the right panel of Fig.~\ref{fig_H2_formation_efficiency} the effects of different gas and dust temperatures are shown for dust grain temperatures between 10 and 60~K for models that include the Langmuir-Hinshelwood and Eley-Rideal \H2 formation mechanism. The ER \H2 formation is efficient for warm gas and dust consistent with \cite{LeBourlot2012A&A...541A..76L}. However, the efficiency drops for dust grains above 100~K when the desorption of physisorbed H-atoms start to overtake the transfer of the atom to a chemisorption site compared to the \H2 formation by the encounter between a physisorbed and a chemisorbed atom. The \H2 formation rate increases with grain surface temperature as expected, although just by a factor 2. This enhancement compared to the standard value may be insufficient to explain observations of abundant hot \H2 gas \citep{Habart2004A&A...414..531H}. 

Our conclusions may change in the future because our results depend critically on numerous rates that are difficult to measure or compute.
\section{Conclusions \& perspectives}~\label{Conclusions}

We have implemented a detailed \H2 and HD formation model in the \ProDiMo code. The model extends the rate-equation treatment of surface reactions to chemisorbed species assuming diffusion-reaction competition surface rates.  We modeled the Langmuir-Hinshelwood and Eley-Rideal mechanisms for physisorbed and chemisorbed H and D atoms. We introduced the concept of chemisorption sites as a pseudo-species, which facilitates the computation of chemisorbed species. We also accounted for the formation of \H2 and HD on  hydrogenated PAH and PAH cations. We implemented a charge exchange chemistry between the PAHs and the other gas-phase species. The Langmuir-Hinshelwood processes (between two physisorbed H-atoms, between a physisorbed and a chemisorbed atom, and between two chemisorbed atoms) are the main \H2 formation route at all dust temperatures. \H2 also form efficiently by the Eley-Rideal mechanism  between 50 and 100~K. The contribution of singly and multiply-hydrogenated PAHs and PAH ions to the \H2 formation is limited when a small cross-section for hydrogen abstraction is adopted. The value of the cross-section varies by orders of magnitude from study to study. More experimental and theoretical works are warranted before a more definitive conclusion on the importance of \H2 formation through hydrogenated PAHs can be drawn.
HD formation proceeds both on the grain surfaces and in the gas-phase. For all dust temperatures below the chemisorption desorption temperature in the case of low photodesorption, the surface-mediated chemisorption processes dominate over the desorption of physisorbed H-atoms. Our  implementation of a warm surface chemistry can be used to model \H2 and HD formation in Photodissociation Regions, protoplanetary disks, and other astrophysical environments where the dust temperature is above 20~K.  Future study will include the formation of \H2 on carbonaceous grain surfaces.

 \begin{acknowledgement}
 We thank Dr. Jake Laas for extensive discussions. IK,WFT, CR, and PW acknowledge funding from the EU FP7- 2011 under Grant Agreement nr. 284405. CR also acknowledges funding by the Austrian Science Fund (FWF), project number P24790. We thank the referee for the useful suggestions. 
 \end{acknowledgement}

\bibliographystyle{aa} 
\bibliography{surface_chemistry}
 
\begin{appendix}
\section{Sticking coefficient}~\label{sticking_coeff}

Although the sticking coefficient can modify the \H2 (HD) formation rate, it does not effect the efficiency of the surface recombination. We discuss here a simple qualitative model, whose can help us understand the behaviour of the sticking coefficient as function of the parameters such as gas and dust temperature as well as the binding energy.
 
The Goodman's version of the modified Baule formula gives the transfer of energy between the incident particle and the surface accounting for surface at temperature $\Td$ \citep{Bonfanti2016QUA:QUA25192}
\begin{equation}
\delta \epsilon = 2.4 \frac{\alpha}{\left(1+\alpha\right)^2}\left(\epsilon+D-\frac{1}{2}k\Td \right),
\end{equation}
where $\alpha$ is the mass ratio between the incident species of mass $m_i$ to the surface mass $m\surf$, $\epsilon$ is the kinetic energy of the incident species at temperature $\Tg$ ($\epsilon=1/2k\Tg$), $\Eb$ is the binding energy, the surface is at temperature $\Td$. $D=3/4 \Eb$ as suggested by 
\cite{Hollenbach1970JChPh}.
The original Baule formulation uses a pre-factor of 4 instead of 2.4. When $\epsilon+D-\frac{1}{2}k\Td < 0$, the species returns to the gas-phase with extra energy acquired from the surface. 
Alternatively, one can use the formula in \cite{Hollenbach1970JChPh} modified to account for the dust surface temperature. First we define a characteristic
frequency of the repulsive collision
\begin{equation}
\omega_0 = \left[ \left(\epsilon+D-0.5k\Td \right)/2m_i b^2\right]^{1/2},
\end{equation}
where $b$ is the experimental determined slope parameter with a value $b=0.30$\AA\ from \citep{Hollenbach1970JChPh}. When $1\lesssim \omega/\omega_0 \leq 1.68$
\begin{equation}
\delta \epsilon = \alpha \left(\epsilon+D-\frac{1}{2}k\Td \right),
\end{equation}
where  $\omega$ is the Debye frequency of the solid $k \theta_{\rm D} = h \omega/2\pi$, where $\theta_{\rm D}$ is the Debye temperature in Kelvin. For crystalline water ice, we use  $\theta_{\rm D}$=222.2~K \citep{Flubacher1960doi:10.1063}, which gives $\omega=2.9 \times 10^{13}$ s$^{-1}$ compared to $\omega_0=4\times10^{13}$ s$^{-1}$ at 78~K \citep{Hollenbach1970JChPh}. The Debye temperature for crystalline olivine is 753~K \citep{Chung1971GeoJ...25..511C}. When $\omega/\omega_0 \geq 1.68$
\begin{equation}
\delta \epsilon = \alpha \left(\epsilon+D-\frac{1}{2}k\Td \right) 2\left[\omega_0^2/(\omega^2-\omega_0^2)\right].
\end{equation}
The sticking coefficient is within about 5\% \cite{Hollenbach1970JChPh}
\begin{equation}
S \approx\frac{\gamma^2+0.8\gamma^3}{1+2.4\gamma+\gamma^2+0.8\gamma^3},
\end{equation}
where $\gamma \equiv \Ec/k\Tg$,  $\Ec= \Omega \left(D\delta \epsilon\right)^{1/2}$. $\Omega$ is a factor
that accounts for the type of surface rebonds: $\Omega^2\simeq1$ for Lambert's law and $\Omega^2=2$ for isotropic scattering above the surface \citep{Hollenbach1970JChPh}. We chosed $\Omega^2=2$, consistent with a rough surface. The sticking coefficient reduces to $S \simeq \gamma^2$ for $\gamma \ll 1$ (inefficient energy transfer) and to $S\simeq 1-3/\gamma^2$ for $\gamma \gg 1$ (efficient energy transfer). The sticking coefficient increases with the binding energy. \cite{Hollenbach1970JChPh} used a value $\delta \epsilon =17$~K for H collision with water ice.


\section{PAH charge exchange chemistry}~\label{PAH_chemistry}

A model of PAH ionization has been presented by \citet{Dartois1997A&A...323..534D}.
We used the circumcoronene (\Circum) as typical PAHs
that are large enough to escape photodissociation in disks around
HerbigAe stars. The circumcoronene can be once negatively-charged
(PAH$^-$) and three times positively charged by absorbing a UV with
energy below 13.6 eV or by charge exchange reactions (PAH$^+$,
PAH$^{2+}$, PAH$^{3+}$, see Table~\ref{tab_PAH}). The effective radius of
a PAH is computed by \citep{Weingartner2001ApJS..134..263W}
\begin{equation}
a_{\PAH}=10^{-7}\left(\frac{N_{\mathrm{C}}}{468}\right)^{1/3}\ \mathrm{cm}, 
\end{equation}
where $N_{\mathrm{C}}$ is the number of carbon atoms in the PAH. The
radius for the circumcoronene is
$a_{\PAH}$(C$_{54}$H$_{18}$)=4.686 $\times$ 10$^{-8}$ cm. The
PAH ionisation potential can either be taken from the literature when
they are measured or estimated \citep{Weingartner2001ApJS..134..263W}
\begin{equation}
IP_{\PAH} = W_0 + (Z_{\PAH}+0.5)\frac{e^2}{a_{\PAH}}+(Z_{\PAH}+2)\frac{e^2}{a_{\PAH}}\frac{0.3\times 10^{-8}}{a_{\PAH}}\ {\mathrm{erg}}, 
\end{equation}
where $W_0$ is the work function assumed to be 4.4 eV
($7.05\times10^{-12}$ erg), and $Z_{\PAH}$ is the charge of
the PAH. The ionisation potentials (I.P.) are listed in
Table~\ref{tab_PAH}.

\begin{table}
\begin{center}
  \caption{Circumcoronene electron affinity and ionisation
    potential. The measured (lit.) and computed (WD2001) values are
    shown.\label{tab_PAH}}
\begin{tabular}{llll}
                  \toprule
                  & E.A. (eV) & I.P. (eV) lit. & I.P WD2001\\
\noalign{\smallskip}   
\hline
C$_{54}$H$_{18}$       & 1.3 & 5.9 & 6.2 \\
C$_{54}$H$_{18}^+$       & ... & 8.8 & 9.4\\
C$_{54}$H$_{18}^{2+}$       & ... & 12.9 & 12.5\\
\noalign{\smallskip}     
\bottomrule
\end{tabular} 
\end{center}
\end{table}
\subsection{PAH photoionisation and PAH$^-$ photodetachment}

Stellar and interstellar ultraviolet photons with energy below 13.6 eV
ionise neutral and ionised PAHs in disk surfaces
\begin{equation}
  \PAH + \UVr \ra\PAH^+ + \elec.
\end{equation}
The photoionisation rates at each disk location were computed by
integrating the product of the photoionisation cross-sections
calculated using the PAH model of \citet{Li2001ApJ...554..778L} with
the internal UV field obtained by solving the continuum dust radiative
transfer and a yield computed according to the prescription of
\citet{Jochims1996A&A...314.1003J}. PAH self-shielding is taken into
account. PAH$^-$ can lose its electron by absorbing a stellar or
interstellar UV photon (photodetachment):
\begin{equation}
  \PAH^{-} + \UVr \ra\PAH + \elec.
\end{equation}
The rates are computed the same way as for the photonionisation.

\subsubsection{Energetic particles induced photoionisation and
  photodetachment}

H$_2$ collisionally excited to Rydberg states by fast 'secondary'
electrons generated by energetic particles (cosmic ray, X-ray, or
radioactive decay) ionisation of hydrogen fluoresces in the
ultraviolet. The fluorescence photons have enough energies to ionise
an electron from a PAH
\begin{equation}
  \PAH + \mathrm{MUV} \ra\PAH^+ + \elec,
\end{equation}
or detach an electron from a PAH anion
\begin{equation}
\PAH^{-} + \mathrm{MUV} \ra\PAH + \elec,
\end{equation}
where MUV stands for UV generated by gas interaction with MeV
particles.  The number of fluorescence photons with energies between
7.1 and 14.6 eV is $f_{\mathrm{Ryd}}\zeta n_{\mathrm{H}}$
\citep{Flower2003MNRAS.343..390F}, where $f_{\mathrm{Ryd}}$ is the
fraction of the secondary electrons that excite the H$_2$, $\zeta $ is
the rate (s$^{-1}$) of total hydrogen ionisation (cosmic ray, X-ray,
and radioactive decay). We adopted a value of 0.15 for
$f_{\mathrm{Ryd}}$ \citep{Flower2003MNRAS.343..390F}.  The rate
coefficient for photoionisation triggered by energetic events is
\begin{equation}
  k_{\mathrm{pi,MeV}}=0.15 \zeta n_{\mathrm{H}} g_{\PAH} y_{\mathrm{pi}}\ \mathrm{cm}^{-3}\ \mathrm{s}^{-1}\ratecoeff, 
\end{equation}
where $g_{\PAH}$ is the fraction of the photons absorbed by
the neutral or positive PAHs compared to the total opacity and
$y_{\mathrm{pi}}$ is the yield of photodetachment. Likewise, the rate
coefficient for photodetachment is
\begin{equation}
  k_{\mathrm{pd,CR}}=0.15 \zeta n_{\mathrm{H}} g_{\PAH^-} y_{\mathrm{pd}}\ \mathrm{cm}^{-3}\ \mathrm{s}^{-1}, 
\end{equation}
where $g_{\PAH^-}$ the fraction of the photons absorbed by the
negative PAHs and $y_{\mathrm{pd}}$ is the yield of photodetachment.
The reactions may be important in the UV-shielded environments
\begin{table}[th]
  \caption{PAH cation recombination rate coefficients measured at room temperature (300~K) by \citet{Binnier2006_B516858A}. Alternative values are also listed in the bottom of the table.}             
  \label{PAH_recombination}
\centering          
\begin{tabular}{llll}     
\hline
\hline
\noalign{\smallskip}        
\multicolumn{1}{c}{Name} & \multicolumn{1}{c}{Formula} & \multicolumn{1}{c}{$k$ [cm$^{3}$ s$^{-1}$]} & \multicolumn{1}{c}{Ref.}\\
\hline
\noalign{\smallskip}   
 naphtalene    & C$_{10}$H$_{8}^+$ & $0.3(\pm 0.1)\times 10^{-6}$ & $a$\\
 azulene       & C$_{10}$H$_{8}^+$ & $1.1(\pm 0.1)\times 10^{-6}$ & \multicolumn{1}{c}{''}\\
 acenaphtalene & C$_{12}$H$_{10}^+$ & $0.5(\pm 0.2)\times 10^{-6}$ & \multicolumn{1}{c}{''}\\
 anthracene    & C$_{14}$H$_{10}^+$ & $2.4(\pm 0.8)\times 10^{-6}$ & \multicolumn{1}{c}{''}\\
 phenanthrene  & C$_{14}$H$_{10}^+$ & $1.7(\pm 0.5)\times 10^{-6}$ & \multicolumn{1}{c}{''}\\
 fluoranthene  & C$_{16}$H$_{10}^+$ & $3.0(\pm 0.9)\times 10^{-6}$ & \multicolumn{1}{c}{''}\\
 pyrene        & C$_{16}$H$_{10}^+$ & $4.1(\pm 1.2)\times 10^{-6}$ & \multicolumn{1}{c}{''}\\
\noalign{\smallskip}   
\hline
\noalign{\smallskip}   
naphtalene    & C$_{10}$H$_{8}^+$ & $0.3(\pm 0.1)\times 10^{-6}$ & $b$ \\
anthracene    & C$_{14}$H$_{10}^+$ & $1.1(\pm 0.5)\times 10^{-6}$ & $c$ \\
\noalign{\smallskip}
\hline                    
\end{tabular}
\tablefoot{$^a$\citet{Binnier2006_B516858A}; $^b$\citet{Abouelaziz1993JChPh..99..237A}; $^c$\citet{Novotny2005JPhCS...4..211N}}
\end{table}
\subsubsection{Electron recombination}

\begin{figure}[!ht] 
  \centering 
  \resizebox{\hsize}{!}{\includegraphics[angle=0,width=10cm,height=7cm,trim=50 65  100 50, clip]{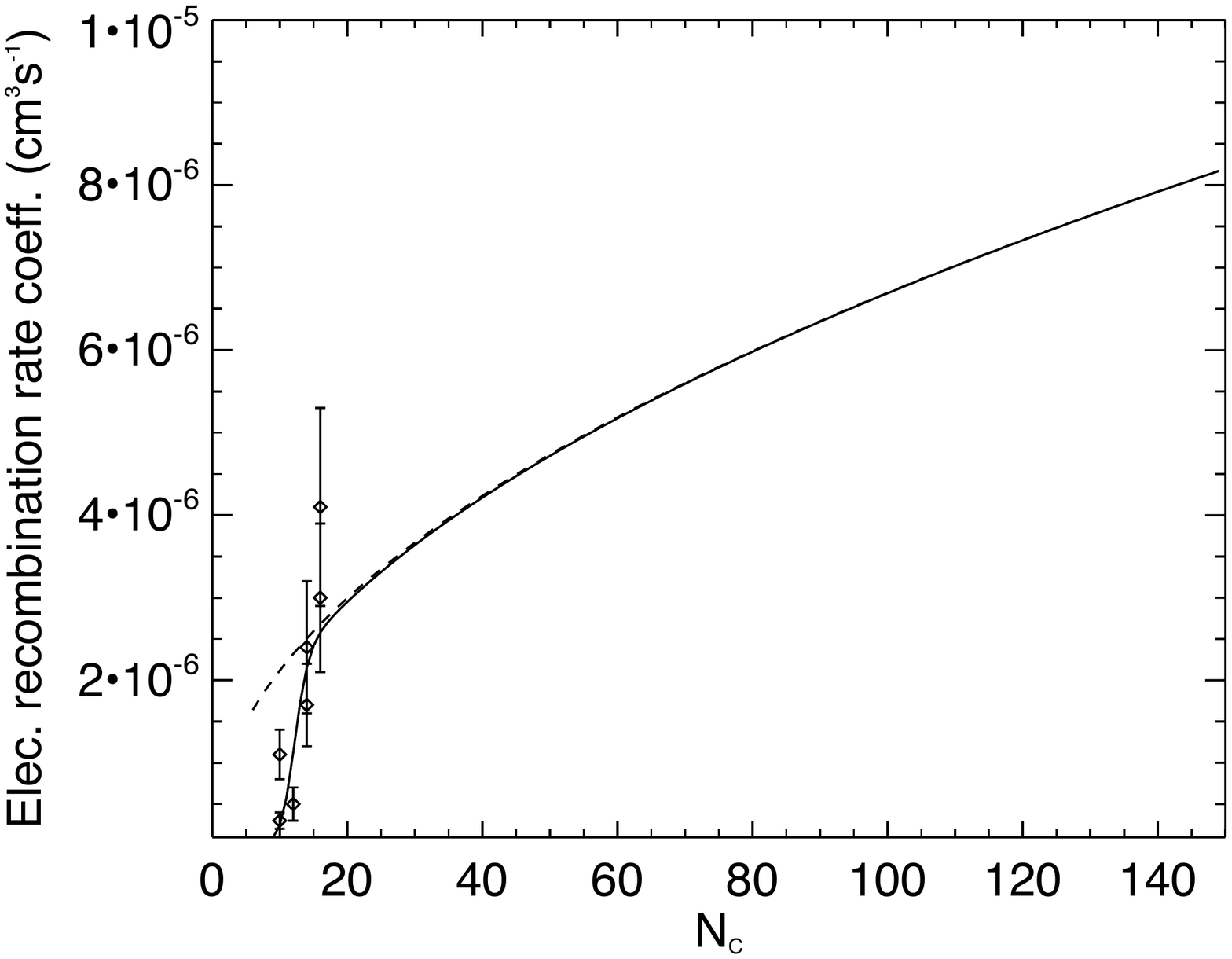}}
  \caption{PAH$^+$ cation electron recombination rate coefficient as
    function of the number of carbon atoms using equation
    \ref{eqn_ker} assuming $\phi_{\PAH}= 0.2$ an $T$=300~K
    (dash-line).  The experimental data from
    \citet{Binnier2006_B516858A} are listed in
    Table~\ref{PAH_recombination}. The solid line includes the
    additional correction $f(a_{\PAH})$ for PAHs of radius
    $a_{\PAH}$ applied to the adopted law to match the
    experimental data.}
  \label{fig_PAH_electron_recombination}          
\end{figure}

\begin{figure}[!htbp]
\centering
 \resizebox{\hsize}{!}{\includegraphics[angle=0,width=10cm,height=7cm,trim=50 65 70 50, clip]{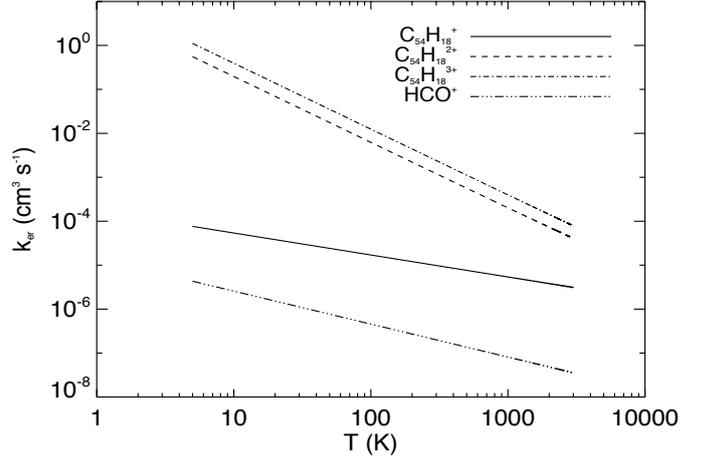}}
\caption{Electron recombination rate coefficients for the PAH cations.} 
  \label{fig_circumcoronene_recomb}          
\end{figure}

Ionised PAHs can recombine with electrons. The electron recombination rate with singly-ionised PAHs
\begin{equation}
\PAH^{+} + \elec \ra \PAH
\end{equation}
are calculated following a classical formalism by assuming that
PAH-cations and electrons interact via a Coulomb potential
\citep{Bakes1994ApJ...427..822B,Tielens2005pcim.book.....T}:
\begin{equation}
  k_{\mathrm{er}} = 4.1\times10^{-5}\phi_{\PAH}f(a_{\PAH})\left(\frac{N_{\mathrm{C}}}{50}\right)^{1/2}\left(\frac{100\mathrm{K}}{\Tg}\right)^{1/2}\ \mathrm{cm}^{3}\ \mathrm{s}^{-1}, \label{eqn_ker}
\end{equation}
where $N_{\mathrm{C}}$ is the number of carbon atoms and
$\phi_{\PAH}$ a correction factor for the disk shape:
$\phi_{\PAH}=\sigma_{\mathrm{disk}}/\sigma_{\mathrm{sphere}}$
between 0.1 and 0.8 \citep{Verstraete1990A&A...237..436V}. The
recombination is essentially not dissociative for the large PAHs
present in disks. An experimental study of the recombination of PAH
cations with electron has been performed by
\citet{Binnier2006_B516858A} at room temperature for seven small PAH
cations.
For multiply-ionised PAHs, the recombination rate is enhanced
\begin{equation}
k_{\mathrm{er}}'=k_{\mathrm{er}}\times\left(1+\frac{W_{\PAH}}{kT_{\mathrm{elec}}}\right),
\end{equation}
where the
$W_{\PAH}=IP_{\PAH}(Z_{\PAH})-IP_{\PAH}(0)$ is the work function and is equal to the difference in ionisation potential between the charged and neutral PAH If no experimental results exist and we assumed $T_{\mathrm{elec}}=T$.
We assumed that the recombination of PAH-H$^+$ cations are dissociative and  have the same rates as for PAH cations.
For PAH with fewer than 20 carbon atoms, an additional correction
factor has to be applied to match the experimental data
\begin{equation}
f(a_{\PAH}) = (1-\exp{(-12\times a_{\PAH}/l_{\elec})})/(1.0+\exp{(12-N_{\mathrm{C}})}),
\end{equation}
where $a_{\PAH}$ is the PAH radius in cm and $l_{\elec}=10^{-7}$ cm
\citep{Weingartner2001ApJS..134..263W}. Alternatively,
\citet{Flower2003MNRAS.343..390F} adopted a size-independent rate
\begin{equation}
  k_{\mathrm{er}} = 3.3\times10^{-6}\left(\frac{300\mathrm{K}}{T}\right)^{1/2}\ \mathrm{cm}^{3}\ \mathrm{s}^{-1}.
\end{equation}
Both rates can be reconciled if we take $\phi_{\PAH}\simeq
0.2$.  For PAHs with 10 carbon atoms or fewer, the recombinations are
dissociative
\citep{Abouelaziz1993JChPh..99..237A,Fournier2013JChPh.138o4201F}. The
comparison between the adopted rates and the laboratory data for
singly-ionised PAH cations is shown in
Fig.~\ref{fig_PAH_electron_recombination}. 
The derived recombination
rates assuming $\phi_{\PAH}\simeq 0.4$ for circumcoronene are
given in Table~\ref{circumcoronene_recombination}. The rate
coefficients as function of the gas temperature are shown in
Fig.~\ref{fig_circumcoronene_recomb}.
\begin{table}[th]
  \caption{Circumcoronene cation electron recombination rate coefficients ($\phi_{\PAH}=
    0.4$). By comparison the recombination rate coefficient for HCO$^+$ is also shown.}             
  \label{circumcoronene_recombination}    
\centering          
\begin{tabular}{ll}     
\hline
\hline
\noalign{\smallskip}        
\multicolumn{1}{c}{Formula} & \multicolumn{1}{c}{$k$ [cm$^{3}$ s$^{-1}$]} \\
\hline
\noalign{\smallskip}   
 C$_{54}$H$_{18}^+$   & $1.7 \times 10^{-5} \left(\frac{100\mathrm{K}}{T}\right)^{1/2}$\\
 C$_{54}$H$_{18}^{++}$ & $1.7 \times 10^{-5} \left(\frac{100\mathrm{K}}{T}\right)^{1/2}\left(1+\frac{36438.8}{T}\right)$ \\
 C$_{54}$H$_{18}^{+++}$ & $1.7 \times 10^{-5} \left(\frac{100\mathrm{K}}{T}\right)^{1/2}\left(1+\frac{72877.6}{T}\right)$\\
 HCO$^+$ & $3.0\pm0.9 \times 10^{-7} \left(\frac{T}{300\mathrm{K}}\right)^{-0.74\pm 0.02}$ \\
\noalign{\smallskip}
\hline                    
\end{tabular}
\end{table}
\subsubsection{Electron attachment}

\begin{figure}[!ht]
\centering
\resizebox{\hsize}{!}{\includegraphics[angle=0,width=10cm,height=7cm,trim=70 65 90 50, clip]{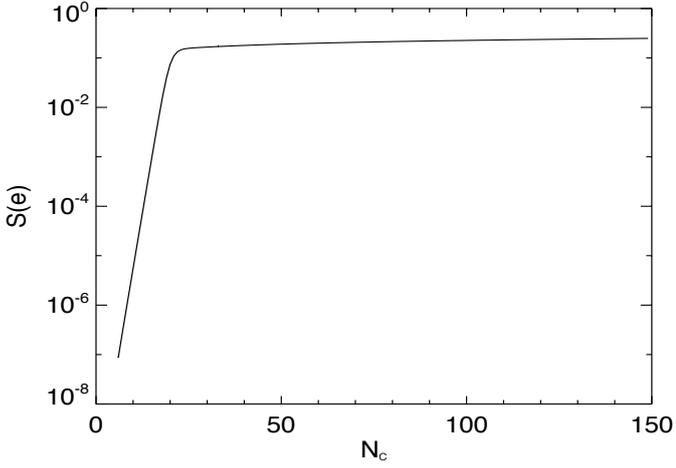}}
\caption{PAH electron sticking coefficient.} 
  \label{fig_PAH_electron_sticking}          
\end{figure}

\begin{figure}[!ht]
\centering  
\resizebox{\hsize}{!}{\includegraphics[angle=0,width=10cm,height=7cm,trim=70 65 90 50, clip]{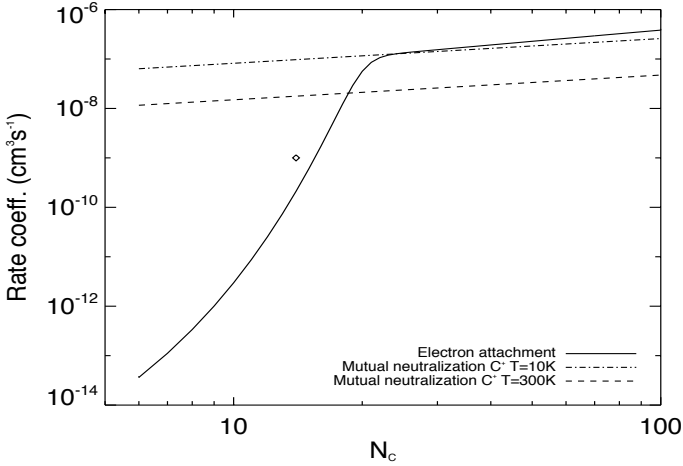}}
\caption{PAH rate coefficients. The solid line is the electron
  attachment rate coefficients. The diamond is the measured PAH
  attachment rate coefficient for anthracene
  \citep{Canosa1994CPL...228...26C}. The dashed- and dashed-dotted
  lines are the mutual neutralisation rates C$^+$ at two kinetic
  temperature.}
  \label{fig_PAH_rates}          
\end{figure}

Electrons can attach on neutral PAHs. Experiments have shown that the
electron attachment cross section is a strong function of the electron
affinity \citep{Tobita1992CP....161..501T}, which changes the electron
sticking coefficient $S_{\PAH}(\elec)$. The electron attachment rates
can be written as \citep{Allamandola1989ApJS...71..733A}
\begin{equation}
k_{\mathrm{ea}} = S_{\PAH}(\elec)k_{\mathrm{f}}\ratecoeff,
\end{equation}
where $k_{\mathrm{f}}$ is the electron capture rate. We adopt the
sticking coefficient analytical formula from
\citet{Weingartner2001ApJS..134..263W}
\begin{equation}
S_{\PAH}(\elec) = (1-e^{-a_{\PAH}/l_{\elec}})/(1+e^{(20-N_{\mathrm{C}})}),
\end{equation}
where $a_{\PAH}$ is the radius of the PAH in cm and $l_{\elec}$ is the
electron escape length equal to 10$^{-7}$ cm (see
Fig.~\ref{fig_PAH_electron_sticking}). The capture rate follows a
Langevin law and thus does not depend on the temperature:
\begin{equation}
k_{\mathrm{f}}=8.5\times10^{-7}\phi_{\PAH}\sqrt{N_{\mathrm{C}}}.
\end{equation}  
Large PAHs ($N_{\mathrm{c}}>$ 20) have electron affinity large enough
($\sim$1 eV) such that the electron sticking coefficient on PAHs
$S(\elec)$ is close to unity. The electron attachment rates are
approximated by
\begin{equation}
k_{\mathrm{ea}} = 8.5\times10^{-7}S_{\PAH}(\elec)\phi_{\PAH}\left(\frac{N_{\mathrm{C}}}{50}\right)^{1/2}\ \mathrm{cm}^{3}\ \mathrm{s}^{-1}.
\end{equation}
The rates are high compared to atomic electronic attachment rate
coefficients, whose values are 10$^{-16}$-10$^{-14}$ cm$^{3}$
s$^{-1}$.  The electron attachment rate as function of the number of
carbon $N_{\mathrm{C}}$ is displayed in Fig.~\ref{fig_PAH_rates}. The
formula is consistent with the measured value of 10$^{-9}$ cm$^{3}$
$^{-1}$ for anthracene (C$_{14}$H$_{10}$) by
\citet{Canosa1994CPL...228...26C}. Other measurements by
\citet{Moustefaoui1998FaDi..109...71M} show that the rates are
temperature independent and lie between 1 and 3$
\times$10$^{-9}$ cm$^{3}$ s$^{-1}$ for anthracene, consistent with our
choice. For $N_{\mathrm{C}}$=30, our rate is 10 times smaller 
than in other studies because they adopt a $N_{\mathrm{C}}^{3/4}$
variation \citep{Wakelam2008ApJ...680..371W,Omont1986A&A...164..159O}.
\subsubsection{PAH mutual neutralisation reactions}

The mutual neutralisation reactions between any atomic cation X$^+$
and PAH
\begin{equation}
\mathrm{X}^+ + \PAH^- \ra\mathrm{X}^*-\PAH \ra\mathrm{X} + \PAH^*,  
\end{equation}
where X$^*$ and PAH$^*$ are excited intermediate species, proceeds at the rate
\begin{equation}
  k_\mathrm{mn} = 2.9\times10^{-7}\phi_{\PAH}\left(\frac{12\ \mathrm{amu}}{m_{\mathrm{X}}}\right)^{1/2}\left(\frac{100\mathrm{K}}{T}\right)^{1/2}\left(\frac{N_{\mathrm{C}}}{50}\right)^{1/2}\ \mathrm{cm}^{3}\ \mathrm{s}^{-1},
\end{equation}
where $m_{\mathrm{X}}$ is the mass of species X and $amu$ is the value
of the atomic mass units. An alternative parametrisation has been proposed to match experimental mutual neutralisation of simple and clustered positive and negative ions \citep{Hickman1979}
\begin{equation}
  k_\mathrm{mn} = 5.33\times10^{-7}(T/300)^{-0.5}\mu^{-0.5}(E.A.)^{-0.4}\   \mathrm{cm}^{3}\ \mathrm{s}^{-1},
\end{equation}
where the reduced mass $\mu$ of the ion pair is in atomic mass units and the electron affinity $E.A.$ (electron detachment energy of the negative ion) is in eV.
The latter formula has been shown to be quite successful in fitting laboratory data \citep*{Miller1980, Smith1978}. 
 For HCO$^+$, $\mu \simeq$ 29 amu. The electron affinity is $\sim$ 1 eV for large PAHs. The temperature-dependence is the same for the two parameterisations.
For a molecular cation AH$^+$ the mutual
neutralisation can be dissociative:
\begin{equation}
\mathrm{AH}^+ + \PAH^- \ra\mathrm{AH}^*-\PAH \ra\mathrm{A} + \mathrm{H} + \PAH,  
\end{equation}
or not
\begin{equation}
\mathrm{AH}^+ + \PAH^- \ra\mathrm{AH}^*-\PAH \ra\mathrm{AH} + \PAH^*. 
\end{equation}
The branching ratio between the two modes is unknown. In the gas
phase, electronic recombinations are dissociative because the
ionisation potentials are higher than the A-H bond energy. In the case
of the recombination with negatively-charged PAHs the excess energy
can be transferred to the PAHs, which have enough vibrational modes to decay rapidly. 
In this work, both branches are assumed
to have the same probability (0.5). All atomic and
molecular ions react with PAH$^-$ and we adopt the first parametrisation. The rate at 10 and 300~K for a
carbon ions and with $\phi_{\PAH}=0.2$ is shown in
Fig.~\ref{fig_PAH_rates}.

\subsubsection{PAH collisional detachment reactions}

Neutral species can detach the electron from negatively-charged PAHs
\begin{equation}
\PAH^{-} + \mathrm{X} \ra\PAH + \mathrm{X} + \elec.
\end{equation}
The rate follows the prescription of \citet{Flower2003MNRAS.343..390F}
\begin{equation}
  k_\mathrm{nd} = 1.5\times10^{-8}\phi_{\PAH}\left(\frac{1\ \mathrm{amu}}{m_{\mathrm{X}}}\right)^{1/2}\left(\frac{N_{\mathrm{C}}}{50}\right)^{1/2}e^{-5500/T}\ \mathrm{cm}^{3}\ \mathrm{s}^{-1},
\end{equation}
with $\phi_{\PAH}$=0.2. The activation barrier is
$E_{\mathrm{a}}/k$=5500~K. Therefore PAH will be more neutralised by
this process in hot disk mid-planes.

\subsubsection{PAH charge exchange reactions}

Neutral and positively-charged PAHs can undergo charge exchanges with
ions.
\begin{equation}
\mathrm{X}^+ + \PAH^{n+} \ra\mathrm{X} + \PAH^{(n+1)+},
\end{equation}
where $n \ge$0. The criterium for the reaction to proceed is that the
reaction is energetically allowed, i.e. the ionisation potential (I.P.)
of species X is higher than that of PAH$^{n+}$.
\citet{Flower2003MNRAS.343..390F} adopted a reaction probability of
0.1 per collision for reactions with $n=0$. We adopt the rate coefficients from
\citet{Tielens2005pcim.book.....T} for $n=$0:
\begin{equation}
k_\mathrm{ce,0} = 2.9\times10^{-8}\phi_{\PAH}\left(\frac{1\ \mathrm{amu}}{m_{\mathrm{X}}}\right)^{1/2}\left(\frac{100\mathrm{K}}{T}\right)^{1/2}\left(\frac{N_{\mathrm{C}}}{50}\right)^{1/2}\ \mathrm{cm}^{3}\ \mathrm{s}^{-1}
\end{equation}
and
\begin{equation}
k_\mathrm{ce,n} = k_\mathrm{ce,0} \times \max\left(0,1-\frac{ne^{2}}{a_{\PAH}kT}\right)\ \mathrm{cm}^{3}\ \mathrm{s}^{-1},
\end{equation}
for positively-charged PAHs ($n>$0) where the effect of the repulsive potential between the two positively-charged species is taken into account.
Assuming that the PAH is a circumcoronene, the minimum temperature required to overcome the repulsion is
\begin{equation}
T_{\mathrm{min}}> \frac{e^2}{a_{\PAH}k} = 16710 \left(\frac{N_{\mathrm{C}}}{468}\right)^{-1/3}\ \mathrm{K}.
\end{equation}
For circumcoronene, there is no realistic gas temperature for which the rate $k_\mathrm{ce,n>0}$ is not null. Alternatively one can use
\begin{equation}
k_\mathrm{ce,n} = k_\mathrm{ce,0} 
\times \mathrm{max} \left(0,1-\frac{W_{\PAH}}{kT}\right)\ \mathrm{cm}^{3}\ \mathrm{s}^{-1}.
\end{equation}
For singly- positively-charge PAH circumcoronene $W_{\PAH}$ = 2.9 eV or 33 652.9 K. Both formulations show that only
large singly-charged PAHs can exchange charge with cations at high gas temperatures. In protoplanetary disk conditions, those reactions do not occur.

 \citet{Tielens2005pcim.book.....T} also considers a double electron
transfer from He$^+$ to neutral PAHs because of the high value of the
helium ioniation potential:
\begin{equation}
\mathrm{He}^+ + \PAH \ra\mathrm{He} + \PAH^{2+}+ \elec,
\end{equation}
with the rate
\begin{equation}
k_{\mathrm{di}}=1.1\times10^{-8}\phi_{\PAH}\left(\frac{N_{\mathrm{C}}}{50}\right)^{1/2}\ \mathrm{cm}^{3}\ \mathrm{s}^{-1}.
\end{equation}

Inversely, positively-charged PAHs can gain an electron from ions
\begin{equation}
\mathrm{X} + \PAH^{n+} \ra\mathrm{X}^+ + \PAH^{(n-1)+}.
\end{equation}
Among the major atomic and molecular species, only sodium (Na) has an
I.P. that is lower than that of the adopted PAH.
    
\subsection{PAH adsorption on grain surfaces}

In addition to the hydrogenation, neutral PAHs can be ionised or condense
(physisorption) onto grain surfaces with an adsorption energy that scales with the number of hydrogen and carbon atoms \citep{Kamp2017A&A...607A..41K}:
\begin{equation}  
  E_{\mathrm{PAH,des}}/k = 482\times (N_{\mathrm{C}}-N_{\mathrm{H}}) +  946\times N_{\mathrm{H}}\ \mathrm{K},\label{desorption_formula}
\end{equation}
where $E_{\mathrm{CC}}$ (=~482 K) is the fitted desorption energy per
graphene-like carbon ($N_{\mathrm{CC}}=N_{\mathrm{C}}-N_{\mathrm{H}}$), and $E_{\mathrm{CH}}$ (=~946 K) is the fitted energy per benzene-like carbon and its adjoining H-atom ($N_{\mathrm{CH}}$=$N_{\mathrm{H}}$). Graphenes are completely de-hydrogenated PAHs. Graphene-like carbons are C-atoms with three covalent bonds with carbons, whereas benzene-like carbon have two covalent bonds with carbons and one bond with a hydrogen atom. For circumcoronene ($N_{\mathrm{CC}}$=36 and $N_{\mathrm{CH}}$=18), the estimated desorption energy is 34380~K. The highest possible value is set by the heat of vaporisation for graphite at $H_{\mathrm{f}}/k\!=\!86240$~K \citep{Pierson1993}. \cite{Michoulier2018C8CP01175C} have calculated the binding energy of PAHs on water ice. For coronene, they found binding energies between 20330 and 28140~K, compared to a value of 17136~K derived using our formula. For simplification, we chose to use the same adsorption energy whatever the type of surface.

\section{Multi-hydrogenated PAHs}~\label{appendix_PAH}

\begin{figure}[!ht]   
\centering  
\resizebox{\hsize}{!}{\includegraphics[angle=0,width=10cm,height=8cm,trim=10 0 10 10, clip]{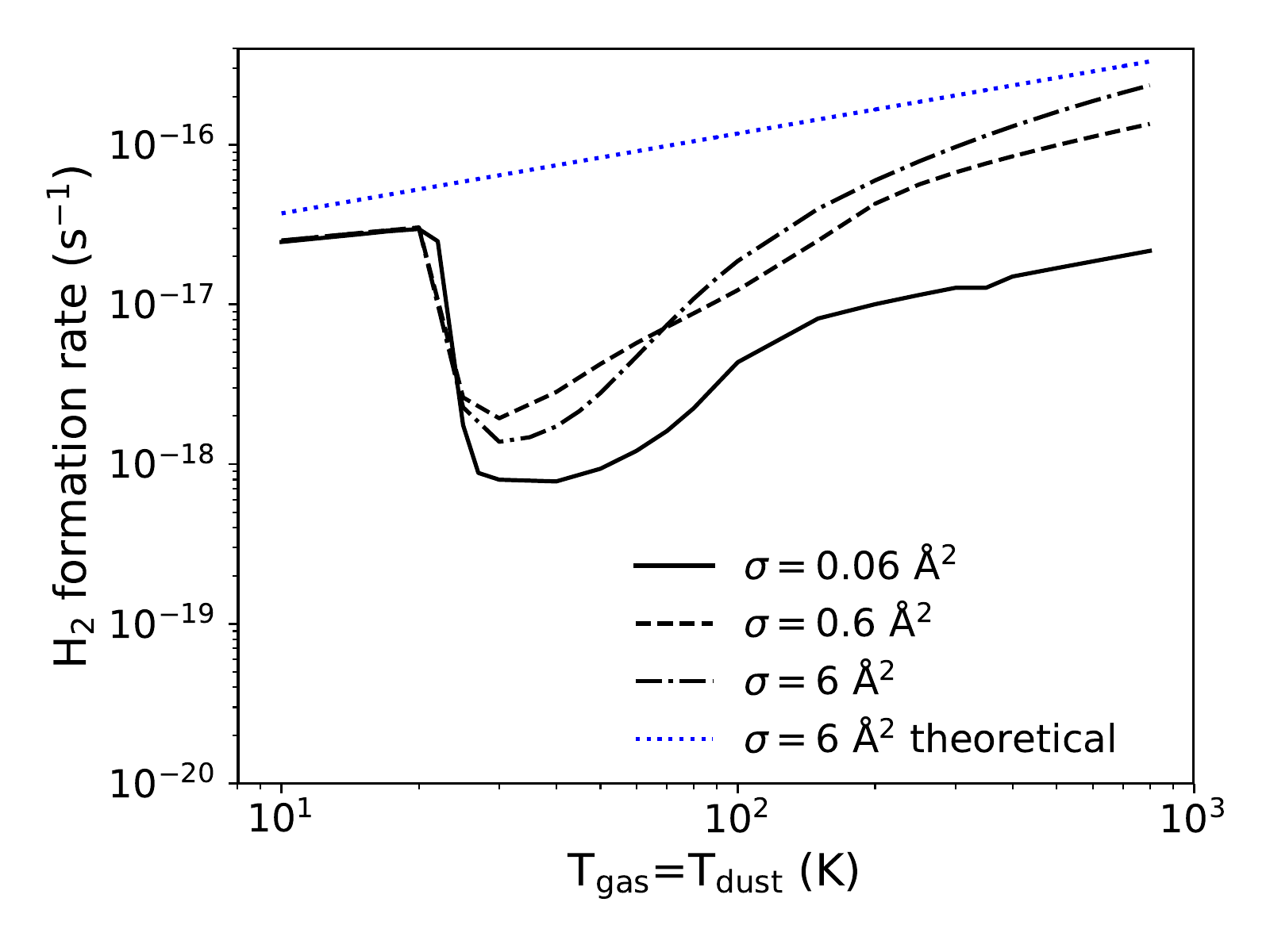}}
\caption{\H2 formation rate with adsorbed H on grains and multi-hydrogenated PAHs (up to PAH-H$_{18}$). The two series show the effect of assuming the abstraction cross-section from \citet{Mennella2012ApJ...745L...2M} and when
assuming a tenfold cross-section.}
  \label{fig_multi_PAH_H2_formation_rates}          
\end{figure}

We ran three series of molecular cloud models with physisorption only and \H2 formation on multi-hydrogenated PAHs (up to PAH-H$_{18}$) and without deuterium chemistry.
One series of models was run with a standard hydrogen abstraction cross-section of $\sigma$=0.06 $\AA^2$ and the other series with a tenfold cross-section ($\sigma$=0.6 $\AA^2$) and hundredfold cross-section ($\sigma$=6 $\AA^2$). Figure~\ref{fig_multi_PAH_H2_formation_rates} shows that unless a large cross-section is assumed, the \H2 formation through hydrogenated PAHs is not as efficient as via chemisorbed H atoms at temperatures greater than 100~K. The \H2 formation rate with large cross-sections are higher than on silicate grain surfaces for temperatures greater than 200~K. The high \H2 formation rate below 20~K is due to  the formation from physisorbed H-atoms on (icy) silicate grains. 
The asymptotic theoretical maximum \H2 formation rate can be estimated by
\begin{equation}
R_{\rm PAH-H_x,max}= \sigma  (k\Tg/(2\pi\,m_{\rm H}))^{1/2} (n_{\rm PAH}/n_{\rm H}) \bar{x},
\end{equation} 
where $\chi_{\rm PAH}=n_{\rm PAH}/n_{\rm H}$ is the PAH abundance and $\bar{x}$ the average
hydrogenation state of the PAHs. Introducing the numerical values, the maximum rate becomes
\begin{equation}	
R_{\rm PAH-H_x,max} \simeq 10^{-16} f_{\rm PAH}\left(\frac{\sigma}{10 \AA^2}\right)\left(\frac{\Tg}{100 {\rm K}}\right)^{1/2}\left(\frac{\bar{x}}{10}\right) \rate.
\end{equation} 
We plotted the theoretical maximum \H2 formation rate using this formula with $f_{\rm PAH}=1$, $x=18$, and $\sigma=6\ \AA^2$ in Fig.~\ref{fig_multi_PAH_H2_formation_rates}.

\section{Chemical network}~\label{appendix_chemical_network}

The species modeled in this study are shown in Table~\ref{tab:standard-species}.
\begin{table*}[!htbp]
\caption{Gas and solid species in the network.}
\label{tab:standard-species}
\vspace*{1mm} 
\begin{tabular}{cp{11cm}r}
\toprule
12 elements & H, He, C, N, O, Ne, Na, Mg, Si, S, Ar, Fe & \\
\noalign{\smallskip}
\hline
\noalign{\smallskip}
(H)         & H, H$^+$, H$^-$, {\bf H$_2$}, H$_2^+$, H$_3^+$, H$_2^{\rm exc}$          &  7 \\
(He)        & He, He$^+$,                                                           &  2 \\
(C-H)       & C, C$^+$, C$^{++}$, CH, CH$^+$, {\bf CH$_2$}, CH$_2^+$, 
              CH$_3$, CH$_3^+$, {\bf CH$_4$}, CH$_4^+$, CH$_5^+$,                     & 12 \\
(C-N)       & CN, CN$^+$, {\bf HCN}, HCN$^+$, HCNH$^+$                               & 5 \\  
(C-O)       & {\bf CO}, CO$^+$, HCO, HCO$^+$, {\bf CO$_2$}, CO$_2^+$, HCO$_2^+$,         & 7 \\
(N-H)       & N, N$^+$, N$^{++}$, NH, NH$^+$, NH$_2$, NH$_2^+$, {\bf NH$_3$}, NH$_3^+$, NH$_4^+$   & 9 \\
(N-N)       & {\bf N$_2$}, N$_2^+$, HN$_2^+$,                                                   & 3 \\
(N-O)       & NO, NO$^+$,                                                             & 2 \\  
(O-H)       & O, O$^+$, O$^{++}$, OH, OH$^+$, {\bf H$_2$O}, H$_2$O$^+$, H$_3$O$^+$,     & 8 \\
(O-O)       & {\bf O$_2$}, O$_2^+$,                                                  & 2 \\
(O-S)       & SO, SO$^+$, {\bf SO$_2$}, SO$_2^+$, HSO$_2^+$                          & 5 \\
(S-H)       & S, S$^+$, S$^{++}$,                                                  & 3 \\
(Si-H)      & Si, Si$^+$, Si$^{++}$, SiH, SiH$^+$, SiH$_2^+$,                     & 6 \\    
(Si-O)      & {\bf SiO}, SiO$^+$, SiOH$^+$,                                    & 3 \\    
(Na)        & Na, Na$^+$, Na$^{++}$,                                           &  3 \\
(Mg)        & Mg, Mg$^+$, Mg$^{++}$,                                           &  3 \\
(Fe)        & Fe, Fe$^+$, Fe$^{++}$,                                           &  3 \\
(Ne)        & Ne, Ne$^+$, Ne$^{++}$,                                           &  3 \\
(Ar)        & Ar, Ar$^+$, Ar$^{++}$,                                           &  3 \\
ice         & CO\#, H$_2$O\#, CO$_2$\#, CH$_4$\#, NH$_3$\#, SiO\#, SO$_2$\#, O$_2$\#, HCN\#, N$_2$\#,  & 10 \\
\noalign{\smallskip}
\hline
\noalign{\smallskip}
additional elements & PAH, * & 2\\
additional species & H\#,   \H2\#, *H\#, D, D+, D-, HD, D\#, HD\#, *D\#, H2D$^+$, HDO, HDO\#, PAH, PAH$^-$, PAH$^+$, PAH\#, PAH-H, PAH-H$^+$, PAH-H\#, PAHD, PAH-D\#, PAH-D$^+$ & 23\\
\noalign{\smallskip}
\hline
\noalign{\smallskip}
 species    & total                                                           & 123 \\
\toprule
\end{tabular}
\tablefoot{Closed-shell molecules are indicated in bold font, ices are indicated by a trailing \#, and chemisorbed species are lead by a * sign. CH$_2$ is a reactive closed-shell species.}
\end{table*}
\begin{table*}
\begin{center}
\caption{Main grain reactions involved in the formation and destruction of \H2  and HD. The energies are expressed in units of Kelvin.  \label{tab_dust_reactions}}
\vspace*{-3mm}          
\begin{tabular}{p{0.3cm}p{1cm}p{0.3cm}p{1cm}p{0.3cm}p{1cm}p{0.3cm}p{1cm}p{8cm}}     
\hline
\hline
\noalign{\smallskip}        
\multicolumn{8}{c}{Reaction} & \multicolumn{1}{c}{Comment}\\
\hline
\noalign{\smallskip}
1 &  & & H   & \rra & H\#    & & &     physisorption, barrierless\\
2 & & & D        & \rra & D\#    & &  &    -- \\
3 & H  & + & *    & \rra & *H\#   & &         & $E\gcH$~=~$\Eactc$~=~(400) 900--15,900~K \\
4 & H\#  &+&  *    & \rra & *H\#   & &        &  $E\pcH$~=~$\Eactc$ \\
5 & D  & + &*       & \rra & *D\# &&    &  $E\gcD$~=~$\Eactc$~+~\ED\\
6 & D\# & + & *      & \rra & *D\#   & &     & $E\pcD$~=~$\Eactc$~+~\ED\\
\noalign{\smallskip}
\hline
\noalign{\smallskip}
7 & & &H\#        & \rra & H    & &       & \EbHp ~=~600~K \\
8 & & &D\#          & \rra & D   & &       & \EbDp~=~\EbHp + \ED\\
9 & H\# & + &     \UV     & \rra & H    & &       & photodesorption \\
10 & D\# & + & \UV          & \rra & D   & &       & -- \\
11 & H\# & + &CR    & \rra & H    & &       &  \\
12 & D\# & + & CR         & \rra & D   & &       & \\
13 & & &*H\#        & \rra &  H   &+& *  & \EbHc~=~10,000--25,000~K \\
14 & & &*D\#        & \rra &  D   &+& *   & \EbDc~=~\EbHc + \ED \\
15 & *H\# & + & \UV     & \rra & H    & + &  *     &  \\
16 & *D\# & + & \UV          & \rra & D   & + & *      & \\
17 & *H\# & + &CR    & \rra & H    & + & *      &  via CR induced UV\\
18 & *D\# & + & CR         & \rra & D   & + & *      & -- \\
\noalign{\smallskip}
\hline
\noalign{\smallskip}
19 & H   &+ &H\#   & \rra & \H2     & &        & Eley-Rideal (ER) mechanism, barrierless \\
20 &  D &+& H\#    & \rra & HD  & &        & --  \\
21 & H  &+& D\#    & \rra & HD   & &           & -- \\
22 & H\#  &+& H\#   & \rra & \H2    & &        & $E\actHpHp$~=~0--250~K  \citep{ Navarro-Ruiz2014_C4CP00819G} \\
23 & H\#  &+& D\#   & \rra & HD   & &     & $E\actHpDp$~=~$E\actHpHp$~+~\ED~=~0--308~K  \\
\noalign{\smallskip}
\hline
\noalign{\smallskip}
24 & H   &+ &*H\#  & \rra & \H2      & + &  *     &   ER mechanism, barrierless \\
25 & H   & + & *D\#   & \rra & HD & +& *   &  -- \\
26 & D   &+ &*H\#   & \rra & HD  &+ &*   & -- \\
27 & H\#   &+ &*H\#  & \rra & \H2      & + & *      &    $E\actHpHc$~=~$\Eactc$\\
28 &  H\# & + &*D\#  & \rra & HD  &+&  *  &  $E\actHpHc$~=~$\Eactc$ + \ED\\
29  &  D\# & + &*H\#  & \rra & HD  &+&  *  &   $E\actDpHc$~=~$\Eactc$~+~\ED\\
30 & *H\# &+&  *H\# & \rra & \H2    & +& 2* & $E\actHcHc$~=~$2 \times \Eactc$\\
31 & *H\# & + &*D\#  & \rra & HD  &+& 2*  & $E\actHcDc$~=~$E\actHcHc$ + \ED\\
\noalign{\smallskip}
\hline
\noalign{\smallskip}
32 & \H2 & + & * & \rra & *H\# & + & H\# & \EdHHchem~=~3481~K \citep{Dino2004_713}\\ 
33 & HD & + & * & \rra & *D\# & + & H\# & \EdHDchem~=~\EdHHchem~+~\ED, 1/2 of the total dissociation rate\\
34 & HD & + & * & \rra & *H\# & + & D\# & \EdHDchem, 1/2 of the total dissociation rate\\
\noalign{\smallskip}
\bottomrule
\end{tabular}
\resizebox{170mm}{!}{
\begin{minipage}{170mm}{\ED=58~K (5 meV),   \ECD=970~K (83 meV), $\EHDHH$~=~415.8~K .}
\end{minipage}}
\end{center}
\end{table*}
\begin{table*}
\begin{center}
\caption{Main gas-phase and {\bf singly-hydrogenated} PAH reactions. The energies are expressed in units of Kelvin.  \label{tab_gas_reactions}}
\vspace*{-3mm}          
\begin{tabular}{p{0.3cm}p{1.5cm}p{0.3cm}p{1.5cm}p{0.3cm}p{1.5cm}p{0.3cm}p{1cm}p{7cm}}     
\hline
\hline
\noalign{\smallskip}        
\multicolumn{8}{c}{Reaction} & \multicolumn{1}{c}{Comment}\\
\hline
\noalign{\smallskip}
35 & & &  PAH        & \rra &  PAH\#    & &  &    physisorption \\
36 & & & PAH-H        & \rra & PAH-H\#    & &  &    -- \\
37 & & & PAH-D        & \rra & PAH-D\#    & &  &    -- \\
\noalign{\smallskip}
\hline
\noalign{\smallskip}
38 & & &  PAH\#         & \rra &  PAH    & &  &    see eq.~\ref{desorption_formula}\\
39 & & & PAH-H\#         & \rra & PAH-H    & &  &    -- \\
40 & & & PAH-D\#         & \rra & PAH-D    & &  &    -- \\
41 & PAH\# &+ & \UV           & \rra &  PAH    & &  &  photodesorption  \\
42 & PAH-H\#  &+ & \UV         & \rra & PAH-H    & &  &    -- \\
43 & PAH-D\#   &+ & \UV         & \rra & PAH-D    & &  &    -- \\
44 & PAH\# &+ & CR          & \rra &  PAH    & &  &    cosmic-ray induced photodesorption\\
45 & PAH-H\#  &+ & CR        & \rra & PAH-H    & &  &    -- \\
46 & PAH-D\#   &+ & CR        & \rra & PAH-D    & &  &    -- \\
\noalign{\smallskip}
\hline
\noalign{\smallskip}
47 &  H & + & PAH          & \rra & PAH-H  & & &  $E\actPAHH$~=~324~K \citep{Boschman2015AA...579A..72B}\\
48 &  D & + & PAH  & \rra & PAH-D  & & &  $E\actPAHD$~=~$E\actPAHH$~+~\ED, D-mass scaling of $1/\sqrt{2}$\\
49 & & & PAH-H   & \rra & H & + & PAH  &  thermal H-detachment with \EbPAHH~=~16,250~K \\
50 & & & PAH-D   & \rra & D & + & PAH  &  \EbPAHD~=~\EbPAHH~+~\ECD \\
51 &PAH-H  & + & \UV    & \rra & H & + & PAH  & photodetachment, $E$(C-H)~=~1.4 eV (16,250~K)\\
52 &PAH-D  & + & \UV    & \rra & D & + & PAH  &  $E$(C-D)~=~4.367~eV (50677.2~K)\\
53 & PAH-H & + & H & \rra & \H2   & + & PAH  & $\sigma=$0.06 $\AA^2$/C atom, $E^{\rm act}$ = 0~K$^a$    \\
54 & PAH-H & + & D & \rra & HD & + & PAH  &  ten-fold H cross-section $\sigma=$ 0.06--0.6 $\AA^2$/C atom \\
55 & PAH-D & + & H & \rra & HD & + & PAH  &  same as for PAH-H \\
56 &  H  & + & PAH$^+$  & \rra & PAH-H$^+$  & & &  $E\actPAHpH$~=~116~K\\
57 &  D  & + & PAH$^+$  & \rra & PAH-D$^+$  & & &  $E\actPAHpD$~=~$E\actPAHpH$~+~\ED\\
58 & PAH-H$^+$ & + & e & \rra & PAH   & + & H  & dissociative recombination  \\
59 & PAH-D$^+$ & + & e & \rra & PAH   & + & D  &   -- \\
60 & PAH-H$^+$ & + & H & \rra & PAH$^+$  & + &  \H2 & Langevin rate \citep{Montillaud2013AA...552A..15M}  \\
61 & PAH-H$^+$ & + & D & \rra & PAH$^+$   & + &  HD & same as for PAH-H$^+$ \\
62 & PAH-D$^+$ & + & H & \rra & PAH$^+$    & + & HD  &  same as for PAH-H$^+$ \\
63 & PAH-H$^+$ & + & HD & \rra & PAH-D$^+$   & + & \H2  &   \\
64 & PAH-D$^+$ & + & \H2 & \rra & PAH-H$^+$   & + & HD  &   endothermic, \Ereac~=~$\EHDHH$~-~\ECD\\
\noalign{\smallskip}
\hline
\noalign{\smallskip}
65 &  \H2  & + & PAH  & \rra & PAH-H  & + & H & \EdH2PAH-H-H~=~3481~K\\
66 &  HD  & + & PAH  & \rra & PAH-D  & + & H & \EdHDPAH~=~\EdH2PAH+$\EHDHH$\\
\noalign{\smallskip}
\hline
\noalign{\smallskip}
67 & PAH & + & \UV   & \rra & PAH$^+$    & + &   e  & photoionisation\\
68 & PAH     & + & e  & \rra & PAH$^-$    & &     & electron attachment\\
69 & PAH$^-$ & + & \UV   & \rra & PAH    &  + &   e  & photodetachement\\
70 & PAH$^+$ & + & e  & \rra & PAH    & &     & electron recombination \\
71 & PAH$^+$ & + & X  & \rra & PAH    & +  &  X$^+$   & charge exchange with species X \\
72 & PAH        & + & X$^+$  & \rra & PAH$^+$    & +  &  X & X=H and C, rates from \cite{Wolfire2008ApJ...680..384W}\\
\noalign{\smallskip}
\hline
\noalign{\smallskip}
73 &  H        & + & e$^{-}$ & \rra & H$^-$ & \UV  &  &radiative attachment\\
74 &  D        & + & e$^{-}$ & \rra & D$^-$ & \UV  &  &radiative attachment\\
75 &  H        & + & H$^-$ & \rra & \H2    & + & e$^-$  &    associative detachment    \\ 
76 &  H        & + & D$^-$ & \rra & HD  & + & e$^-$    &  associative detachment   \\ 
77 &  D        & + & H$^-$ & \rra & HD  & + & e$^-$    &  associative detachment    \\ 
78 &  H        & + &  D       & \rra & HD  & + &  \UV     &  radiative association \citep{Stancil1997ApJ...490...76S} \\  
79 & \H2   & + & D$^+$       & \rra & HD  &+ &H$^+$    &   \cite{Honvault2013JPCA..117.9778H}\\
80 & HD & + & H$^+$       & \rra & \H2    &+ &D$^+$    &   \Ereac~=~-416~K   \\
81 & \H2   & + & D        & \rra & HD  &+ &H    & \cite{Simbotin2011PCCP...1319148S}           \\
82 & HD    &+& H       & \rra & \H2    &+ &D    &  $\Delta E$~=~-416~K \\
83 &  H + H  &+& H        & \rra & \H2    &+ & H       & three-body reactions\\
84 &  H + H   &+& \H2   & \rra & \H2    &+ & \H2     & -- \\
85 & \H2   & + & \UV   & \rra & H    & + & H  & photodissociation including self-shielding   \\
86 &  HD      & + & \UV    & \rra & H    & +  & D & photodissociation including self-shielding  \\
87 & \H2 & + & CR& \rra & H & + & H & by secondary electrons and CR-generated UV photons\\ 
88 & HD & + & CR& \rra & D & + & D & -- \\
\noalign{\smallskip}
\bottomrule 
\end{tabular}
\resizebox{170mm}{!}{
\begin{minipage}{170mm}{$^a$ \cite{Mennella2012ApJ...745L...2M,Rauls2008ApJ...679..531R}. \ED=58~K (5 meV),   \ECD=970~K (83 meV), $\EHDHH$~=~415.8~K.}
\end{minipage}}
\end{center}
\end{table*}
\section{Transmission function}~\label{transmission_function}

\begin{figure}[!htbp]
  \centering
  \includegraphics[angle=0,width=9.0cm,height=7.5cm,trim=25 70  70 300, clip]{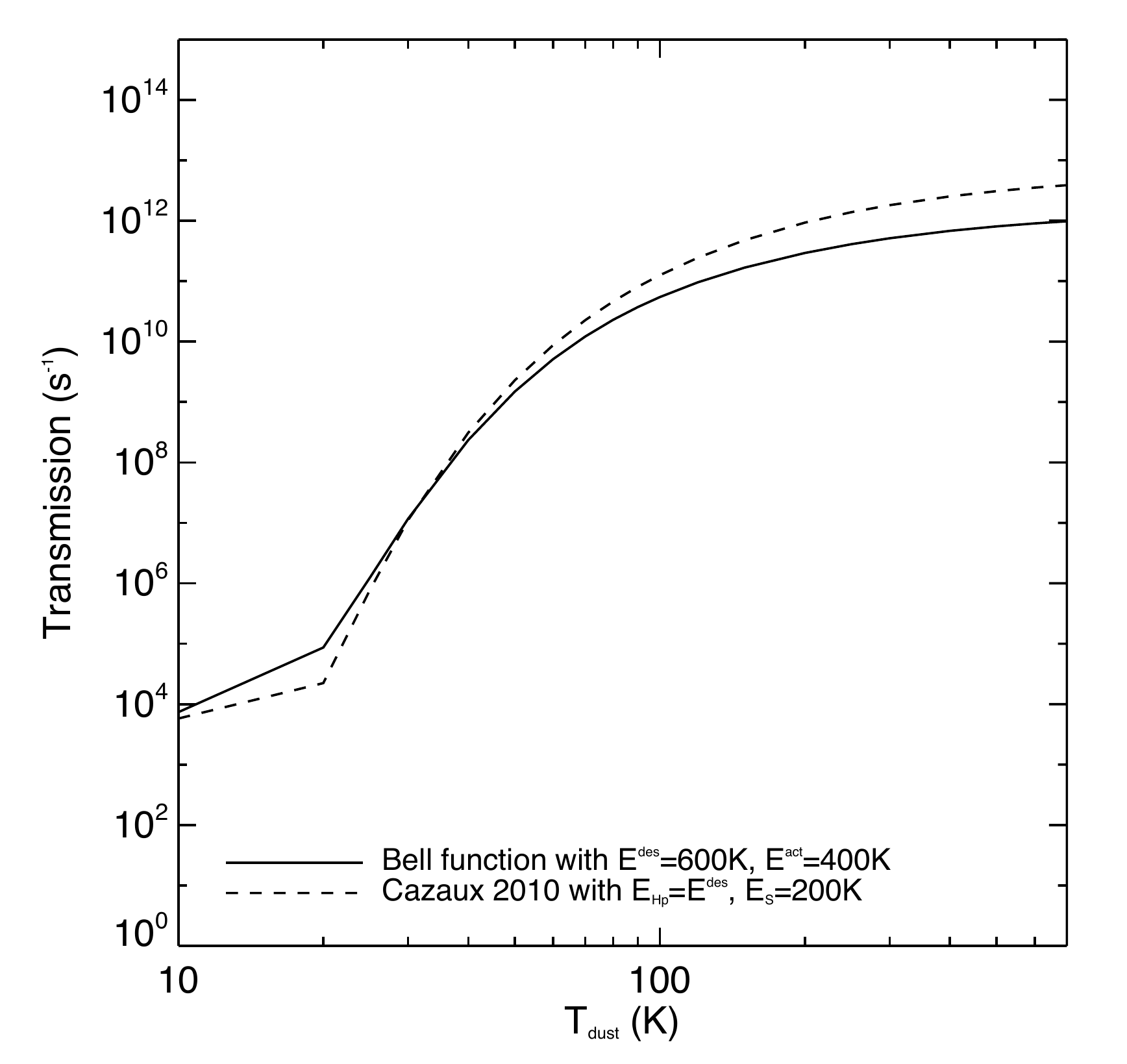}  
  \caption{Transmission function $\Rpc$ using the Bell formulation or the formulation in \cite{Cazaux2010ApJ...715..698C}. The adopted values are given in the figure, the chosen value for the quantum width is 2.5 \AA\ for silicate.}
  \label{fig_transmission_function}          
\end{figure}

Part of the difference in the \H2 formation efficiency can be explained by the choice of different transmission function. Fig.~\ref{fig_transmission_function} shows the transmission function for $\Rpc$ using the Bell formulation or the transmission function in \cite{Cazaux2010ApJ...715..698C}. The Bell formulation gives a higher transmission in the quantum tunnelling part below 20~K while the exponential pre-factor in the Cazaux formulation is higher than the Bell pre-factor at high temperatures.

\section{Tables of variable symbols}\label{appendix_table_symbols}

The main variable symbols are summarized in Table~\ref{tab_variables} and ~\ref{tab_variables2}.

\begin{table*}
\begin{center}
\caption{Variables used in the manuscript. \label{tab_variables}}
\vspace*{-0.5mm}          
{\begin{tabular}{lll}     
\toprule
Variable & Symbol & Units \& remarks\\ 
\noalign{\smallskip}     
\hline
\noalign{\smallskip}
Species $i$ mass &  $\mi$ & grams \\
Dust grain radius squared &  $\a2$ & cm$^2$, input parameter \\
Dust number density & $\nd$ & cm$^{-3}$\\
Number density of gas-phase species $i$&  $n_i$ & cm$^{-3}$ \\
Surface density of adsorption sites &  $\Ns$ & 1.5$\times$10$^{15}$ cm$^{-2}$, input parameter\\
Number of adsorption sites per ice monolayer & $\nsite$  & no units, $\nsite =4\pi \Ns \a2$ \\
Surface site cross-section & $\sigma_{\mathrm{surf,site}}$ &  $\sigma_{\mathrm{surf,site}}=1/\Ns$ cm$^{2}$\\
Total number density of chemisorption sites & $n\surfchem$ & $n\surfchem=\nsite \nd=4\pi \Ns \a2 \nd$ cm$^{-3}$\\
Number density of physisorbed species $i$ &  $\nice$ & cm$^{-3}$\\
Number density of chemisorbed species $i$&  $\nchem$ & cm$^{-3}$\\
Number density of unoccupied chemisorption sites & $n_{*}$ &  $n_{*}=n\surfchem - \!\sum_i \nchem$ cm$^{-3}$\\
Number density of chemically active physisorbed species $i$  &  $n\act$ & cm$^{-3}$\\
Number of chemically active physisorption layers & $\Nact$ & no units, an input parameter with typical value 1--10\\ 
Total number of density of physisorbed species (ice) &  $\nicetot$ & $\nicetot\!=\!\sum_i n_{\varhash,i}$ cm$^{-3}$\\
Number of physisorbed ice layers per grain &  $N\layer$ & no units, $N\layer  = \nicetot/(n_{\mathrm{d}}\ \nsite)$\\
Fraction of available adsorption sites &  $f_{\rm avail}$ & 1 for physisorption\\
     & &  for chemisorption $=n_*/n\surfchem$ if $N\layer<1$,$=0$ if $N\layer\ge1$\\
Gas temperature  & $\Tg$        &   K \\
Dust temperature & $\Td$        &   K \\
General activation barrier width &  $a_{\rm{\AA}}$ & \AA \\
Activation barrier &  $E_{\rm act}$ & K \\
Species mass in atomic mass units &  $m_{\rm amu}$ & amu\\
General adsorption rate & $R^{\rm ads}_i$ & s$^{-1}$ \\
General sticking coefficient &  $S_i $ & no units, between 0 and 1\\
General Bell's formula & $Q\Bell$ & no units\\
Physisorption rate &  $R\gp$ & s$^{-1}$ \\
Physisorption sticking coefficient & $S_{\rm phys}$ & no units, between 0 and 1\\
Species $i$ thermal speed & $v^{\rm th} _i$ & cm s$^{-1}$\\
Chemisorption activation energy &  $E\gc$ & ergs \\
Chemisorption rate &  $R\gc$ &s$^{-1}$  \\
Chemisorption Bell's function &  $Q\gc$ & no units \\
Chemisorption sticking coefficient &  $S_{\rm chem}$ & no units, between 0 and 1\\
Chemisorption activation barrier width &  $a\gc$ & cm, assuming a rectangular-shaped barrier \\
Desorption energy & $E\des$ & ergs\\
Binding energy  & $\Eb$ & ergs, physisorption and chemisorption\\
Desorption activation energy & $E\actdes$ & ergs, non-null for desorption from chemisorption sites\\
Surface vibration frequency & $\freq$ &  $\freq=\sqrt{(2 \Ns \Eb)/(\pi^2 m_i)}$ Hz \\
Activation barrier for desorption &  $E\actdes$ & ergs, non null for chemisorption\\
Desorption energy &  $E\des$ & ergs, for either the physisorbed (=$\Eb$) chemisorbed (=$\Eb+E\actdes$) \\
Total desorption rate from a physisorption site &  $R\despg$ & s$^{-1}$ \\
Thermal desorption rate from a physisorption site & $R\pgtherm$ & s$^{-1}$\\
Photodesorption rate from a physisorption site&  $R\pgph$ &  s$^{-1}$  \\
Photodesorption yield & $Y_i$ & no units\\
UV flux enhancement w.r.t. to the ISM value & $ \chi F_{\rm Draine}$ & 1 $\chi F_{\rm Draine}=$1.9921$\times$ 10$^8$ photons/cm$^2$/s\\
Cosmic-Ray induced desorption rate from a physisorption site &  $R\pgCR$ & s$^{-1}$ \\
Activattion barrier width for desorption from a chemisorption site & $a\actdes$ & cm\\
Total desorption rate from a chemisorption site &  $R\descg$ & s$^{-1}$ \\
Thermal desorption from a chemisorption site &  $R\cgtherm$ & s$^{-1}$ \\
Photodesorption rate from a chemisorption site&  $R\cgph$ & s$^{-1}$ \\
Cosmic-ray induced desorption from a chemisorption site &  $R\cgCR$ & $\simeq 0$ s$^{-1}$\\
Fraction of time a grain remains at 70~K up upon a cosmic-ray hit & $f(70K)$ & no units\\
Thermal desorption rate for a dust grain at 70~K & $R\pgtherm(70K)$ & s$^{-1}$\\
Thermal diffusion rate for a dust grain at 70~K &  $R\therdiff(70K)$ & s$^{-1}$\\
Thermal surface diffusion rate & $ R\therdiff$ & s$^{-1}$, includes quantum tunnelling effect\\
Cosmic-Ray induced surface diffusion rate & $ R\CRdiff$ & s$^{-1}$ \\
Bell's formula for the diffusion processes & $Q\diff$ & no units\\
Diffusion activate energy & $E\diff$ & ergs\\
Diffusion activation barrier width & $a\diff$ & cm\\
Total surface diffusion rate & $R\diff$ &  $R\diff  = R\therdiff +R\CRdiff$ s$^{-1}$,  $P\diff=R\diff/\freq$ \\
Activation barrier width for diffusion & $a\diff$ & cm\\
\noalign{\smallskip}
\bottomrule
\end{tabular}}
\vspace*{-2mm} 
\tablefoot{The index $i$ means that the variable applies to species $i$. Depending on the process, the species $i$ is a gas-phase, physisorbed, or chemisorbed species.}
\end{center}
\end{table*}
\begin{table*}
\begin{center}
\caption{Variables used in the manuscript, part 2. \label{tab_variables2}}
\vspace*{-0.5mm}          
{\begin{tabular}{lll}     
\toprule
Variable & Symbol & Units \& remarks\\ 
\noalign{\smallskip}     
\hline
\noalign{\smallskip} 
Hydrogen diffusion barrier between physisorption sites &  $E\diffH$ & ergs\\
Hydrogen diffusion barrier between chemisorption sites &  $E\diffHc$ & ergs \\
Deuterium diffusion barrier between physisorption sites &  $E\diffD$ & ergs \\
Deuterium diffusion barrier between chemisorption sites&  $E\diffDc$ & ergs \\
Physisorbed hydrogen desorption energy &  $E\desHp$ & ergs \\
Physisorbed deuterium desorption energy &  $E\desDp$ & ergs \\
Chemisorbed hydrogen desorption energy &  $E\desHc$ & ergs \\
Chemisorbed deuterium desorption energy &  $E\desHc$ & ergs \\
Surface reaction rate coefficient between species $i$ and $j$ & $k_{ij} $ & cm$^3$ s$^{-1}$ \\
Surface reaction probability &  $\kappa_{ij}$ & no units, between 0 and 1\\
Surface reaction activation barrier & $E\act$ & ergs\\
Surface reaction activation barrier width & $a^r_{ij}$ & cm\\
Hydrogen thermal desorption rate from a physisorption site & $R\pgH$ & s$^{-1}$\\
Hydrogen diffusion rate between physisorption sites & $R\diffH$ & s$^{-1}$\\
Hydrogen transfer rate from a physisorption to a chemisorption site & $R\pcH$ & s$^{-1}$\\
Acatavation barrier energy for H transfer from a physisorption to a chemisorption site & $E^{\rm act}\Hp$ & ergs\\
Bell's function for H transfer from a physisorption to a chemisorption site & $Q\pcH$ & no units \\ 
Hydrogen transfer rate from a chemisorption to a physisorption site & $R\cpH$ & s$^{-1}$\\ 
Eley-Rideal \H2 formation rate (physisorption site) & $R^{\rm gp}_{\rm H_2}$ & s$^{-1}$\\ 
Eley-Rideal \H2 formation rate (chemisorption site) &$R^{\rm gc}_{\rm H_2}$ & s$^{-1}$\\ 
\H2 formation rate after encounter between a physisorbed and a chemisorption H-atom & $R\HpHc$ & cm$^3$ s$^{-1}$\\
Analytical \H2 formation rate & $R_{\rm H_2}^{\rm Cazaux}$ & s${-1}$\\
 \noalign{\smallskip}   
 \hline
 \noalign{\smallskip}   
PAH effective radius & $a_{\PAH}$ & cm\\
PAH number of carbon atoms &  $N_C$ & no units\\
PAH number of hydrogen atoms &  $N_H$ & no units\\
PAH ionization potential &$ IP_{\PAH} $ & ergs\\ 
Hydrogenated PAH  hydrogen association rate coefficient &  $k_{\mathrm{PAH-H_x,H}} $ & cm$^3$ s$^{-1}$ \\
Hydrogen association on PAH-H$_x$ activation energy &  $E\actPAHH$ & ergs\\
Hydrogenated PAH hydrogen abstraction rate coefficient &  $k_{\mathrm{PAH-H_x}}$ & cm$^3$ s$^{-1}$\\
Effective temperature upon absorption of a photon of energy $\UVr$ &  $T_e$ & K\\
PAH internal temperature & $T_{\mathrm{PAH}}$ & K\\
Unimolecular PAH thermal dissociation rate at $T_e$ &  $R_{\mathrm{PAH-H_x,T_e}}$ &  s$^{-1}$ \\
Hydrogen binding energy on PAH-H$_x$  & $E_0$ & ergs, $\sim$ 1--2 eV\\
Yield for PAH-H$_x$ photodissociation  &  $Y_{\mathrm{PAH-H_x,UV}}$ & no units, =0 for $\UVr<E_0$ \\
Typical PAH IR photon emission rate     &  $R_{\mathrm{IR}}$ & $\sim$1 s$^{-1}$ \\
Thermal unimolecular dissociation rate &  $R_{\mathrm{PAH-H_x,therm}} $ & s$^{-1}$  \\
Hydrogen association rate coefficient for  (hydrogenated) PAHs cations & $k_{\mathrm{PAH-H_x^{n+},H}}$ & cm$^3$ s$^{-1}$\\
Activation energy for hydrogen association for PAH cations & $E\actPAHpH$ & ergs\\
Hydrogenated ionized PAH hydrogen abstraction rate coefficient &$k_{\mathrm{(PAH-H_x)^{n+}}}$ & cm$^3$ s$^{-1}$ \\
Energetic particle induced photoionisation rate coefficient &  $k_{\mathrm{pi,MeV}}$ & cm$^3$ s$^{-1}$\\
Energetic particle induced photodetachment rate coefficient &  $k_{\mathrm{pd,CR}}$ & cm$^3$ s$^{-1}$\\
Singly-ionized PAH cations electron recombination rate coefficient & $k_{\mathrm{er}}$ & cm$^3$ s$^{-1}$\\
Multiply-ionized PAH cations electron recombination rate coefficient & $k_{\mathrm{er}}'$ & cm$^3$ s$^{-1}$\\
PAH shape correction factor & $\phi_{\PAH}$ & $\phi_{\PAH}=\sigma_{\mathrm{disk}}/\sigma_{\mathrm{sphere}}$\\
PAH electron attachment rate coefficient & $k_{\mathrm{ea}}$ & cm$^3$ s$^{-1}$\\
Electron sticking coefficient on PAHs & $S_{\PAH}(\elec)$ & no units\\
Mutual neutralisation rate coefficient & $k_\mathrm{mn} $ & cm$^3$ s$^{-1}$\\
Electron collisional detachment rate coefficient &   $k_\mathrm{nd}$  & cm$^3$ s$^{-1}$\\
Neutral PAH charge-exchange reactions & $k_\mathrm{ce,0} $ & cm$^3$ s$^{-1}$\\
Charged PAH charge-exchange reactions & $k_\mathrm{ce,n}$ & cm$^3$ s$^{-1}$\\
Double charge transfer reaction & $k_{\mathrm{di}}$ & cm$^3$ s$^{-1}$\\
PAH adsorption/desorption energy & $E_{\mathrm{PAH,des}}$ & cm$^3$ s$^{-1}$\\
 \noalign{\smallskip}   
 \hline
 \noalign{\smallskip}   
Total \H2 formation rate according to \cite{Cazaux2002ApJ...575L..29C,Cazaux2004ApJ...604..222C} &$R_{\rm H_2}^{\rm Cazaux}$ & s$^{-1}$\\
\H2 recombination efficiency & $\epsilon$ & \citet{Cazaux2002ApJ...575L..29C}\\
Total \H2 formation rate according to \cite{Jura1974ApJ...191..375J,Jura1975ApJ...197..581J,Jura1975ApJ...197..575J} & $R_{\mathrm{H_2}^{\mathrm{Jura}}} $ & s$^{-1}$\\
\noalign{\smallskip}
\bottomrule
\end{tabular}}
\vspace*{-2mm} 
\tablefoot{See notes for Table~\ref{tab_variables}.}
\end{center}
\end{table*}

\end{appendix}

\end{document}